\documentclass{amsart}
\usepackage[utf8]{inputenc}

\usepackage{xcolor}
\usepackage{amsmath}
\usepackage{amssymb}
\usepackage{comment}
\usepackage{bbm}
\usepackage{mathrsfs}
\usepackage{graphicx}
\usepackage{algorithm}
\usepackage{algpseudocode}

\usepackage{geometry}
\usepackage{hyperref}

\usepackage{tikz}
\usetikzlibrary{positioning, calc}
\usepackage{float}

\allowdisplaybreaks

\title{Deep Learning for Operator Maps in Optimal Liquidation}
\author{Ryan Donnelly, Luca Galimberti, Brian Healy, Siqi Zhang}
\address{Luca Galimberti \\ 
King's College London\\
Department of Mathematics\\
Strand Building, WC2R 2LS, London, UK}
\email[]{luca.galimberti\@@kcl.ac.uk}
\address{Ryan Donnelly \\ 
King's College London\\
Department of Mathematics\\
Strand Building, WC2R 2LS, London, UK}
\email[]{ryan.f.donnelly\@@kcl.ac.uk}
\address{Brian Healy \\ 
Decision Science Limited\\
London, UK}
\email[]{brian\@@decisionscience.net}
\address{Siqi Zhang \\ 
King's College London\\
Department of Mathematics\\
Strand Building, WC2R 2LS, London, UK}
\email[]{siqi.5.zhang\@@kcl.ac.uk}

\newtheorem{theorem}{Theorem}[section]
\newtheorem*{theorem*}{Theorem}

\newtheorem{lemma}[theorem]{Lemma}
\newtheorem{proposition}[theorem]{Proposition}

\newtheorem{remark}[theorem]{Remark}

\newtheorem{Assumption}{Assumption}

\newcommand{\R}{\mathbb R}

\newcommand{\N}{\mathbb N}

\renewcommand{\P}{\mathbb P} 
\newcommand{\EE}{\mathbb E} 

\newcommand{\norm}[1]{\left\lVert#1\right\rVert}
\newcommand{\abs}[1]{\left |#1\right|}
\newcommand{\X}{\mathfrak X}

\DeclareMathOperator{\Supp}{supp}

\newcommand{\1}{\mathbbm{1}}

\newcommand{\PP}{\mathscr P}
\newcommand{\MM}{\mathscr M}

\providecommand{\keywords}[1]
{
  \small	
  \textbf{\textit{Keywords---}} #1
}

\date{\today}

\begin{document}
\begin{abstract}

We analyze the solution map of a optimal execution problem in which price dynamics are subject to a general transient impact kernel. Given the structure of the optimal trading strategy derived by Abi Jaber and Neuman in \cite{https://doi.org/10.1111/mafi.12465}, we show that the strategy is locally Lipschitz continuous with respect to the underlying model parameters governing temporary and transient price impact as well as the fundamental price process. Continuity of this map justifies the application of neural-net based approximations which take model parameters as inputs. We propose a method generating samples of the transient price impact kernel for the purposes of creating training data, and a method of projecting the kernel to a finite dimensional representation appropriate for input to a neural net, and a dimensional reduction based on the relationship between inventory and trade signal. We demonstrate the effectiveness of the trained neural net and use it to quantify the potential financial costs associated with misspecifying the transient impact kernel.

\end{abstract}


\maketitle

\tableofcontents

\section{Introduction}

The market trading phenomenon known as price impact can affect the value of asset holdings, depending on the schedule of submitted trades. Indeed, present and future prices of an asset will typically be lower if a trader makes a sequence of large sell actions than they would have been otherwise. Such effects manifest as a cost to the trader when employing a liquidation program. Therefore, in order to ensure efficient trade prices, these effects need to be incorporated into market model assumptions when designing trade algorithms.

Extensive empirical analysis of price impact effects and associated dynamics has been carried out in works such as \cite{biais1995empirical}, \cite{almgren2005direct}, \cite{cont2014price}, and 
\cite{bechler2017order}. Each of these works analyzes trade and quote data, or tick-by-tick data, to estimate models that capture the relationship between trade volumes and price dynamics. Further empirical analysis and modeling that consider phenomena beyond price impact can be found in \cite{sirignano2019deep}, \cite{lehalle2017limit}, and \cite{cartea2018enhancing}.

Preliminary work on investigating the effects of price impact on optimal trade scheduling appears in \cite{bertsimas1998optimal} and \cite{almgren2001optimal}. These works show how to balance the tradeoff between the cost of trading quickly due to price impact and the risk of trading slowly due to random future price changes. These findings come from framing the agent's liquidation objective as a dynamic stochastic control problem and then computing the optimal strategy for a risk-neutral or risk-averse trader. In the last few decades, the literature on optimal liquidation has extended significantly to include additional types of price impact and market features beyond price impact. See, for example, \cite{cartea2016incorporating} and \cite{lehalle2019incorporating} which incorporate a signal into the dynamics of the price process. The stochasticity of price impact and volatility is considered in, for example, \cite{almgren2012optimal}, \cite{cheridito2014optimal}, \cite{chan2015optimal}, and \cite{barger2019optimal}.

Transient price impact and resilience are effects that quantify how prices revert back to pre-trade values after having been subjected to impact due to market activity. Adapting optimal trading strategies to account for transient price impact has been explored in works such as \cite{bank2019optimal}, \cite{chen2019portfolio}, and \cite{ekren2019portfolio}. A very general form of transient price impact is considered in \cite{https://doi.org/10.1111/mafi.12465} on which the model of the present work is based. In that paper, the authors characterize the optimal strategy in terms of a backward stochastic differential equation (BSDE) and an operator-valued Riccati equation. In this work, we derive several regularity properties of the solution map that justify the application of a type of universal approximation theorem, thereby facilitating the training of a recurrent neural network (RNN) to learn how to compute the optimal trading strategy as a function of the infinite-dimensional model parameters.

Other work has applied machine learning techniques to algorithmic trading problems. Each of \cite{ning2021double}, \cite{cartea2021deep}, \cite{mounjid2024improving}, and \cite{macri2024reinforcement} considers a reinforcement learning (RL) paradigm in the context of optimal trade execution, either to take advantage of statistical arbitrage or to liquidate a large block of shares. Similar techniques are applied in \cite{gueant2019deep} to devise optimal market making strategies. In \cite{chen2024parametric}, instead of approaching the problem via RL, a neural net (NN) is trained to take model parameters and states as inputs. This NN represents the agent's value function, which is used in a dynamic programming sense to compute the optimal strategy, unlike our work in which the RNN represents the agent's trading strategy directly.

A closely related approach is developed in \cite{meng2025incontextoperatorlearninglinear}, where they consider a similar optimal liquidation framework with transient impact. Their method first uses an in-context operator network to infer an unknown price-impact operator from observed trajectories and then trains a separate neural-network policy for the inferred environment. Their numerical experiments consider exponential and power-law kernels. By contrast, in this paper we assume that the kernel is given and train a single RNN to approximate the solution directly across a broad class of admissible kernels.

A precise specification of price impact is difficult to deduce, and so any specification will naturally be subject to error. In particular, the transient price impact considered in \cite{https://doi.org/10.1111/mafi.12465} and the present work depends on a kernel function which must be estimated before the optimal strategy can be computed. After training an RNN to take any kernel as input, we quantify how a misspecification of the kernel can affect the resulting performance of the liquidation programme, which in essence determines how much value is gained by a more accurate transient impact kernel.

The remainder of the paper is organized as follows: Section \ref{sec:model} recalls the model with a general transient impact kernel of Abi Jaber and Neuman found in \cite{https://doi.org/10.1111/mafi.12465}, restates their results about the optimal trading strategy, and contains our main theoretical results. The proofs of these theoretical results are contained in Section \ref{sec: Main results}. Section \ref{sec:neural_network} describes the architecture and training methodology for the neural network that represents the solution map. Section \ref{sec:numerical_results} contains several numerical experiments to show the effectiveness of the neural network and quantify the effect that knowledge of transient impact has on trading performance. Section \ref{sec:conclusion} concludes, and some supplemental formulae are contained in the Appendix, Section \ref{sec: Appendix}.

\vspace{10mm}

\section*{Acknowledgements}
The numerical experiments in this work were performed using King's Computational Research, Engineering and Technology Environment (CREATE) at King's College London~\cite{king2022king}.

\section{Model setup and problem formulation}\label{sec:model}

We now recall the model considered by Abi Jaber and Neuman \cite{https://doi.org/10.1111/mafi.12465}, and formulate our problem in more detail. Let $0<T<\infty$ denote a fixed deterministic time horizon, and consider a stochastic basis $(\Omega,\mathcal{F},\{\mathcal{F}_t\}_{t\in[0,T]},\P)$. We fix a progressively measurable process $(P_t)_{t\in[0,T]}$, which plays the role of the unaffected price of the risky asset, and which satisfies
\begin{equation}\label{eq:integrability for P}
    \EE \left[
\sup_{0\le t\le r\le T} (\EE[P_r|\mathcal F_t])^2
\right]< \infty.
\end{equation}

This assumption in particular implies
\[
\EE \left[
\sup_{0\le t\le T} P_t^2
\right]< \infty,\quad\quad \EE \int_0^T P_t^2\,dt
< \infty.
\]

We then consider a trader with an initial position of $q\in\R$ shares in the risky asset. At time $0\le t\le T$, the number of shares held by the trader is
\[
Q^u_t := q - \int_0^t u_s\,ds, 
\]
where $(u_s)_{s\in[0,T]}$ is the trading speed of the trader, with a positive speed corresponding to selling, which is chosen from the set of admissible strategies
\begin{equation}\label{eq: A}
    \mathcal{A}:=\left\{
u; \, u\text{ progressively measurable s.t. } 
\EE \left[
\sup_{0\le t\le T} u_t^2\right]< \infty
\right\}.
\end{equation}
We endow this space with the norm
\[
\norm{u}_{\mathcal A} : = \left(\EE \left[
\sup_{0\le t\le T} u_t^2\right]\right)^{1/2},\quad u\in\mathcal A. 
\]
This makes $\mathcal{A}$ a normed space with respect to the natural norm $\norm{u}_{\mathcal A}:=\left(\EE 
\sup_{0\le t\le T} u_t^2\right)^{1/2}$, which embeds continuously in $L^2(\Omega\times [0,T])$. It is assumed that the trader's activity causes a price impact on the execution price of the risky asset. More precisely, the actual price $S^u$ induced by the trading rate $u$ at which the orders are executed is
\begin{equation}\label{eq: actual price}
    S^u_t:= P_t -\lambda u_t - Z^u_t,\quad 0\le t \le T,
\end{equation}
where
\begin{equation}\label{eq: Zu}
    Z^u_t:= h_0(t) + \int_0^t G(t,s)u_s\,ds,\quad 0\le t\le T,
\end{equation}
and $h_0:[0,T]\to\R$ is a continuous deterministic function and $G$ is a Volterra kernel. In the following, we mainly focus on kernels of the form
\[
G(t,s) = \mathbbm{1}_{s<t}H(t-s),
\]
with $H:(0,\infty)\to [0,\infty)$. Therefore, the trader's strategy not only instantaneously affects the execution price $S^u$ in an adverse manner via a linear temporary price impact $\lambda>0$, but it also induces a longer-lasting effect $Z^u$ due to linear transient price impact. Therefore, by taking into account these two kinds of impact as well as the general price of the asset, the trader's goal is to unwind her initial position $q\in\R$ through the maximization of the performance functional

\begin{equation}\label{eq: max J}
\max_{u\in\mathcal{A}}J(u),    
\end{equation}
where
\begin{equation}\label{eq: J(u)}
    J(u):= \EE \left[
    \int_0^TS^u_tu_t\,dt + Q^u_TP_T -\phi\int_0^T(Q^u_t)^2dt - \varrho(Q^u_T)^2
    \right].
\end{equation}
When it is necessary to emphasise the dependence on the model inputs, we write the performance functional as
\[
J^{h_0,H,\lambda,P,q}(u).
\]
Otherwise we retain the shorter notation $J(u)$.
We observe that $-\infty<J(u)<\infty$ for any $u\in\mathcal A$, and we recall that the first two terms describe the trader's terminal wealth, namely her final cash position, which is composed of the accrued trading costs and the risky asset position's final book value. The term involving the parameter $\phi\ge 0$ implements a penalty on the trader's running inventory, while the one involving the parameter $\varrho\ge 0$ represents a penalty on the terminal inventory. 

Despite the path-dependency of the model, the maximization problem \eqref{eq: max J} can be solved explicitly under very mild assumptions on the price impact kernel $G$. More precisely, we recall that a square integrable kernel $G:[0,T]^2\to [0,\infty)$ is a Volterra kernel if $G(t,s)=0$ for $s\ge t$; a Volterra kernel is nonnegative definite if for any $f\in L^2([0,T])$ we have
\[
\int_0^T\int_0^T (G(t,s)+G(s,t))f(s)f(t)\,dsdt\ge 0.
\]
In \cite{https://doi.org/10.1111/mafi.12465}, the following class $\mathcal G$ of admissible kernels is considered: a nonnegative definite Volterra kernel $G$ is in $\mathcal G$ if and only if 
\[
\sup_{t\le T}\int_0^T\abs{G(t,s)}^2ds + \sup_{s\le T}\int_0^T\abs{G(t,s)}^2dt<\infty,
\]
and
\[
\lim_{h\to 0}\int_0^T\abs{G(t+h,s)-G(t,s)}^2ds=0,\quad t\le T.
\]
We observe that a kernel of the form $G(t,s) = \mathbbm{1}_{s<t}H(t-s)$ with $H$ square integrable, $H \ge 0$ and satisfying
\begin{equation}\label{eq: nonnegative}
    \int_0^T\int_0^T H(\abs{t-s}) f(s)f(t)\,dsdt\ge 0,\quad f\in L^2([0,T]),
\end{equation}
is automatically in $\mathcal{G}$.

We then have the following:

\begin{theorem*}[Theorem 4.4 in \cite{https://doi.org/10.1111/mafi.12465}]
    Assume $G\in\mathcal{G}$, $h_0\in C([0,T])$, $\lambda>0$ and $\phi,\varrho\ge 0$. Then there exists a unique $u^\ast\in\mathcal{A}$ such that 
    \[
    J(u^\ast)=\max_{u\in\mathcal{A}}J(u).
    \]
    Furthermore, the optimal strategy $u^\ast$ is explicitly given by the solution to a linear Volterra equation of the form
    \begin{equation}\label{eq: uast}
        u^\ast_t=a_t+\int_0^tB(t,s)u^\ast_s\,ds,
    \end{equation}
    where $(a_t)_{t\in[0,T]}$ is a stochastic process that depends linearly on the price process $P$, and $B$ is a deterministic kernel, whose explicit expression will be given below; refer to Subsection \ref{sub: a and B}.
\end{theorem*}

In light of this last result, a natural question is to study the dependence of the optimal strategy $u^\ast$ on the parameters which specify the impact propagator model \eqref{eq: actual price} and \eqref{eq: Zu}, namely the kernel $G$, the deterministic function $h_0$, and the temporary impact parameter $\lambda$, as well as on the number of shares $q$ and the unaffected price $P$.

More precisely, after the usual identification of functions that agree almost everywhere, we consider the closed convex cone 
\begin{equation}\label{eq: H}
    \mathcal{H}:=\left\{H\in L^2([0,T]) \,\big|\;H\ge 0 \text{ and \eqref{eq: nonnegative} holds}  \right\} \subseteq L^2([0,T]).
\end{equation}
From what we have recalled above, clearly $\mathbbm 1_{s<t}H(t-s)\in\mathcal{G}$ for any $H\in\mathcal H$.

Consider the space $\X:= C([0,T])\times L^2([0,T])\times \R$ endowed with the natural norm 
\[
\norm{(h_0,H,\lambda)}_\X:= \max\{
\norm{h_0}_\infty, \norm{H}_{L^2},  \abs{\lambda}
\},\quad (h_0,H,\lambda)\in \X.
\]
Clearly, this is a Banach space. We define the space of the impact propagator models $\mathscr M$ as

\begin{equation}\label{eq: impact propagator model}
    \mathscr M:= C([0,T])\times \mathcal H\times (0,\infty) \subseteq \X,
\end{equation}
and we endow it with the topology inherited by $\X$. We also define the space of price processes 
\begin{equation}\label{eq: space of Prices}
    \mathscr P:=\left\{P=(P_t)_{t\in [0,T]}: \text{adapted cadlag process s.t. } \EE \left[
\sup_{0\le t\le r\le T} (\EE[P_r|\mathcal F_t])^2
\right]< \infty\right\},
\end{equation}
endowed with the norm
\[
\norm{P}_{\mathscr P}:= \left(\EE \left[
\sup_{0\le t\le r\le T} (\EE[P_r|\mathcal F_t])^2
\right]\right)^{1/2}, \quad P\in \mathscr P.
\]
\begin{remark}
    The original setting of \cite{https://doi.org/10.1111/mafi.12465} considers more in general progressively measurable processes. Here, we have restricted ourselves to the subset of adapted and cadlag processes for the following technical reason: in order to make the space $\mathscr P$ a normed space, we are required as usual to identify objects $\mathbb P\otimes dt$-a.e. $(\omega,t)$. However, should we wish to keep working in a progressively measurable setting, this identification would preclude us later to carry out pointwise evaluation, like for instance $P_T$ in the cost functional $J$. On the other hand, in an adapted and cadlag setting, one can always choose the original cadlag representative of the class, and therefore safely carry out pointwise operations.
\end{remark}

We finally consider the space
\[
\mathscr Y :=\mathscr M\times \mathscr P\times \R \subset \X\times\mathscr P\times\R,
\]
endowed with the product topology induced by the product norm
\[
\norm{((h_0,H,\lambda),P,q)}_{\X\times\mathscr P\times\R}:= \max\{
\norm{h_0}_\infty, \norm{H}_{L^2},  \abs{\lambda},
\norm{P}_{\mathscr P},\abs{q}
\}.
\]
We define the solution operator $\mathscr S$ as
\begin{equation}\label{eq: solution operator}
    \mathscr S: \mathscr Y \to \mathcal A, \quad ((h_0,H,\lambda),P,q) \mapsto u^\ast,
\end{equation}
where $u^\ast$ is the optimal trading strategy for the impact propagator model with parameters $(h_0,\mathbbm 1_{s<t}H(t-s),\lambda)$, unaffected price $P$ and initial position $q$: trivially, $\mathscr S$ is well-defined by Theorem 4.4 in \cite{https://doi.org/10.1111/mafi.12465}. In this definition, we have chosen to separate the role of the exogenous variables $(h_0,H,\lambda)$ and $P$ from the number of shares $q$, which is under the trader's control. In addition to these theoretical reasons, this choice is also motivated by the fact that the following proofs and arguments will become more straightforward.

\begin{remark}
    Given an arbitrary metric space $(\mathcal Z,d)$ and a point $z\in\mathcal Z$, $B_r^\mathcal{Z}(z)$ will denote the open ball of radius $r>0$ around $z$.
\end{remark}

Our first result shows locally Lipschitz continuity of $ \mathscr S: \mathscr M\times \mathscr P\times \R \to \mathcal A$ with $(P,q)\in\mathscr P\times \R$ fixed. 

\begin{theorem}\label{thm: main result}
    Assume $\phi,\varrho\ge 0$ and fix $(P,q)\in \mathscr P\times \R$. Then for every $(\hat{h}_0,\hat{H},\hat{\lambda})\in\mathscr M$ there exist $B_r^\MM(\hat{h}_0,\hat{H},\hat{\lambda})\subset\MM$ and a positive constant $L_\X=L_\X(\hat{h}_0,\hat{H},\hat{\lambda},P,q,r)$ such that
    \[
    \norm{\mathscr S(h_0',H',\lambda',P,q)-\mathscr S(h_0,H,\lambda,P,q) }_{\mathcal A}\le L_\X
\norm{(h_0',H',\lambda')-(h_0,H,\lambda)}_\X,
    \]
    for all $(h_0',H',\lambda'),(h_0,H,\lambda)\in B_r^\MM(\hat{h}_0,\hat{H},\hat{\lambda})$, and in particular $r<\frac{\hat{\lambda}}{5+4\sqrt{T}}$.
\end{theorem}
An expression for the constant $L_\X$ can be written by tracing its dependence through the proof of Theorem \ref{thm: main result} in the Appendix. We do not explicitly write this expression because our work only relies on the existence of such a constant and we do not study its dependence on underlying parameters.

The next two results reveal that if we freeze the variables $(h_0,H,\lambda)$, the operator $\mathscr S$ is globally Lipschitz continuous in $P$ and $q$.

\begin{theorem}\label{thm: stability wrt P}
    Assume $\phi,\varrho\ge 0$, and fix $(h_0,H,\lambda)\in\mathscr M$ and $q\in \R$. Then, for any $P,P'\in\mathscr P$ we have
    \[
\norm{\mathscr S(h_{0},H,\lambda,P,q) - \mathscr S(h_{0},H,\lambda,P',q)}_{\mathcal{A}} \le L_\PP(H,\lambda) \norm{P-P'}_{\mathscr P}.
\]
The precise expression of $L_\PP(H,\lambda)$ is provided by \eqref{eq: LP}. In particular, for any $\lambda_0,R>0$,
    \begin{equation}\label{eq: stability wrt P}
        \sup_{\mathrel{\substack{ \norm{H}_{L_2}\le R, \, \lambda\ge \lambda_0 \\ h_0\in C([0,T]),\,q\in\R    }}}    
    \norm{\mathscr S(h_{0},H,\lambda,P,q) - \mathscr S (h_{0},H,\lambda,P',q)}_{\mathcal{A}} \le C \norm{P-P'}_{\mathscr P},
    \end{equation}
    with $C=C(\phi,\varrho,T,\lambda_0,R)$. 
    
\end{theorem}

Similarly to Theorem \ref{thm: stability wrt P}, we have the following stability result for the initial number of shares $q$: 

\begin{theorem}\label{thm: stability wrt q}
    Assume $\phi,\varrho\ge 0$, and fix $(h_0,H,\lambda)\in\MM$ and $P\in\PP$. Then, for any $q,q'\in\R$ we have
    \[
    \norm{\mathscr S(h_{0},H,\lambda,P,q) - \mathscr S (h_{0},H,\lambda,P,q')}_{\mathcal{A}} \le L_\R(H,\lambda) \abs{q-q'}. 
\]
    The precise expression of $L_\R(H,\lambda)$ is provided by \eqref{eq: LR}. In particular, for any $\lambda_0,R>0$,
    \begin{equation}\label{eq: stability wrt q}
        \sup_{\mathrel{\substack{ \norm{H}_{L_2}\le R, \, \lambda\ge \lambda_0 \\ h_0\in C([0,T]),\,P\in\mathscr P    }}}    
    \norm{\mathscr S(h_{0},H,\lambda,P,q) - \mathscr S (h_{0},H,\lambda,P,q')}_{\mathcal{A}} \le C \abs{q-q'} ,
    \end{equation}
    with $C=C(\phi,\varrho,T,\lambda_0,R)$. 
    
\end{theorem}

By combining Theorems \ref{thm: main result}, \ref{thm: stability wrt P}, and \ref{thm: stability wrt q}, we also achieve local Lipschitz continuity of $\mathscr S$ in all its variables: 

\begin{theorem}\label{thm: joint continuity}
The nonlinear operator 
\[\mathscr Y\ni ((h_0,H,\lambda),P,q)\mapsto \mathscr S(h_0, H,\lambda,P,q)\in\mathcal A,
\] 
is locally Lipschitz continuous. Namely, for 
any arbitrary point $\hat{y}=((\hat{h}_0,\hat{H},\hat{\lambda}),\hat{P},\hat{q})\in \mathscr Y$, by choosing $r<\frac{\hat{\lambda}}{5+4\sqrt{T}}$, we have that $\mathscr S$ restricted to $B^\mathscr Y_r(\hat{y})\subset\mathscr Y$ is Lipschitz continuous with constant $C(\hat{y},r)$. The explicit expression of this constant can be recovered by means of the auxiliary functions defined in Subsection \ref{sub: list}.
\end{theorem}

In light of all the above, we also show that the performance functional $J$ evaluated at $u^\ast$ is locally Lipschitz continuous with respect to $(h_0,H,\lambda,P,q)\in \mathscr M\times\mathscr P\times \R$, namely,

\begin{theorem}\label{thm: continuity of J}
    Define $J^{\ast}:\mathscr M\times\mathscr P\times \R \to \R$ as
    \[
    J^{\ast}(h_0,H,\lambda,P,q) :=J^{h_0,H,\lambda,P,q}\circ\mathscr S(h_0,H,\lambda,P,q).
    \]
    Then for every $\hat{y}=((\hat{h}_0,\hat{H},\hat{\lambda}),\hat{P},\hat{q})\in \mathscr Y$, by choosing $r<\frac{\hat{\lambda}}{5+4\sqrt{T}}$, there exists a constant $C(\hat{y},r)$ such that 
    for all $y'=((h_0',H',\lambda'),P',q')$ and $y=((h_0,H,\lambda),P,q)\in B_r^\mathscr Y(\hat{y})$,
    \[
    \abs{J^\ast(h_0',H',\lambda',P',q') - J^\ast(h_0,H,\lambda,P,q)}\le C(\hat{y},r)\norm{y'-y}_\mathscr Y.
    \]

\end{theorem}

\subsection{Integral operators}

In this subsection, we are going to recall some notions and properties about operators, in particular, integral operators. We define $L^2([0,T]^2,\R^{2\times 2})$ as the space of measurable kernels $\Lambda:[0,T]^2\to\R^{2\times 2}$ such that
\[
\int_0^T\int_0^T \abs{\Lambda(t,s)}^2dsdt<\infty,
\]
where $\abs{\cdot}$ stands for any matrix norm. For convenience in subsequent sections, we use the operator norm of a matrix. Any kernel $\Lambda\in L^2([0,T]^2,\R^{2\times 2}) $ naturally induces an integral operator $\boldsymbol{\Lambda}$ via
\[
(\boldsymbol{\Lambda}g)(s):=\int_0^T \Lambda(s,u)g(u)\,du,\quad  g\in L^2([0,T],\R^2).
\]
It holds that
\[
\norm{\boldsymbol{\Lambda}g}_{L^2([0,T],\R^2)} \le \norm{g}_{L^2([0,T],\R^2)} \norm{\Lambda}_{L^2([0,T]^2,\R^{2\times 2})},\quad g\in L^2([0,T],\R^2).
\]
and thus $\boldsymbol{\Lambda}$ is a linear operator bounded from $L^2([0,T],\R^2)$ into itself with $\norm{\boldsymbol{\Lambda}}_{op}\le \norm{\Lambda}_{L^2([0,T]^2,\R^{2\times 2})}$ ($\norm{\cdot}_{op}$ denotes the operator norm).

For any $\Lambda_1,\Lambda_2\in L^2([0,T]^2,\R^{2\times 2})$, we define the $\star$-product as
\[
(\Lambda_1\star\Lambda_2)(s,u):=\int_0^T\Lambda_1(s,z)\Lambda_2(z,u)\,dz,\quad (s,u)\in [0,T]^2,
\]
which is a well-defined kernel in $L^2([0,T]^2,\R^{2\times 2})$. It can be proven that the integral operator induced by $\Lambda_1\star\Lambda_2$ is $\boldsymbol{\Lambda_1}\boldsymbol{\Lambda_2}$, namely the composition of $\boldsymbol{\Lambda}_1$ and $\boldsymbol{\Lambda}_2$. From this fact, it then follows that
\begin{equation}\label{eq: composition of integral operators}
\norm{\boldsymbol{\Lambda_1}\boldsymbol{\Lambda_2}}_{op}\le\norm{\boldsymbol{\Lambda_1} }_{op} \norm{\boldsymbol{\Lambda_2}}_{op}\le \norm{\Lambda_1}_{L^2([0,T]^2,\R^{2\times 2})}\norm{\Lambda_2}_{L^2([0,T]^2,\R^{2\times 2})},
\end{equation}
and
\begin{equation}\label{eq: composition of integral operators 2}
    \norm{\boldsymbol{\Lambda_1}\boldsymbol{\Lambda_2}}_{op}\le \norm{\Lambda_1\star\Lambda_2}_{L^2([0,T]^2,\R^{2\times 2})}.
\end{equation}
We denote by $\Lambda^\ast$ the adjoint kernel of $\Lambda$ for the scalar product of $L^2([0,T],\R^2)$, namely
\[
\Lambda^\ast(s,u)=\Lambda(u,s)^\top,\quad (s,u)\in [0,T]^2,
\]
and by $\boldsymbol{\Lambda}^\ast$ the corresponding adjoint integral operator. Trivially, $\norm{\boldsymbol{\Lambda}^\ast}\le \norm{\Lambda^\ast}_{L^2([0,T],\R^{2\times 2})}$. 

Mutatis mutandis, all these results clearly hold for a scalar kernel $\Lambda\in L^2([0,T]^2)$ as well. As we recalled above, a square integrable kernel $\Lambda:[0,T]^2\to [0,\infty)$ is a Volterra kernel if $\Lambda(t,s)=0$ for $s\ge t$; in addition, it is nonnegative definite if for any $f\in L^2([0,T])$ we have
\[
2\langle \boldsymbol{\Lambda}f,f \rangle_{L^2}=
\int_0^T\int_0^T (\Lambda(t,s)+\Lambda(s,t))f(s)f(t)\,dsdt\ge 0.
\]
We also recall that given a Hilbert space $(E,\langle\cdot,\cdot\rangle)$, $\ast$ is a linear isometry from $\mathcal{L}(E)$ to itself, where $\mathcal{L}(E)$ denotes the space of linear bounded operators from $E$ to $E$ endowed with $\norm{\cdot}_{op}$.

Finally, we observe that in many instances, we will be required to evaluate suprema of expressions of the form
\[
\sup_{0\le t\le T}\int_0^T\varphi(t,s)^2\,ds,
\]
where $\varphi\in L^2([0,T]^2)$ only (and so, technically speaking, an equivalence class). However, from \cite{https://doi.org/10.1111/mafi.12465} we know that for any of these expressions, it is always possible to find a representative in the equivalence class for which the supremum above is actually finite. In the following, we will adopt this convention throughout and tacitly.

\subsection{The process $a$ and the kernel $B$}\label{sub: a and B}

We provide the explicit expression for the stochastic process $a$ and the deterministic integral kernel $B$ that define the optimal strategy $u^\ast$ in \eqref{eq: uast}. For further details, we refer to \cite{https://doi.org/10.1111/mafi.12465}.

We set:
\begin{itemize}
    \item $\Tilde{h}_0(t):= h_0(t) -2\varrho q,\, 0\le t\le T$, where we recall that $q$ is the initial number of shares. 
    \item $K_t(s)\equiv K(s,t)\equiv (\Tilde{G}(s,t), -\mathbbm{1}_{t\le s} )^\top\equiv (2\varrho\mathbbm{1}_{t<s} + G(s,t), -\mathbbm{1}_{t\le s} )^\top \in\R^2,\,0\le s,t\le T$.
    \item $\tilde{\boldsymbol{G}}_t$ is the operator induced by the kernel $(s,u)\mapsto \Tilde{G}(s,u)\mathbbm 1_{u\ge t}$ (so here $u$ is the variable of integration (see \cite[before eq. (4.2)]{https://doi.org/10.1111/mafi.12465}).
    \item $\boldsymbol{1}_t$ is the integral operator induced by the kernel $(u,s)\mapsto \mathbbm 1_{u\ge s}\mathbbm{1}_{s\ge t}$ (so here $s$ is the variable of integration (see \cite[eq. (4.3)]{https://doi.org/10.1111/mafi.12465}).
    \item $\boldsymbol{D}_t:= 2\lambda \operatorname{id} + (\tilde{\boldsymbol{G}}_t + \tilde{\boldsymbol{G}}_t^\ast) +2\phi \boldsymbol{1}_t^\ast\boldsymbol{1}_t  $ where $\operatorname{id}$ is the identity operator on $L^2([0,T])$. If $\lambda>0$ and $\varrho,\phi\ge 0$, then $\boldsymbol{D}_t$ is a positive definite, self-adjoint, and invertible scalar operator from $L^2([0,T])$ into itself.
    \item The matrix operator
    \[
    \boldsymbol{\Gamma}_t^{-1} := 
    \begin{pmatrix}
        \boldsymbol{D}^{-1}_t & -2\phi \boldsymbol{D}^{-1}_t \boldsymbol{1}_t^\ast\\
        -2\phi \boldsymbol{1}_t\boldsymbol{D}_t^{-1} & -2\phi\operatorname{id} + 4\phi^2\boldsymbol{1}_t\boldsymbol{D}_t^{-1}\boldsymbol{1}_t^\ast
    \end{pmatrix}.
    \]
    \item We define $\Theta=\{\Theta_t(s);\,t\in[0,s],s\in[0,T]\}$ as
    \[
    \Theta_t(s):= - \left(
    \boldsymbol{\Gamma}_t^{-1} \1_{\cdot\ge t}\, \EE[P_\cdot - P_T|\mathcal F_t]\, \underline{e}_1
    \right)(s)\in\R^2
    \]
    where $\underline{e}_1=(1,0)^\top$.    
\end{itemize}
We then define ($\mathbbm{1}_t(s):=\mathbbm{1}_{s\ge t}$)
\begin{equation}\label{eq: a}
    a_t:= \frac{1}{2\lambda} \left(\EE[P_t-P_T |\mathcal F_t] -\tilde{h}_0(t) + \langle \Theta_t,K_t \rangle_{L^2} + \langle \boldsymbol{\Gamma}_t^{-1}K_t, \mathbbm{1}_{t} (\tilde{h}_0,q)^\top \rangle_{L^2}\right),
\end{equation}
and
\begin{equation}\label{eq: B}
    B(t,s) := \1_{s<t}\,\frac{1}{2\lambda} \left(
    \langle \boldsymbol{\Gamma}^{-1}_tK_t,\1_t(\tilde{G}(\cdot,s),-1)^\top\rangle_{L^2} - \tilde{G}(t,s)
    \right).
\end{equation}

We remark that both $a$ and $B$ depend on the exogenous variables $\lambda$, $G$, and $h_0$. From Lemma 7.1 in \cite{https://doi.org/10.1111/mafi.12465}, we have that
\[
\EE \left[\sup_{0\le t\le T} a^2_t\right]<\infty,\quad \quad
\sup_{0\le t\le T}\int_0^TB(t,s)^2ds<\infty.
\]

\section{Proof of main results}\label{sec: Main results}

In this section we prove all results that have been stated above.

\subsection{Strategy of proof of Theorem \ref{thm: main result}} The proof of Theorem \ref{thm: main result} is based on the two Propositions below plus a Gr\"onwall type argument. In the whole section, $(P,q)\in \mathscr{P}\times \R$ is fixed once and for all.

\begin{proposition}\label{prop: 1}
   For any $\hat{x}=(\hat{h}_0,\hat{H},\hat{\lambda})\in  \mathscr M$, by choosing $0<r<\frac{\hat{\lambda}}{5+4\sqrt{T}}$ the following holds: for any two arbitrary points $x=(h_0,H,\lambda)$ and $x'=(h'_0,H',\lambda')$ in the open ball $B^\MM_r(\hat{h}_0,\hat{H},\hat{\lambda}) \subset\MM$, we have
   \[
\EE\sup_{t\le T}\abs{a'_t-a_t}^2 \le 
\mathscr L_1(h_0,H,\lambda,h_0',H',\lambda',P,q) \norm{x'-x}^2_\X,
\]
where the function $\mathscr{L}_1$ is defined in \eqref{eq: L1} and
\[
 a'_{t}:= \frac{1}{2\lambda'} \left(\EE[P_t-P_T |\mathcal F_t] -\tilde{h}'_{0}(t) + \langle \Theta'_{t},K'_{t} \rangle_{L^2} + \langle (\boldsymbol{\Gamma}'_{t})^{-1}K'_{t}, \mathbbm{1}_{t} (\tilde{h}'_{0},q)^\top \rangle_{L^2}\right),
\]
and all terms in this expression involving any of the symbols $\lambda$, $H$, and $h_0$ are obtained by replacing these variables with $\lambda'$, $H'$, and $h'_{0}$ respectively.

\end{proposition}

\begin{proposition}\label{prop: 2}
Assume the same setting of the previous Proposition. Then
\[
\sup_{t\le T}\int_0^T\abs{B'(t,s)-B(t,s)}^2ds \le \mathscr L_{\Delta B}(H,\lambda,\lambda')\norm{x'-x}^2_\X,
\]
where $\mathscr L_{\Delta B}$ is defined in \eqref{eq: LDeltaB} and
\[
B'(t,s) := \1_{s<t}\,\frac{1}{2\lambda'} \left(
    \langle (\boldsymbol{\Gamma}'_t)^{-1}K'_{t},\1_t(\tilde{G}'(\cdot,s),-1)^\top\rangle_{L^2} - \tilde{G}'(t,s)
    \right),
\]
with the same convention as Proposition \ref{prop: 1}.

\end{proposition}
    
\subsubsection{Proof of Proposition \ref{prop: 1}} In this subsection, we prove Proposition \ref{prop: 1}, whose proof hinges upon several technical lemmas which provide some stability estimates in the terms appearing in the definitions of $a'_{t}$ and $a_t$. In the whole section, we are assuming the following:

\begin{Assumption}\label{hp: 1}
We fix an arbitrary $\hat{x}=(\hat{h}_0,\hat{H},\hat{\lambda})\in\mathscr M$ and we consider two arbitrary points $x=(h_0,H,\lambda)$ and $x'=(h'_0,H',\lambda')$ in the open ball $B^\MM_r(\hat{h}_0,\hat{H},\hat{\lambda}) \subset \MM$ of radius 
\[
0<r< \frac{\hat{\lambda}}{5+4\sqrt{T}}.
\]
\end{Assumption}

Additionally, we frequently (and most of the time, tacitly) make use of the elementary inequality
\begin{equation}\label{eq: elementary ineq}
    (c_1+\cdots +c_m)^2 \le m(c_1^2+\cdots +c_m^2),\quad c_1,\cdots,c_m\in\R,m\in\N.
\end{equation}

\begin{lemma}\label{lemma: 1}
Under Assumption \ref{hp: 1} it holds that
    \[
    \EE\sup_{t\le T}\abs{\tilde{h}'_{0}(t)-\tilde{h}_0(t)}^2=\norm{h_0'-h_0}_\infty^2.
    \] 
\end{lemma}
\begin{proof}
    Trivial.
\end{proof}

\begin{lemma}\label{lemma: 2}
Under Assumption \ref{hp: 1} it holds that
\[
\sup_{t\le T}\int_0^T \1_{s<t}\abs{H'(t-s)-H(t-s)}^2ds\le \int_0^T \abs{H'(r)-H(r)}^2dr,
\]
and
\[
\sup_{t\le T}\int_0^T \abs{K'(s,t)-K(s,t)}^2ds\le \int_0^T \abs{H'(r)-H(r)}^2dr,
\]
where $K'(s,t):= (\Tilde{G}'(s,t), -\mathbbm{1}_{t\le s} )^\top\equiv (2\varrho\mathbbm{1}_{t<s} + \1_{t<s} H'(s-t), -\mathbbm{1}_{t\le s} )^\top \in\R^2,\,0\le s,t\le T$.

\end{lemma}
\begin{proof}
    It holds uniformly in $0\le t \le T$ that
    \[
    \begin{split}
        \int_0^T \1_{s<t}\abs{H'(t-s)-H(t-s)}^2ds &= \int_0^t \abs{H'(t-s)-H(t-s)}^2ds\\
        &= \int_0^t \abs{H'(r)-H(r)}^2dr\\
        &\le \int_0^T \abs{H'(r)-H(r)}^2dr.
    \end{split}
    \]
    For the second statement, we observe that $\abs{K'(s,t)-K(s,t)}^2 = \1_{t<s}[H'(s-t)-H(s-t)]^2$, and so
    \[
    \begin{split}
        \int_0^T\abs{K'(s,t)-K(s,t)}^2ds &= \int_t^T\abs{H'(s-t)-H(s-t)}^2ds
        =\int_0^{T-t}\abs{H'(r)-H(r)}^2dr \\
        &\le \int_0^T \abs{H'(r)-H(r)}^2dr,
    \end{split}
    \]
    uniformly in $t$.    
\end{proof}

\begin{lemma}\label{lemma: 3}  
Under Assumption \ref{hp: 1}, upon defining 
\[
\boldsymbol{D}'_{t}:= 2\lambda' \operatorname{id} + (\tilde{\boldsymbol{G}'}_{t} + \tilde{\boldsymbol{G}'}_{t}^\ast) +2\phi \boldsymbol{1}_t^\ast\boldsymbol{1}_t,\quad 0\le t\le T,
\]
it holds that
\[
\sup_{t\le T}\norm{\boldsymbol{D}'_{t} - \boldsymbol{D}_t}_{op} \le 2\abs{\lambda'-\lambda} + 2\sqrt{T}\norm{H'-H}_{L^2}  .
\]
\end{lemma}
\begin{proof}
    By definition, it holds that
    \[
    \boldsymbol{D}'_{t} - \boldsymbol{D}_t = 2(\lambda' -\lambda)\operatorname{id} + (  \tilde{\boldsymbol{G}'}_{t} - \tilde{\boldsymbol{G}}_{t}) + 
    (\tilde{\boldsymbol{G}'}_{t}^\ast - \tilde{\boldsymbol{G}}_{t}^\ast).
    \]
    
    Therefore, uniformly in $t$, we have
    \[
    \begin{split}
        \norm{\boldsymbol{D}'_{t} - \boldsymbol{D}_t }_{op} &\le 2\abs{\lambda'-\lambda} + \norm{\tilde{\boldsymbol{G}'}_{t} - \tilde{\boldsymbol{G}}_{t}}_{op} + \norm{\tilde{\boldsymbol{G}'}_{t}^\ast - \tilde{\boldsymbol{G}}_{t}^\ast}_{op}\\
        &\le 2\abs{\lambda'-\lambda} + 2\norm{\tilde{\boldsymbol{G}'}_{t} - \tilde{\boldsymbol{G}}_{t}}_{op}\\
        &\le 2\abs{\lambda'-\lambda} + 2\left[
        \int_0^T\int_0^T\abs{\tilde{G}'(s,u)-\Tilde{G}(s,u)}^2\1_{u\ge t}\,dsdu
        \right]^{1/2}\\
        &\le 2\abs{\lambda'-\lambda} + 2\left[
        \int_0^T\int_0^T\abs{\tilde{G}'(s,u)-\Tilde{G}(s,u)}^2dsdu
        \right]^{1/2}\\
        &\le 2\abs{\lambda'-\lambda} + 2\left[
        \int_0^T\int_0^T\abs{G'(s,u)-G(s,u)}^2dsdu
        \right]^{1/2}\\
        &\le 2\abs{\lambda'-\lambda} + 2\left[
        \int_0^T\int_0^T\1_{u<s} \abs{H'(s-u)-H(s-u)}^2dsdu
        \right]^{1/2}\\
        &\le 2\abs{\lambda'-\lambda} + 2\sqrt{T}\norm{H'-H}_{L^2}.
    \end{split}
    \]
\end{proof}

\begin{lemma}\label{lemma: 4}
    Assume the same setting of Lemma \ref{lemma: 3}. Then it holds that
    \[
    \sup_{t\le T}\norm{(\boldsymbol{D}')^{-1}_{t} - \boldsymbol{D}^{-1}_t}_{op} 
    \le \frac{1}{\lambda^2} \left[\abs{\lambda'-\lambda} + \sqrt{T}\norm{H'-H}_{L^2}\right].
    \]
\end{lemma}
\begin{proof}
    We recall this general fact: if $E$ is a Hilbert space (say), $A:E\to E$ is a linear bounded and invertible operator, $B:E\to E$ is another linear bounded operator such that $\norm{B}_{op}<\norm{A^{-1}}^{-1}_{op}$, then it holds that also $A+B$ is invertible and 
    \begin{equation}\label{eq: A+B - A}
        \norm{(A+B)^{-1} - A^{-1}}_{op} \le \frac{\norm{B}_{op}\norm{A^{-1}}^2_{op} }{1-\norm{B}_{op}\norm{A^{-1}}_{op} }.
    \end{equation}
    From the proof of Lemma 4.1 \cite{https://doi.org/10.1111/mafi.12465}, we have that $\boldsymbol{D}_t$ is self-adjoint and that $\tilde{\boldsymbol{G}}_t + \tilde{\boldsymbol{G}}_t^\ast$ is non-negative definite. Therefore, we deduce that $\boldsymbol{D}_t \ge 2\lambda \operatorname{id}$, uniformly in $t\le T$. From this, and again by the fact that $\boldsymbol{D}_t$ is self-adjoint, we obtain from Lemma \ref{lemma: bound for the inverse} that $ \norm{\boldsymbol{D}_t^{-1}}_{op} \le 1/2\lambda,\, t\le T$, and so 
    $0<2\lambda \le \inf_{t\le T} \norm{\boldsymbol{D}_t^{-1}}_{op}^{-1}$.
    
    As by construction $r$ is such that $0<r<\frac{\hat{\lambda}}{5+4\sqrt{T}}$, it follows from Lemma \ref{lemma: 3} and the triangle inequality that 
    \[
    \sup_{t\le T}\norm{\boldsymbol{D}'_{t} - \boldsymbol{D}_t}_{op} \le 2\abs{\lambda'-\lambda} + 2\sqrt{T}\norm{H'-H}_{L^2} \le 4r(1+\sqrt{T})<\hat{\lambda}-r<\lambda.
    \]
    Then, by construction, for all $t\le T$ we have $\norm{\boldsymbol{D}'_{t} - \boldsymbol{D}_t}_{op} < \lambda < 2\lambda \le \norm{\boldsymbol{D}_t^{-1}}_{op}^{-1}$, and thus, in virtue of \eqref{eq: A+B - A},
    \[
    \begin{split}
    \norm{(\boldsymbol{D}')^{-1}_{t} - \boldsymbol{D}^{-1}_t}_{op} &\le \frac{\norm{\boldsymbol{D}'_{t} - \boldsymbol{D}_t}_{op}\norm{ \boldsymbol{D}_t^{-1}}^2_{op} }{1-\norm{\boldsymbol{D}'_{t} - \boldsymbol{D}_t}_{op}\norm{ \boldsymbol{D}_t^{-1}}_{op} }\\
    &\le \frac{1}{4\lambda^2}\, \frac{\norm{\boldsymbol{D}'_{t} - \boldsymbol{D}_t}_{op}}{1-\norm{\boldsymbol{D}'_{t} - \boldsymbol{D}_t}_{op}\norm{ \boldsymbol{D}_t^{-1}}_{op} } ,\quad t\le T.
    \end{split}
    \]
    But $\norm{\boldsymbol{D}'_{t} - \boldsymbol{D}_t}_{op}\norm{ \boldsymbol{D}_t^{-1}}_{op} < \lambda \cdot  \frac{1}{2\lambda}=\frac12$ for $t\le T$, and so
    \[
    \norm{(\boldsymbol{D}')^{-1}_{t} - \boldsymbol{D}^{-1}_t}_{op} <
    \frac{1}{4\lambda^2}\cdot 2 \norm{\boldsymbol{D}'_{t} - \boldsymbol{D}_t}_{op},\quad t\le T,
    \] 
    i.e. 
    \[
    \begin{split}
        \sup_{t\le T}\norm{(\boldsymbol{D}')^{-1}_{t} - \boldsymbol{D}^{-1}_t}_{op} &< \frac{1}{2\lambda^2} \sup_{t\le T} \norm{\boldsymbol{D}'_{t} - \boldsymbol{D}_t}_{op} \\
        &\le \frac{1}{\lambda^2} \left[\abs{\lambda'-\lambda} + \sqrt{T}\norm{H'-H}_{L^2}\right].
    \end{split}
    \]    
\end{proof}

Before proceeding, let us re-write now the stochastic process $\Theta$ in a more convenient form: since $\underline{e}_1=(1,0)^\top$, it holds for $0\le t\le s\le T$ that
\begin{equation}\label{eq: new form for theta}
    \Theta_t(s)= -
\begin{pmatrix}
    \left(\boldsymbol{D}^{-1}_t \mathbbm{1}_{\cdot\ge t}\, \EE[P_\cdot - P_T|\mathcal F_t]\right)(s) \\
    \left(-2\phi\boldsymbol{1}_t\boldsymbol{D}_t^{-1} \mathbbm{1}_{\cdot\ge t}\, \EE[P_\cdot - P_T|\mathcal F_t]\right)(s)
\end{pmatrix} 
.
\end{equation}

\begin{lemma}\label{lemma: 5}
    Assume the same setting of Lemma \ref{lemma: 3}. Then  
    \[
 \EE\sup_{t\le T}\abs{\langle \Theta'_{t} ,K'_{t}\rangle_{L^2} - \langle \Theta_{t} ,K_{t}\rangle_{L^2}}^2 
  \le [f_1(\lambda,H',P) + f_2(\lambda,P)] \norm{x-x'}^2_\X,
\]
where 
    \begin{equation}\label{eq: new form for theta n}
        \Theta'_{t}(s)= -
\begin{pmatrix}
    \left((\boldsymbol{D}')^{-1}_{t} \mathbbm{1}_{\cdot\ge t}\, \EE[P_\cdot - P_T|\mathcal F_t]\right)(s) \\
    \left(-2\phi\boldsymbol{1}_t(\boldsymbol{D}')_{t}^{-1} \mathbbm{1}_{\cdot\ge t}\, \EE[P_\cdot - P_T|\mathcal F_t]\right)(s)
\end{pmatrix}
,
    \end{equation}
with $K'_{t}(s)\equiv K'(s,t)$ defined in \ref{lemma: 2}, and where $f_1$ and $f_2$ are defined in \eqref{eq: f1} and \eqref{eq: f2} respectively.

Additionally, the following a priori estimate holds:
\begin{equation}\label{eq: apriori est theta K}
    \EE\sup_{t\le T}\abs{ \langle \Theta_{t} ,K_{t}\rangle_{L^2}}^2 
  \le f_3(H,\lambda,P),
\end{equation}
where $f_3$ is defined in \eqref{eq: f3}. 
\end{lemma}
\begin{proof}
    It holds that
\[
\begin{split}
    \frac{1}{2} \,&\EE\sup_{t\le T}\abs{\langle \Theta'_{t} ,K'_{t}\rangle_{L^2} - \langle \Theta_{t} ,K_{t}\rangle_{L^2}}^2 
    \\ & \le \EE\sup_{t\le T}\abs{\langle \Theta'_{t} ,K'_{t}\rangle_{L^2} - \langle \Theta_{t} ,K'_{t}\rangle_{L^2}}^2 + \EE\sup_{t\le T}\abs{\langle \Theta_{t} ,K'_{t}\rangle_{L^2} - \langle \Theta_{t} ,K_{t}\rangle_{L^2}}^2
    \\ & = \EE\sup_{t\le T}\abs{\langle \Theta'_{t} - \Theta_t ,K'_{t}\rangle_{L^2}}^2 + \EE\sup_{t\le T}\abs{\langle \Theta_{t} ,K'_{t} - K_t\rangle_{L^2}}^2
    \\ &  = \EE \sup_{t\le T}\left(\int_t^T [\Theta'_{t}(s)- \Theta_t(s)]\cdot K'(s,t)\,ds   \right)^2 
    \\ & \quad\quad
    + \EE \sup_{t\le T}\left(\int_t^T \Theta_t(s)\cdot [K'(s,t)-K(s,t)]\,ds   \right)^2 
    \\ & \le \EE \sup_{t\le T} \left(
    \int_t^T \abs{\Theta'_{t}(s)- \Theta_t(s)}^2ds\int_t^T\abs{K'(s,t)}^2ds
    \right) 
    \\ & \quad\quad+  \EE \sup_{t\le T} \left(
    \int_t^T \abs{ \Theta_t(s)}^2ds\int_t^T\abs{K'(s,t)-K(s,t)}^2ds
    \right) 
    \\ & \le \EE \sup_{t\le T} \left(
    \int_t^T \abs{\Theta'_{t}(s)- \Theta_t(s)}^2ds
    \right) \sup_{t\le T} \int_t^T\abs{K'(s,t)}^2ds 
    \\ & \quad\quad+  \EE \sup_{t\le T} \left(
    \int_t^T \abs{ \Theta_t(s)}^2ds 
    \right)  \sup_{t\le T} \int_t^T\abs{K'(s,t)-K(s,t)}^2ds
    \\ & := (I) + (II).
\end{split}
\]
By means of \eqref{eq: new form for theta} and \eqref{eq: new form for theta n}, we can write 
\[
\begin{split}
    &\int_t^T \abs{\Theta'_{t}(s)- \Theta_t(s)}^2ds\\ & \quad \le \int_0^T  \abs{\left( [(\boldsymbol{D}')^{-1}_{t} - \boldsymbol{D}^{-1}_t] \,[\1_{\cdot\ge t}\EE[P_\cdot - P_T | \mathcal F_t]]\right)(s)   }^2ds 
    \\ & \quad \quad + 4\phi^2\int_0^T \abs{\left( \boldsymbol{1}_t[(\boldsymbol{D}')^{-1}_{t} - \boldsymbol{D}^{-1}_t] \,[\1_{\cdot\ge t}\EE[P_\cdot - P_T | \mathcal F_t]]\right)(s)   }^2ds    
    \\ & \quad \le \norm{(\boldsymbol{D}')^{-1}_{t} - \boldsymbol{D}^{-1}_t  }_{op}^2 \int_t^T \EE[P_r - P_T | \mathcal F_t]^2dr 
    \\ & \quad \quad + 4\phi^2 \norm{\boldsymbol{1}_t[(\boldsymbol{D}')^{-1}_{t} - \boldsymbol{D}^{-1}_t]}_{op}^2 \int_t^T \EE[P_r - P_T | \mathcal F_t]^2dr. 
\end{split}
\]
As $\norm{\boldsymbol{1}_t[(\boldsymbol{D}')^{-1}_{t} - \boldsymbol{D}^{-1}_t]}_{op}\le \norm{\boldsymbol{1}_t}_{op} \norm{(\boldsymbol{D}')^{-1}_{t} - \boldsymbol{D}^{-1}_t}_{op}\le \frac{T-t}{\sqrt{2}} \norm{(\boldsymbol{D}')^{-1}_{t} - \boldsymbol{D}^{-1}_t}_{op} $, we conclude that
\[
\int_t^T \abs{\Theta'_{t}(s)- \Theta_t(s)}^2ds \le (1+2\phi^2 T^2) \norm{(\boldsymbol{D}')^{-1}_{t} - \boldsymbol{D}^{-1}_t}_{op}^2 \int_t^T \EE[P_r - P_T | \mathcal F_t]^2dr,
\]
and, by setting $R:= (1+2\phi^2 T^2) \sup_{t\le T}\norm{(\boldsymbol{D}')^{-1}_{t} - \boldsymbol{D}^{-1}_t}_{op}^2 $,
\[
\sup_{t\le T}\int_t^T \abs{\Theta'_{t}(s)- \Theta_t(s)}^2ds \le R\, \sup_{t\le T} \int_t^T \EE[P_r - P_T | \mathcal F_t]^2dr.
\]
We observe that by expanding the conditional expectation, we obtain
\[
\begin{split}
  \int_t^T \EE[P_r - P_T | \mathcal F_t]^2dr &= \int_t^T \left(\EE[P_r | \mathcal F_t] - \EE[P_T | \mathcal F_t]\right)^2dr 
   \le 2  \int_t^T \left(\EE[P_r | \mathcal F_t]^2 + \EE[P_T | \mathcal F_t]^2\right)dr
  \\ & \le 2\int_t^T \EE[P_r|\mathcal F_t]^2
\,dr + 2T\EE[P_T|\mathcal F_t]^2,
\end{split}
\]
and so
\[
\begin{split}
    &\sup_{t\le T}\int_t^T \abs{\Theta'_{t}(s)- \Theta_t(s)}^2ds 
    \le 2R\left\{
    \sup_{t\le T}\int_t^T \EE[P_r|\mathcal F_t]^2\,dr + 
    T\sup_{t\le T}\EE[P_T|\mathcal F_t]^2
    \right\}
    \\&\quad\le 2R\left\{
    \sup_{t\le T}\left[(T-t) \sup_{t\le r\le T}
    \EE[P_r|\mathcal F_t]^2\right] + 
    T\sup_{0\le t\le r\le T}\EE[P_r|\mathcal F_t]^2
    \right\}
    \\&\quad\le 2R\left\{
     T \sup_{0\le t\le r\le T}
    \EE[P_r|\mathcal F_t]^2  + 
    T\sup_{0\le t\le r\le T}\EE[P_r|\mathcal F_t]^2
    \right\}.
\end{split}
\]
By taking the expectation, we have
\[
    \EE \sup_{t\le T}\int_t^T \abs{\Theta'_{t}(s)- \Theta_t(s)}^2ds 
    \le 4R
    T\EE\left[
    \sup_{0\le t\le r\le T}
    \EE[P_r|\mathcal F_t]^2\right] = 4RT\norm{P}^2_{\mathscr P}.
\]
Thus,
\[
(I) \le 4RT \norm{P}^2_{\mathscr P} \sup_{t\le T} \int_t^T\abs{K'(s,t)}^2ds \le  4RT \norm{P}^2_{\mathscr P} \left[2\norm{H'}^2_{L^2} + (1+8\rho^2)T\right],
\]
because, arguing as in Lemma \ref{lemma: 2}, it holds that $\sup_{t\le T} \int_t^T\abs{K'(s,t)}^2ds\le \left[2\norm{H'}^2_{L^2} + (1+8\varrho^2)T\right]$. Thus, by means of Lemma \ref{lemma: 4} we obtain 
\[
(I)\le \frac12 f_1(\lambda,H',P)
\norm{x'-x}_\X^2.
\]
It remains to deal with $(II)$: analogously to what we have just done, we can bound  
\[
(II)\le \frac12 f_2(\lambda,P) \norm{x'-x}^2_\X,
\]
and eventually
\[
 \EE\sup_{t\le T}\abs{\langle \Theta'_{t} ,K'_{t}\rangle_{L^2} - \langle \Theta_{t} ,K_{t}\rangle_{L^2}}^2 
  \le [f_1(\lambda,H',P) + f_2(\lambda,P)] \norm{x-x'}^2_\X.
\]
In a similar way, one can show \eqref{eq: apriori est theta K} and the Lemma is proven.     
\end{proof}

We now deal with the last term in \eqref{eq: a}, namely with $\langle \boldsymbol{\Gamma}_t^{-1}K_t, \mathbbm{1}_{t} (\tilde{h}_0,q)^\top \rangle_{L^2}$. Let us expand it first. It holds that
\[
\begin{split}
    \boldsymbol{\Gamma}_t^{-1}K_t &= \boldsymbol{\Gamma}_t^{-1}
    \begin{pmatrix}
        2\varrho\mathbbm{1}_{t<\cdot} + G(\cdot,t), \\
        -\mathbbm{1}_{t\le \cdot} 
    \end{pmatrix}\\
    & = 
    \begin{pmatrix}
        \boldsymbol{D}^{-1}_t2\varrho\1_{t<\cdot} + \boldsymbol{D}^{-1}_tG(\cdot,t) + 2\phi\boldsymbol{D}^{-1}_t\boldsymbol{1}^\ast_t \1_{t\le\cdot} \\
        -4\phi\varrho\boldsymbol{1}_t\boldsymbol{D}^{-1}_t\1_{t<\cdot} - 2\phi\boldsymbol{1}_t\boldsymbol{D}^{-1}_tG(\cdot,t) + 2\phi\operatorname{id}\1_{t\le\cdot} - 4\phi^2\boldsymbol{1}_t\boldsymbol{D}^{-1}_t\boldsymbol{1}^\ast_t\1_{t\le\cdot}
    \end{pmatrix}.
\end{split}
\]
Therefore, the $L^2$-scalar product can be expanded as 
\begin{equation}\label{eq: expansion of L2 scalar product}
    \begin{split}
     &\langle \boldsymbol{\Gamma}_t^{-1}K_t, \mathbbm{1}_{t} (\tilde{h}_0,q)^\top \rangle_{L^2} \\
     & \quad\quad =\int_t^T\left[
     (\boldsymbol{D}^{-1}_t2\varrho\1_{t<\cdot})(s) + (\boldsymbol{D}^{-1}_tG(\cdot,t))(s) + (2\phi\boldsymbol{D}^{-1}_t\boldsymbol{1}^\ast_t \1_{t\le\cdot})(s)
     \right]\tilde{h}_0(s)\,ds \\
     & \quad\quad\quad +q\int_t^T\left[
     (-4\phi\varrho\boldsymbol{1}_t\boldsymbol{D}^{-1}_t\1_{t<\cdot})(s)  + (- 2\phi\boldsymbol{1}_t\boldsymbol{D}^{-1}_tG(\cdot,t))(s)\right]ds \\
     &\quad\quad\quad + 
     q\int_t^T\left[  (2\phi\operatorname{id}\1_{t\le\cdot})(s)+ (- 4\phi^2\boldsymbol{1}_t\boldsymbol{D}^{-1}_t\boldsymbol{1}^\ast_t\1_{t\le\cdot})(s)
     \right]ds,
    \end{split}
\end{equation}
and we remind that this quantity is deterministic, and therefore there is no need for $\EE$. We are now ready to prove 

\begin{lemma}\label{lemma: 6}
    Assume the same setting of Lemma \ref{lemma: 3}. Then,
    \[
    \begin{split}
    &\sup_{t\le T} \abs{\langle (\boldsymbol{\Gamma}')_{t}^{-1}K'_{t}, \mathbbm{1}_{t} (\tilde{h}'_{0},q)^\top \rangle_{L^2} - \langle \boldsymbol{\Gamma}_t^{-1}K_t, \mathbbm{1}_{t} (\tilde{h}_0,q)^\top \rangle_{L^2}  }^2
    \\&\quad\le   
    \left\{
g_1(\lambda',\lambda,H') + g_2(\lambda,h_0,q) + g_3(\lambda,h_0',q) + g_4(\lambda,H',q) + g_5(\lambda,q)
\right\}^2 \norm{x'-x}^2_\X,
    \end{split}
    \]
where $g_i,i=1,\dots,5$ are defined in equations \eqref{eq: g1},$\dots$,\eqref{eq: g5}.
Additionally, the following a priori estimate holds:
\begin{equation}\label{eq: apriori gamma K h0}
 \sup_{t\le T} \abs{\langle \boldsymbol{\Gamma}_t^{-1}K_t, \mathbbm{1}_{t} (\tilde{h}_0,q)^\top \rangle_{L^2}  }^2 \le g_6(h_0,H,\lambda,q),
\end{equation}
where $g_6$ is defined in \eqref{eq: g6}. 

\end{lemma}

\begin{proof}
    Using the expansion above,
    \begin{equation*}
        \begin{split}
            \mathcal{I}&:= \abs{\langle \boldsymbol{(\Gamma}'_{t})^{-1}K'_{t}, \mathbbm{1}_{t} (\tilde{h}'_{0},q)^\top \rangle_{L^2} - \langle \boldsymbol{\Gamma}_t^{-1}K_t, \mathbbm{1}_{t} (\tilde{h}_0,q)^\top \rangle_{L^2}  }\\
            &\quad\le \abs{\int_t^T ((\boldsymbol{D}')^{-1}_{t}2\varrho\1_{t<\cdot})(s)\tilde{h}'_{0}(s) - (\boldsymbol{D}^{-1}_t2\varrho\1_{t<\cdot})(s)\tilde{h}_0(s)\,ds}\\
            &\quad\quad + \abs{\int_t^T ((\boldsymbol{D}')^{-1}_{t}G'(\cdot,t))(s)\tilde{h}'_{0}(s) - (\boldsymbol{D}^{-1}_tG(\cdot,t))(s)\tilde{h}_0(s)\,ds}\\
            &\quad\quad + \abs{\int_t^T(2\phi(\boldsymbol{D}')^{-1}_{t}\boldsymbol{1}^\ast_t \1_{t\le\cdot})(s)\tilde{h}'_{0}  -(2\phi\boldsymbol{D}^{-1}_t\boldsymbol{1}^\ast_t \1_{t\le\cdot})(s)\tilde{h}_0(s)\,ds }\\
            &\quad\quad + 
            \abs{q\int_t^T (-4\phi\varrho\boldsymbol{1}_t(\boldsymbol{D}')^{-1}_{t}\1_{t<\cdot})(s) - (-4\phi\varrho\boldsymbol{1}_t\boldsymbol{D}^{-1}_t\1_{t<\cdot})(s)\,ds }\\
            &\quad\quad + \abs{q\int_t^T (- 2\phi\boldsymbol{1}_t(\boldsymbol{D}')^{-1}_{t}G'(\cdot,t))(s) -(- 2\phi\boldsymbol{1}_t\boldsymbol{D}^{-1}_tG(\cdot,t))(s)\,ds} \\
            &\quad\quad + \abs{q\int_t^T (- 4\phi^2\boldsymbol{1}_t(\boldsymbol{D}')^{-1}_{t}\boldsymbol{1}^\ast_t\1_{t\le\cdot})(s) - (- 4\phi^2\boldsymbol{1}_t\boldsymbol{D}^{-1}_t\boldsymbol{1}^\ast_t\1_{t\le\cdot})(s)\,ds}\\
            &\quad =: \mathcal{I}_1 + \cdots +\mathcal{I}_6.
        \end{split}
    \end{equation*}
Let us analyze each term separately. By means of H\"older's inequality, 
\[
\begin{split}
    \mathcal{I}_1 &\le 2\varrho \norm{(\boldsymbol{D}')^{-1}_{t} \1_{t<\cdot} }_{L^2} \norm{\tilde{h}'_{0}-\tilde{h}_0}_{L^2} + 2\varrho \norm{((\boldsymbol{D}')^{-1}_{t} - \boldsymbol{D}^{-1}_{t}) \1_{t<\cdot}}_{L^2}\norm{\tilde{h}_{0}}_{L^2}\\
    &\le 2\varrho \norm{(\boldsymbol{D}')^{-1}_{t}}_{op}\norm{\1_{t<\cdot}}_{L^2} \norm{h'_{0}- h_0}_{\infty} + 2\varrho
    \norm{((\boldsymbol{D}')^{-1}_{t} - \boldsymbol{D}^{-1}_{t})}_{op}
     \norm{\1_{t<\cdot}}_{L^2} \norm{\tilde{h}_{0}}_{\infty}\\
    &\le 
    2\varrho \sqrt{T}\norm{(\boldsymbol{D}')^{-1}_{t}}_{op}  \norm{h'_{0}- h_0}_{\infty} + 2\varrho \sqrt{T}
    \norm{((\boldsymbol{D}')^{-1}_{t} - \boldsymbol{D}^{-1}_{t})}_{op}
     \norm{\tilde{h}_{0}}_{\infty}.
\end{split} 
\]
As $ \norm{(\boldsymbol{D}')_t^{-1}}_{op} \le 1/2\lambda',\, t\le T$ (refer to the proof of Lemma \ref{lemma: 4}), we deduce, always from that Lemma, that
\[
    \mathcal{I}_1 \le \varrho\sqrt{T}\frac{1}{\lambda'}\norm{h'_0-h_0}_\infty + 2\varrho\sqrt{T} \frac{1}{\lambda^2} \left[\abs{\lambda'-\lambda} + \sqrt{T}\norm{H'-H}_{L^2}\right]\left[\norm{h_0}_{\infty}+2\varrho \abs{q}\right].
\]
Moreover,
\[
\begin{split}
    \mathcal{I}_2&\le \norm{(\boldsymbol{D}')^{-1}_{t}G'(\cdot,t) }_{L^2} \norm{\tilde{h}'_{0}-\tilde{h}_0}_{L^2} + \norm{(\boldsymbol{D}')^{-1}_{t}G'(\cdot,t) - \boldsymbol{D}^{-1}_{t}G(\cdot,t) }_{L^2} \norm{\tilde{h}_0}_{L^2}\\
    &\quad\le \norm{(\boldsymbol{D}')^{-1}_{t}}_{op} \norm{G'(\cdot,t)}_{L^2} \norm{\tilde{h}'_{0}-\tilde{h}_0}_{L^2} + \norm{((\boldsymbol{D}')^{-1}_{t}- \boldsymbol{D}^{-1}_{t})G'(\cdot,t) }_{L^2} \norm{\tilde{h}_0}_{L^2}\\
    &\quad\quad + \norm{(\boldsymbol{D}^{-1}_{t}(G'(\cdot,t) -G(\cdot,t)) }_{L^2} \norm{\tilde{h}_0}_{L^2}\\
    &\quad\le \norm{(\boldsymbol{D}')^{-1}_{t}}_{op} \norm{H'}_{L^2} \norm{\tilde{h}'_{0}-\tilde{h}_0}_{L^2} + \norm{(\boldsymbol{D}')^{-1}_{t}- \boldsymbol{D}^{-1}_{t}}_{op}\norm{H'}_{L^2} \norm{\tilde{h}_0}_{L^2}\\
    &\quad\quad +\norm{\boldsymbol{D}^{-1}_{t}}_{op}
    \norm{H' -H}_{L^2} \norm{\tilde{h}_0}_{L^2}
    \\&\quad\le
       \frac{\norm{H'}_{L^2}}{2\lambda'}\norm{h'_0-h_0}_\infty + \frac{1}{\lambda^2} \left[\abs{\lambda'-\lambda} + \sqrt{T}\norm{H'-H}_{L^2}\right]\left[\norm{h_0}_{\infty}+2\varrho \abs{q}\right].
\end{split}
\] 
Similarly, as $\norm{\boldsymbol{1}^\ast_t}_{op}=\norm{\boldsymbol{1}_t}_{op}\le \frac{T}{\sqrt{2}}$, we have
\[
\begin{split}
    \mathcal{I}_3&\le 2\phi \norm{(\boldsymbol{D}')^{-1}_{t}\boldsymbol{1}^\ast_t\1_{t\le \cdot} - \boldsymbol{D}^{-1}_t\boldsymbol{1}^\ast_t\1_{t\le } }_{L^2}\norm{\tilde{h}'_{0}}_{L^2}  +
    2\phi \norm{ \boldsymbol{D}^{-1}_t\boldsymbol{1}^\ast_t\1_{t\le } }_{L^2}\norm{\tilde{h}'_{0}-\tilde{h}_0}_{L^2}\\
    & \le  2\phi \norm{((\boldsymbol{D}')^{-1}_{t}  - \boldsymbol{D}^{-1}_t)\boldsymbol{1}^\ast_t  }_{op} \norm{\1_{t\le\cdot}}_{L^2}     \norm{\tilde{h}'_{0}}_{L^2}  +
    2\phi \norm{ \boldsymbol{D}^{-1}_t\boldsymbol{1}^\ast_t  }_{op}\norm{\1_{t\le\cdot}}_{L^2}
    \norm{\tilde{h}'_{0}-\tilde{h}_0}_{L^2}\\
    & \le  2\phi \norm{(\boldsymbol{D}')^{-1}_{t}  - \boldsymbol{D}^{-1}_t }_{op} \frac{T}{\sqrt{2}}\sqrt{T}      \norm{\tilde{h}'_{0}}_{L^2}  +
    2\phi \norm{ \boldsymbol{D}^{-1}_t   }_{op}  \frac{T}{\sqrt{2}}\sqrt{T}   
    \norm{\tilde{h}'_{0}-\tilde{h}_0}_{L^2}
    \\&\le \frac{\sqrt{2}\phi T^{3/2}}{\lambda^2}\left[\abs{\lambda'-\lambda} + \sqrt{T}\norm{H'-H}_{L^2}\right]\left[\norm{h_0'}_{\infty}+2\varrho \abs{q}\right] + \frac{\phi T^{3/2}}{\sqrt{2}\lambda}\norm{h_0'-h_0}_\infty.
\end{split}
\]
\[
\begin{split}
    \mathcal{I}_4 &\le 4\phi\varrho\abs{q}\sqrt{T}\norm{\boldsymbol{1}_t(\boldsymbol{D}')^{-1}_{t}\1_{t<\cdot} - \boldsymbol{1}_t\boldsymbol{D}^{-1}_{t}\1_{t<\cdot} }_{L^2} \\
    &\le 4\phi\varrho\abs{q}\sqrt{T}\norm{\boldsymbol{1}_t((\boldsymbol{D}')^{-1}_{t} - \boldsymbol{D}^{-1}_{t})}_{op}\norm{\1_{t<\cdot} }_{L^2}\\
    & \le 4\phi\varrho \abs{q} T \norm{\boldsymbol{1}_t}_{op}\norm{(\boldsymbol{D}')^{-1}_{t}-\boldsymbol{D}^{-1}_t}_{op}\\
    & \le 4\phi\varrho\abs{q}\frac{T^2}{\sqrt{2}}\norm{(\boldsymbol{D}')^{-1}_{t}-\boldsymbol{D}^{-1}_t}_{op}\\
    & \le 4\phi\varrho\abs{q}\frac{T^2}{\sqrt{2}} \frac{1}{\lambda^2}\left[\abs{\lambda'-\lambda} + \sqrt{T}\norm{H'-H}_{L^2}\right].
\end{split}
\]
\[
\begin{split}
    \mathcal{I}_5 & \le  2\phi\abs{q}\sqrt{T}\norm{\boldsymbol{1}_t((\boldsymbol{D}')^{-1}_{t}G'(\cdot,t) - \boldsymbol{D}^{-1}_tG(\cdot,t)) }_{L^2} \\
    & \le 2\phi \abs{q}\sqrt{T} \norm{\boldsymbol{1}_t}_{op} \norm{   (\boldsymbol{D}')^{-1}_{t}G'(\cdot,t) - \boldsymbol{D}^{-1}_tG(\cdot,t) }_{L^2}\\
    & \le 2\phi\abs{q}\frac{T^{3/2}}{\sqrt{2}} \left\{
    \norm{(\boldsymbol{D}')^{-1}_{t}- \boldsymbol{D}^{-1}_{t}}_{op}\norm{H'}_{L^2} 
 +\norm{\boldsymbol{D}^{-1}_{t}}_{op}
    \norm{H' -H}_{L^2}   
    \right\}\\
    &\le 2\phi\abs{q}\frac{T^{3/2}}{\sqrt{2}} \left\{
    \frac{1}{\lambda^2}\left[\abs{\lambda'-\lambda} + \sqrt{T}\norm{H'-H}_{L^2}\right] \norm{H'}_{L^2}
    +\frac{1}{2\lambda}\norm{H'-H}_{L^2}
    \right\}.
\end{split}
\]
\[
\begin{split}
    \mathcal{I}_6 &\le  4\phi^2\abs{q} \sqrt{T}\norm{\boldsymbol{1}_t ((\boldsymbol{D}')^{-1}_{t} \boldsymbol{1}^\ast_t\1_{t\le\cdot} - \boldsymbol{D}^{-1}_{t} \boldsymbol{1}^\ast_t\1_{t\le\cdot})}_{L^2}\\
    &\le 4\phi^2\abs{q} \sqrt{T}\norm{\boldsymbol{1}_t}_{op}\norm{(\boldsymbol{D}')^{-1}_{t} - \boldsymbol{D}^{-1}_t }_{op} \norm{\boldsymbol{1}^\ast_t}_{op}\norm{\1_{t\le\cdot}}_{L^2}\\
    &\le 4\phi^2\abs{q} \frac{T^{3}}{\sqrt{2}} \norm{(\boldsymbol{D}')^{-1}_{t} - \boldsymbol{D}^{-1}_t }_{op}\\
    &\le 
     4\phi^2\abs{q} \frac{T^{3}}{\sqrt{2}}
     \frac{1}{\lambda^2}\left[\abs{\lambda'-\lambda} + \sqrt{T}\norm{H'-H}_{L^2}\right].
\end{split}
\]
Then, long but elementary computations reveal that
\[
\mathcal{I}\le \left\{
g_1(\lambda',\lambda,H') + g_2(\lambda,h_0,q) + g_3(\lambda,h_0',q) + g_4(\lambda,H',q) + g_5(\lambda,q)
\right\}\cdot \norm{x'-x}_\X,
\]
and thus, as the right-hand side does not depend on $t$, 
\[
\sup_{t\le T}\mathcal{I}^2\le \left\{
g_1(\lambda',\lambda,H') + g_2(\lambda,h_0,q) + g_3(\lambda,h_0',q) + g_4(\lambda,H',q) + g_5(\lambda,q)
\right\}^2 \norm{x'-x}^2_\X.
\]
In a similar fashion, starting from \eqref{eq: expansion of L2 scalar product} and using \eqref{eq: elementary ineq} with $m=7$, one can also prove the a priori estimate. This concludes the Lemma. 
\end{proof}

For future reference, we record here the following a priori estimate for the stochastic process $a$ in terms of its constituent parts $h_0,H,\lambda,P$ and $q$:

\begin{lemma}\label{lemma: apriori estimate for a}
    Given arbitrary $((h_0,H,\lambda),P,q)\in \mathscr M\times\PP\times\R$, it holds that
    \[
    \EE\sup_{t\le T}a^2_t \le \frac{1}{\lambda^2} \left\{
    4\norm{P}^2_\PP+2\norm{h_0}_\infty^2+8\varrho^2q^2+ f_3(H,\lambda,P) + g_6(h_0,H,\lambda,q)
    \right\},
    \]
    where $f_3$ and $g_6$ are defined in \eqref{eq: f3} and \eqref{eq: g6} respectively.
\end{lemma}
\begin{proof}
    The result simply follows by squaring \eqref{eq: a} and using Lemma \ref{lemma: 5} and \ref{lemma: 6} to bound the respective terms.
\end{proof}

We are finally ready to prove Proposition \ref{prop: 1}.
\begin{proof}[Proof of Proposition \ref{prop: 1}]
    We set for convenience 
    \[
    A'_t :=\lambda' a'_t,\quad A_t:=\lambda a_t.
    \]
Then,
\[
\begin{split}
    &\EE\sup_{t\le T}\abs{ A'_t  - A_t }^2  \\
    &\quad\le 3\EE\sup_{t\le T}\abs{\tilde{h}'_{0}(t) - \tilde{h}_0(t)}^2 + 
3\EE\sup_{t\le T}\abs{\langle \Theta'_{t} ,K'_{t}\rangle_{L^2} - \langle \Theta_{t} ,K_{t}\rangle_{L^2} }^2\\ &\quad\quad +
3\EE\sup_{t\le T}\abs{\langle (\boldsymbol{\Gamma}')_t^{-1}K'_{t}, \mathbbm{1}_{t} (\tilde{h}'_{0},q)^\top \rangle_{L^2} - \langle \boldsymbol{\Gamma}_t^{-1}K_t, \mathbbm{1}_{t} (\tilde{h}_0,q)^\top \rangle_{L^2}}^2
\\&\quad\le
3\norm{h_0'-h_0}_\infty^2 + [f_1(\lambda,H',P) + f_2(\lambda,P)] \norm{x-x'}^2_\X 
\\&\quad\quad +\left\{
g_1(\lambda',\lambda,H') + g_2(\lambda,h_0,q) + g_3(\lambda,h_0',q) + g_4(\lambda,H',q) + g_5(\lambda,q)
\right\}^2 \norm{x'-x}^2_\X,
\end{split}
\] 
in virtue of Lemmas \ref{lemma: 1}, \ref{lemma: 5} and \ref{lemma: 6}. Therefore, since
\[
\EE\sup_{t\le T}\abs{a'_t-a_t}^2=\EE\sup_{t\le T}\abs{\frac{A'_t}{\lambda'} - \frac{A_t}{\lambda}  }^2 \le \frac{2}{(\lambda')^2}\EE\sup_{t\le T}\abs{A'_t-A_t}^2 
+2\frac{\abs{\lambda'-\lambda}^2}{(\lambda')^2} \EE\sup_{t\le T}a_t^2,
\]
we obtain, by means of the trivial estimates $\norm{h_0'-h_0}_\infty\le \norm{x'-x}_\X$ and $\abs{\lambda'-\lambda}\le \norm{x'-x}_\X$  and the apriori estimate of Lemma \ref{lemma: apriori estimate for a} that
\[
\EE\sup_{t\le T}\abs{a'_t-a_t}^2 \le \mathscr{L}_1(h_0,H,\lambda,h_0',H',\lambda',P,q) \norm{x'-x}^2_\X,
\]
where $\mathscr{L}_1$ is defined in \eqref{eq: L1}.

\end{proof}

\subsubsection{Proof of Proposition \ref{prop: 2}} In this part, we aim to prove Proposition \ref{prop: 2}. Similar to Proposition \ref{prop: 1}, we first need this technical lemma.

\begin{lemma}\label{lemma: 7}
     Under Assumption \ref{hp: 1} it holds that
\[
   \begin{split}
       &\sup_{t\le T} \int_0^T \mathbbm{1}_{s<t}
    \abs{\langle \boldsymbol{(\Gamma}')_{t}^{-1}K'_{t}, \mathbbm{1}_{t} (\tilde{G}'(\cdot,s),-1)^\top \rangle_{L^2} - \langle \boldsymbol{\Gamma}_t^{-1}K_t, \mathbbm{1}_{t} (\tilde{G}(\cdot,s),-1)^\top \rangle_{L^2}  }^2ds\\
    &\quad \le
    Tj_0^2(H,\lambda,\lambda')\norm{x'-x}^2_\X,
   \end{split}
    \]
where $j_0$ is defined in \eqref{eq: j0} and where evidently $\tilde{G}'(s,t)=2\varrho\1_{t<s}+ \1_{t<s}H'(s-t)$.
   
\end{lemma}

\begin{proof}

Expanding $\langle \boldsymbol{\Gamma}_t^{-1}K_t, \mathbbm{1}_{t} (\tilde{G}(\cdot,s),-1)^\top \rangle_{L^2}$ by means of the definition of $\boldsymbol{\Gamma}^{-1}_t,K_t$ and $\tilde{G}$ in the same spirit of Lemma \ref{lemma: 6}, we see that
\begin{equation}\label{eq: lemma 7 (1)}
\begin{split}
    &\langle \boldsymbol{\Gamma}_t^{-1}K_t, \mathbbm{1}_{t} (\tilde{G}(\cdot,s),-1)^\top \rangle_{L^2} =\\
    &\quad = \int_t^T \left[
    (\boldsymbol{D}^{-1}_t2\varrho\1_{t<\cdot })(r) + (\boldsymbol{D}^{-1}_tG(\cdot,t))(r) + (2\phi \boldsymbol{D}^{-1}_t\boldsymbol{1}^\ast_t \1_{t\le \cdot })(r)
    \right]\cdot \left[
    2\varrho\1_{s<r}+G(r,s)    \right]\,dr\\
    &\quad\quad +\int_t^T\left[ (4\phi\varrho\boldsymbol{1}_t\boldsymbol{D}^{-1}_t\1_{t<\cdot})(r) + (2\phi\boldsymbol{1}_t\boldsymbol{D}^{-1}_t G(\cdot,t))(r)- (2\phi\operatorname{id} \1_{t\le \cdot})(r)\right] dr\\
    &\quad\quad + \int_t^T
(4\phi^2\boldsymbol{1}_t\boldsymbol{D}^{-1}_t\boldsymbol{1}^\ast_t \1_{t\le\cdot})(r)
    \,dr.
\end{split}
\end{equation}
Thus, 
\[
\begin{split}
   \mathcal{J}&:=\abs{\langle (\boldsymbol{\Gamma}')_{t}^{-1}K'_{t}, \mathbbm{1}_{t} (\tilde{G}'(\cdot,s),-1)^\top \rangle_{L^2} - \langle \boldsymbol{\Gamma}_t^{-1}K_t, \mathbbm{1}_{t} (\tilde{G}(\cdot,s),-1)^\top \rangle_{L^2}  }\\
    &\le  \mathcal{J}_1+\cdots + \mathcal{J}_9, 
\end{split}
\]
where:
\begin{align*}
    \mathcal J_1 &:= 4\varrho^2\abs{\int_t^T\left[((\boldsymbol{D}')^{-1}_{t}\1_{t<\cdot })(r) - (\boldsymbol{D}^{-1}_t\1_{t<\cdot })(r)\right]\1_{s<r}\,dr}\,,\\
    \mathcal J_2 &:= 2\varrho\abs{\int_t^T((\boldsymbol{D}')^{-1}_{t}\1_{t<\cdot })(r)\,G'(r,s) - (\boldsymbol{D}^{-1}_t\1_{t<\cdot })(r)\,G(r,s)\,dr}\,,\\
    \mathcal J_3 &:= 2\varrho\abs{\int_t^T\left[((\boldsymbol{D}')^{-1}_{t}G'(\cdot,t))(r)-(\boldsymbol{D}^{-1}_tG(\cdot,t))(r)\right]\1_{s<r}\,dr}\,,\\
    \mathcal J_4 &:= \abs{\int_t^T((\boldsymbol{D}')^{-1}_{t}G'(\cdot,t))(r)\,G'(r,s)-(\boldsymbol{D}^{-1}_tG(\cdot,t))(r)\,G(r,s)\,dr}\,,\\
    \mathcal J_5 &:= 4\phi\varrho\abs{\int_t^T\left[((\boldsymbol{D}')^{-1}_{t}\boldsymbol{1}_t^\ast\1_{t\le\cdot })(r) - (\boldsymbol{D}^{-1}_t\boldsymbol{1}^\ast_t\1_{t\le \cdot })(r)\right]\1_{s<r}\,dr}\,,\\
    \mathcal J_6 &:= 2\phi\abs{\int_t^T((\boldsymbol{D}')^{-1}_{t}\boldsymbol{1}^\ast_t\1_{t\le\cdot })(r)\,G'(r,s) - (\boldsymbol{D}^{-1}_t\boldsymbol{1}^\ast_t\1_{t\le\cdot })(r)\,G(r,s)\,dr}\,,\\
    \mathcal J_7 &:= 4\phi\varrho\abs{\int_t^T (\boldsymbol{1}_t(\boldsymbol{D}')^{-1}_{t}\1_{t<\cdot})(r) - (\boldsymbol{1}_t\boldsymbol{D}^{-1}_t\1_{t<\cdot})(r)\,dr}\,,\\
    \mathcal J_8 &:= 2\phi\abs{\int_t^T(\boldsymbol{1}_t(\boldsymbol{D}')^{-1}_{t} G'(\cdot,t))(r) - (\boldsymbol{1}_t\boldsymbol{D}^{-1}_t G(\cdot,t))(r)\,dr}\,,\\
    \mathcal J_9 &:= 4\phi^2\abs{\int_t^T( \boldsymbol{1}_t(\boldsymbol{D}')^{-1}_{t}\boldsymbol{1}^\ast_t \1_{t\le\cdot})(r)-(\boldsymbol{1}_t\boldsymbol{D}^{-1}_t\boldsymbol{1}^\ast_t \1_{t\le\cdot})(r)\,dr}\,.
\end{align*}

Let us bound each of these terms separately. Similarly to what has been done in the course of the proof of Lemma \ref{lemma: 6}, it holds:
\[
\mathcal{J}_1 \le 4\varrho^2 \norm{(\boldsymbol{D}')^{-1}_t - \boldsymbol{D}^{-1}_{t}}_{op} \norm{\1_{t<\cdot}}_{L^2} \norm{\1_{s<\cdot}}_{L^2} \le 4\varrho^2 T \norm{(\boldsymbol{D}')^{-1}_t - \boldsymbol{D}^{-1}_{t}}_{op}.
\]
Adding and subtracting the quantity $((\boldsymbol{D}')^{-1}_t\1_{t<\cdot})(r)\,G(r,s)$ inside the second integral leads to 
\[
\begin{split}
    \mathcal{J}_2&\le 2\varrho \norm{(\boldsymbol{D}')^{-1}_t}_{op} \norm{\1_{t<\cdot}}_{L^2} \norm{G'(\cdot,s)-G(\cdot,s)}_{L^2} 
    + 2\varrho\norm{(\boldsymbol{D}')^{-1}_t-\boldsymbol{D}^{-1}_{t}}_{op} \norm{\1_{t<\cdot}}_{L^2} \norm{ G(\cdot,s)}_{L^2}\\
    &\le 2\varrho T^{1/2}\left(\norm{(\boldsymbol{D}')^{-1}_t}_{op}  \norm{G'(\cdot,s)-G(\cdot,s)}_{L^2} + \norm{(\boldsymbol{D}')^{-1}_t-\boldsymbol{D}^{-1}_{t}}_{op}  \norm{ G(\cdot,s)}_{L^2}\right)\\
    &\le 2\varrho T^{1/2}\left(\norm{(\boldsymbol{D}')^{-1}_t}_{op}  \norm{H'-H}_{L^2} + \norm{(\boldsymbol{D}')^{-1}_t-\boldsymbol{D}^{-1}_{t}}_{op}  \norm{ H }_{L^2}\right).
\end{split}
\] 
Similarly, adding and subtracting $((\boldsymbol{D}')^{-1}_tG(\cdot,t))(r)$ in the third integral gives 
\[
\begin{split}
    \mathcal{J}_3&\le 2\varrho\left( \norm{(\boldsymbol{D}')^{-1}_t}_{op}  \norm{G'(\cdot,t)-G(\cdot,t)}_{L^2} 
    + \norm{(\boldsymbol{D}')^{-1}_t-\boldsymbol{D}^{-1}_{t}}_{op}  \norm{ G(\cdot,t)}_{L^2}  \right)\norm{\1_{s<\cdot}}_{L^2}\\
    &\le 2\varrho T^{1/2}\left(\norm{(\boldsymbol{D}')^{-1}_t}_{op}  \norm{H'-H}_{L^2} + \norm{(\boldsymbol{D}')^{-1}_t-\boldsymbol{D}^{-1}_{t}}_{op}  \norm{ H }_{L^2}\right).
\end{split}
\]
In the same spirit,
\[
\begin{split}
    \mathcal{J}_4&\le \int_t^T\abs{((\boldsymbol{D}')^{-1}_tG'(\cdot,t))(r)}\,\abs{G'(r,s) - G(r,s)}\,dr \\
    &\quad + \int_t^T\abs{((\boldsymbol{D}')^{-1}_t     
    [ G'(\cdot,t) -G(\cdot,t) ])(r)}\,\abs{G(r,s)}\,dr\\
    &\quad + \int_t^T\abs{([(\boldsymbol{D}')^{-1}_t - \boldsymbol{D}^{-1}_{t}]     
    G(\cdot,t) )(r)}\,\abs{G(r,s)}\,dr\\
    & \le \norm{(\boldsymbol{D}')^{-1}_t}_{op}  
    \norm{G(\cdot,t)}_{L^2} 
    \norm{G'(\cdot,s)-G(\cdot,s)}_{L^2}\\
    &\quad + \norm{(\boldsymbol{D}')^{-1}_t}_{op}   
    \norm{G'(\cdot,t)-G(\cdot,t)}_{L^2} \norm{G(\cdot,s)}_{L^2} \\
    &\quad + \norm{(\boldsymbol{D}')^{-1}_t-\boldsymbol{D}^{-1}_{t}}_{op}   
    \norm{G(\cdot,t)}_{L^2} \norm{G(\cdot,s)}_{L^2}\\
    &\le 2\norm{(\boldsymbol{D}')^{-1}_t}_{op} \norm{H}_{L^2}\norm{H'-H}_{L^2} + \norm{(\boldsymbol{D}')^{-1}_t-\boldsymbol{D}^{-1}_{t}}_{op}   
    \norm{H}^2_{L^2}.
\end{split}
\]
Then,
\[
    \mathcal{J}_5\le 4\phi\varrho  \norm{(\boldsymbol{D}')^{-1}_t-\boldsymbol{D}^{-1}_{t}}_{op} \norm{\boldsymbol{1}^\ast_t}_{op} \norm{\1_{t\le\cdot}}_{L^2} \norm{\1_{s<\cdot}}_{L^2}\le 4\phi\varrho T^{5/2}\norm{(\boldsymbol{D}')^{-1}_t-\boldsymbol{D}^{-1}_{t}}_{op}.
\]
By adding and subtracting the quantity $((\boldsymbol{D}')^{-1}_t\boldsymbol{1}^\ast_t\1_{t\le \cdot})(r)G(r,s)$ inside the sixth integral, we obtain once more
\[
    \mathcal{J}_6\le
    2\phi T^{3/2}\left(\norm{(\boldsymbol{D}')^{-1}_t}_{op}  \norm{H'-H}_{L^2} + \norm{(\boldsymbol{D}')^{-1}_t-\boldsymbol{D}^{-1}_{t}}_{op}  \norm{ H }_{L^2}\right).
\]
For the terms $\mathcal{J}_7$ and $\mathcal{J}_9$ we obtain respectively
\[
\mathcal{J}_7\le 4\phi\varrho T^{2} \norm{(\boldsymbol{D}')^{-1}_t-\boldsymbol{D}^{-1}_{t}}_{op},
\]
and
\[
\mathcal{J}_9\le 4\phi^2 T^{3} \norm{(\boldsymbol{D}')^{-1}_t-\boldsymbol{D}^{-1}_{t}}_{op}.
\]
Finally,
\[
\mathcal{J}_8\le 2\phi T^{3/2}\left( \norm{(\boldsymbol{D}')^{-1}_t}_{op} \norm{H'-H}_{L^2}+ \norm{(\boldsymbol{D}')^{-1}_t-\boldsymbol{D}^{-1}_{t}}_{op} \norm{H}_{L^2}  \right).
\]

Therefore, with the help of all these bounds and by means of Lemma \ref{lemma: 4} once more, after tedious but elementary computations, we arrive at
\[
\begin{split}
    \mathcal J &\le \left\{
    4\varrho^2T+4\phi T^2(\varrho+\phi T)+4\phi\varrho T^{5/2}+\left[\norm{H}_{L^2}+ 4\sqrt{T}(\varrho^2+\phi T)\right]\norm{H}_{L^2}
    \right\} 
    \\
    &\quad\quad\cdot \frac{1}{\lambda^2}\left[\abs{\lambda'-\lambda}+\sqrt{T}\norm{H'-H}_{L^2}\right] 
    \\&\quad\quad+ \left(2\sqrt{T}(\varrho^2+\phi T)+\norm{H}_{L^2}\right)\frac{1}{\lambda'}\norm{H'-H}_{L^2}. 
\end{split}
\]
Observe that the right-hand side of this last expression does not depend on $s$ nor on $t$. Majorizing once more $\abs{\lambda'-\lambda}\le\norm{x'-x}_\X$ and $\norm{H'-H}_{L^2}\le\norm{x'-x}_\X$, taking the square on both sides, integrating in $ds$ over $[0,T]$ and finally taking the supremum for $t\le T$ will lead to the thesis, namely 
\[
\sup_{t\le T}\int_0^T \mathbbm{1}_{s<t}\mathcal{J}^2ds\le 
Te^2_1(H,\lambda,\lambda') \norm{x'-x}_\X^2.
\]
\end{proof}

Before proving Proposition \ref{prop: 2}, we observe that it clearly holds true that
\begin{equation}\label{eq: G tilde - G}
    \sup_{t\le T} \int_0^T\1_{s<t}
    \abs{\tilde{G}'(t,s) - \tilde{G}(t,s)}^2ds \le \norm{H'-H}^2_{L^2}.
\end{equation}

Analogously to Lemma \ref{lemma: apriori estimate for a}, we now provide the following a priori estimate for the kernel $B$ in terms of its constituent parts $H$ and $\lambda$:

\begin{lemma}\label{lemma: apriori estimate for B}
Given arbitrary $((h_0,H,\lambda),P,q)\in \mathscr M\times\PP\times\R$, it holds that
    \[
    \sup_{t\le T}\int_0^T \abs{B(t,s)}^2ds\le 
    \frac{T}{2\lambda^2}\left\{
j_1(\lambda) + j_2(H,\lambda)
\right\}^2 + \frac{1}{\lambda^2}\left\{
4\varrho^2 T + \norm{H}^2_{L^2}
\right\},
\]
where $j_1$ and $j_2$ are defined in \eqref{eq: j1} and \eqref{eq: j2} respectively.

\end{lemma}
\begin{proof}
    We need to bound the two terms appearing in \eqref{eq: B}. We start with the quantity
    \[
    \langle \boldsymbol{\Gamma}^{-1}_tK_t,\1_t(\tilde{G}(\cdot,s),-1)^\top\rangle_{L^2}.
    \]
    We argue as in the proof of Lemma \ref{lemma: 7}. Starting from \eqref{eq: lemma 7 (1)} we have:
    \[
    \begin{split}
        &\abs{\langle \boldsymbol{\Gamma}^{-1}_tK_t,\1_t(\tilde{G}(\cdot,s),-1)^\top\rangle_{L^2}}\\
        &\quad\le 4\varrho^2 T\norm{\boldsymbol{D}^{-1}_t}_{op} + 2\varrho\norm{\boldsymbol{D}^{-1}_t}_{op}\sqrt{T}\norm{H}_{L^2} +2\varrho\norm{\boldsymbol{D}^{-1}_t}_{op}\sqrt{T}\norm{H}_{L^2}\\
        &\quad\quad+\norm{\boldsymbol{D}^{-1}_t}_{op}\norm{H}^2_{L^2} + 4\phi\varrho T^{5/2}\norm{\boldsymbol{D}^{-1}_t}_{op} + 2\phi\norm{\boldsymbol{D}^{-1}_t}_{op} T^{3/2}\norm{H}_{L^2}\\
        &\quad\quad+ 4\phi\varrho T^2 \norm{\boldsymbol{D}^{-1}_t}_{op} + 2\phi\norm{\boldsymbol{D}^{-1}_t}_{op}T^{3/2}\norm{H}_{L^2} +2\phi T+ 2\phi^2T^3 \norm{\boldsymbol{D}^{-1}_t}_{op}\\
        &\quad\le C(\phi,\varrho,T) \left[1+\norm{\boldsymbol{D}^{-1}_t}_{op} + \norm{\boldsymbol{D}^{-1}_t}_{op}\norm{H}^2_{L^2}\right].
    \end{split}
    \]
Elementary but tedious computations will lead to
\[
\abs{\langle \boldsymbol{\Gamma}^{-1}_tK_t,\1_t(\tilde{G}(\cdot,s),-1)^\top\rangle_{L^2}}\le j_1(\lambda) + j_2(H,\lambda).
\]
Thus, as we did in Lemma \ref{lemma: 7}, we square the last quantity, integrate in $s$ over $[0,t]$, and take $\sup_{t\le T}$ to eventually obtain
\[
\sup_{t\le T}\int_0^T\1_{s<t}\abs{\langle \boldsymbol{\Gamma}^{-1}_tK_t,\1_t(\tilde{G}(\cdot,s),-1)^\top\rangle_{L^2}}^2ds \le T \left\{
j_1(\lambda) + j_2(H,\lambda)
\right\}^2.
\]
Moreover, it holds that
\[
\sup_{t\le T}\int_0^T\1_{s<t} \tilde{G}(t,s)^2ds \le 8\varrho^2 T + 2\norm{H}^2_{L^2},
\]
and thus
\[
    \sup_{t\le T}\int_0^T \abs{B(t,s)}^2ds\le 
    \frac{T}{2\lambda^2}\left\{
j_1(\lambda) + j_2(H,\lambda)
\right\}^2 + \frac{1}{\lambda^2}\left\{
4\varrho^2 T + \norm{H}^2_{L^2}
\right\}.
\]

\end{proof}

\begin{proof}[Proof of Proposition \ref{prop: 2}]

We define for convenience $\mathcal B'(t,s)=2\lambda'B'(t,s)$ and $\mathcal B(t,s)=2\lambda B(t,s)$. Then, by Lemma \ref{lemma: 7} and \eqref{eq: G tilde - G}, we deduce
\[
\sup_{t\le T}\int_0^T\abs{\mathcal B'(t,s)-\mathcal B(t,s)}^2ds \le 2Tj_0^2(H,\lambda,\lambda')\norm{x'-x}^2_\X + 2\norm{x'-x}^2_\X,
\]
and thus
\[
\begin{split}
    &\sup_{t\le T}\int_0^T\abs{B'(t,s)-B(t,s)}^2ds =
    \sup_{t\le T}\int_0^T\abs{\frac{1}{2\lambda'}\mathcal B'(t,s)-\frac{1}{2\lambda}\mathcal B(t,s)}^2ds \\
    &\quad = \sup_{t\le T}\int_0^T\abs{\frac{1}{2\lambda'}\mathcal B'(t,s)-\frac{1}{2\lambda'}\mathcal B(t,s)+\frac{1}{2\lambda'}\mathcal B(t,s)-\frac{1}{2\lambda}\mathcal B(t,s)}^2ds \\
    &\quad \le \frac{1}{2(\lambda')^2}\left\{
    2Tj_0^2(H,\lambda,\lambda')+2
    \right\}\norm{x'-x}^2_\X \\
    &\quad\quad+ 2\left\{
    \frac{T}{2\lambda^2}[j_1(\lambda)+j_2(H,\lambda)]^2 +\frac{1}{\lambda^2}[4\varrho^2T+\norm{H}^2_{L^2}]
    \right\}\frac{\abs{\lambda'-\lambda}^2}{(\lambda')^2} \\
    &\quad \le \mathscr L_{\Delta B}(H,\lambda,\lambda')\norm{x'-x}^2_\X.
\end{split}
\]

\end{proof}

The last ingredient required to prove Theorem \ref{thm: main result} is the following a priori estimate, valid for any solution $u^\ast$, based on a Gr\"onwall-type argument:

\begin{proposition}\label{prop: apriori est. for u}
    Let $u^\ast=\mathscr S(h_{0},H,\lambda,P,q)$ be the optimal trading strategy for the impact propagator model with parameters $(h_0,H,\lambda)\in\MM$, unaffected price $P\in\PP$ and initial position $q\in\R$. Then   
    \[
    \EE \sup_{0\le s\le T}\abs{u^\ast_s}^2 \le M(h_0,H,\lambda,P,q),
    \]
    where $M$ is defined in \eqref{eq: M}.     
\end{proposition}

\begin{proof}
Fix $t$, then directly by \eqref{eq: uast} we have $u^\ast_s = a_s +\int_0^sB(s,r)u^\ast_r\,dr,0\le s\le t$, which, when squared, leads to, by means of H\"older's inequality,
\begin{equation}\label{eq: Groenwall 01}
\begin{split}
    \abs{u^\ast_s}^2 &\le 2\abs{a_s}^2 + 2\left[\int_0^sB(s,r)u^\ast_r\,dr\right]^2\\
    &\le 2\sup_{0\le s\le T}\abs{a_s}^2 + 2\sup_{0\le s\le t} \int_0^s\abs{B(s,r)}^2\,dr \cdot \sup_{0\le s\le t} \int_0^s\abs{u^\ast_r}^2dr.
\end{split}
\end{equation}
Define $g(t):= \EE \sup_{0\le s\le t}\abs{u^\ast_s}^2,\,0\le t\le T$. By construction, $g(t_1)\le g(t_2)$ for $t_1<t_2$ and $g
(t)\le g(T)<\infty$, because $u^\ast$ belongs to $\mathcal{A}$. Namely, $g$ is bounded and monotone, and thus integrable over $[0,T]$, a fact that will allow us to apply Gr\"onwall's inequality afterward. By taking first $\sup_{0\le s\le t}$ on the left-hand side of \eqref{eq: Groenwall 01} and subsequently taking the expectation on both sides, we arrive at
\begin{equation*}
\begin{split}
    g(t)     
    &\le 2\EE\sup_{0\le s\le T}\abs{a_s}^2 + 2\sup_{0\le s\le t} \int_0^s\abs{B(s,r)}^2\,dr \cdot \EE \sup_{0\le s\le t} \int_0^s\abs{u^\ast_r}^2dr\\
    & \le 2\EE\sup_{0\le s\le T}\abs{a_s}^2 + 2\sup_{0\le s\le T} \int_0^T\abs{B(s,r)}^2\,dr \cdot \EE  \int_0^t\abs{u^\ast_r}^2dr\\
    & \le 2\EE\sup_{0\le s\le T}\abs{a_s}^2 + 2\sup_{0\le s\le T} \int_0^T\abs{B(s,r)}^2\,dr \cdot \EE \int_0^t\sup_{0\le \gamma\le r} \abs{u^\ast_\gamma}^2dr.
\end{split}
\end{equation*}
We observe that the process $(\omega,r)\mapsto \sup_{0\le\gamma\le r}\abs{u^\ast_\gamma}^2$ is progressively measurable, and hence measurable. Therefore, by Tonelli's theorem, 
\[
g(t)\le 2\EE\sup_{0\le s\le T}\abs{a_s}^2 + 2\sup_{0\le s\le T} \int_0^T\abs{B(s,r)}^2\,dr \cdot \int_0^t g(r)dr.
\]

By Gr\"onwall's lemma
\[
g(t)\le 2\EE\sup_{0\le s\le T}\abs{a_s}^2\exp\{2t \sup_{0\le s\le T} \int_0^T\abs{B(s,r)}^2\,dr\},\quad 0\le t\le T,
\]
and by means of Lemma \ref{lemma: apriori estimate for a} and \ref{lemma: apriori estimate for B} we finally obtain
\[
g(T)\le M(h_0,H,\lambda,P,q).
\]
\end{proof}

\subsubsection{Proof of \ref{thm: main result}}

\begin{proof}

The argument relies on a variation of the Gr\"onwall-type argument just presented, coupled with all the a priori and stability estimates we have obtained so far. As in Assumption \ref{hp: 1}, we fix an arbitrary $\hat{x}=(\hat{h}_0,\hat{H},\hat{\lambda})\in\mathscr M$ and we consider two arbitrary points $x=(h_0,H,\lambda)$ and $x'=(h'_0,H',\lambda')$ in the open ball $B^\MM_r(\hat{h}_0,\hat{H},\hat{\lambda}) \subset \MM$ of radius 
\[
0<r< \frac{\hat{\lambda}}{5+4\sqrt{T}}.
\]

For brevity, we set $(u')^{\ast}_{t}:=\mathscr S(h'_{0},H',\lambda',P,q)_t$ and $u^\ast_{t}:=\mathscr S(h_{0},H,\lambda,P,q)_t$ with $0\le t \le T$. Fix $t$, and then directly by \eqref{eq: uast}, we have
\[
\begin{split}
    (u')^\ast_{s}-u^\ast_s &= a'_{s}-a_s +\int_0^s[B'(s,r)-B(s,r)]u^\ast_r\,dr\\
    &\quad \quad+ \int_0^sB'(s,r)[(u')^\ast_{r}-u^\ast_r]\,dr,\quad 0\le s\le t.
\end{split}
\]
By raising this to the square and by means of H\"older's inequality,
\begin{equation*}
\begin{split}
    \frac13 \abs{(u')^\ast_{s}-u^\ast_s}^2 &\le \abs{a'_{s}-a_s}^2 + \left[\int_0^s[B'(s,r)-B(s,r)]u^\ast_r\,dr\right]^2\\
    &\quad\quad + \left[\int_0^sB'(s,r)[(u')^\ast_{r}-u^\ast_r]\,dr\right]^2\\
    &\le \sup_{0\le s\le T}\abs{a'_{s}-a_s}^2 \\
    &\quad\quad + \sup_{0\le s\le t} \int_0^s\abs{B'(s,r)-B(s,r)}^2\,dr \cdot \sup_{0\le s\le t} \int_0^s(u^\ast_r)^2dr\\
    &\quad\quad + \sup_{0\le s\le t} \int_0^sB'(s,r)^2\,dr\cdot \sup_{0\le s\le t} \int_0^s\abs{(u')^\ast_{r}-u^\ast_r}^2\,dr.
\end{split}
\end{equation*}
Define $g(t):= \EE \sup_{0\le s\le t}\abs{(u')^\ast_{s}-u^\ast_s}^2,\,0\le t\le T$. As before, by construction, $g(t_1)\le g(t_2)$ for $t_1<t_2$ and $g
(t)\le g(T)<\infty$, because both $(u')^\ast$ and $u^\ast$ belong to $\mathcal{A}$. Namely, $g$ is bounded and monotone, and thus integrable over $[0,T]$. Arguing as before, we arrive at
\begin{equation*}
\begin{split}
    \frac13 g(t)     
    &\le \EE\sup_{0\le s\le T}\abs{a'_{s}-a_s}^2 + \sup_{0\le s\le t} \int_0^s\abs{B'(s,r)-B(s,r)}^2\,dr \cdot \EE \sup_{0\le s\le t} \int_0^s(u^\ast_r)^2dr\\
    &\quad\quad + \sup_{0\le s\le t} \int_0^sB'(s,r)^2\,dr\cdot \EE\sup_{0\le s\le t} \int_0^s\abs{(u')^\ast_{r}-u^\ast_r}^2\,dr\\
    & \le \EE\sup_{0\le s\le T}\abs{a'_{s}-a_s}^2 + \sup_{0\le s\le T} \int_0^T\abs{B'(s,r)-B(s,r)}^2\,dr \cdot \EE \int_0^t\sup_{0\le \gamma\le r} (u^\ast_\gamma)^2dr\\
    &\quad\quad + \sup_{0\le s\le T} \int_0^TB'(s,r)^2\,dr\cdot \EE\int_0^t\sup_{0\le \gamma\le r} \abs{(u')^\ast_{\gamma}-u^\ast_\gamma}^2\,dr,
\end{split}
\end{equation*}
because $\sup_{0\le s\le t}\int_0^s(u^\ast_r)^2dr\le \int_0^t\sup_{0\le\gamma\le r}(u^\ast_\gamma)^2dr$ (and similarly for the term involving $\abs{(u')^\ast_{r}-u^\ast_r}^2$). Once more, we have progressive measurability, and thus, 
by Tonelli's theorem, 
\begin{equation*}
\begin{split}
    \frac13 g(t)        
    & \le \EE\sup_{0\le s\le T}\abs{a'_{s}-a_s}^2 + \sup_{0\le s\le T} \int_0^T\abs{B'(s,r)-B(s,r)}^2\,dr \cdot  \int_0^t\EE\sup_{0\le \gamma\le r} (u^\ast_\gamma)^2dr\\
    &\quad\quad + \sup_{0\le s\le T} \int_0^TB'(s,r)^2\,dr\cdot \int_0^t g(r)\,dr.
\end{split}
\end{equation*}
In virtue of Proposition \ref{lemma: apriori estimate for a} and Lemma \ref{lemma: apriori estimate for B} applied to $((h_0',H',\lambda'),P,q)$, we can write
\[
\begin{split}
    \frac13 g(t)        
    & \le \EE\sup_{0\le s\le T}\abs{a'_{s}-a_s}^2 + \sup_{0\le s\le T} \int_0^T\abs{B'(s,r)-B(s,r)}^2\,dr \cdot  
    TM(h_0,H,\lambda,P,q)\\
    &\quad\quad + \left\{\frac{T}{2(\lambda')^2}\left[
j_1(\lambda') + j_2(H',\lambda')
\right]^2 + \frac{1}{(\lambda')^2}\left[
4\varrho^2 T + \norm{H'}^2_{L^2}
\right]\right\}\cdot \int_0^t g(r)\,dr.
\end{split}
\]
By means of Propositions \ref{prop: 1} and \ref{prop: 2},
\[
\begin{split}
    g(t)        
    & \le \mathscr L_{\Delta} (h_0,H,\lambda,h_0',H',\lambda',P,q)\norm{x'-x}^2_\X +
    \\
    &\quad\quad + 3\left\{\frac{T}{2(\lambda')^2}\left[
j_1(\lambda') + j_2(H',\lambda')
\right]^2 + \frac{1}{(\lambda')^2}\left[
4\varrho^2 T + \norm{H'}^2_{L^2}
\right]\right\}\cdot \int_0^t g(r)\,dr,
\end{split}
\]
where $\mathscr L_{\Delta}$ is defined in \eqref{eq: LDelta}. By means of Gr\"onwall's inequality, we obtain once more, after taking the square root, 
\begin{equation}\label{eq: preGronwall}
    \begin{split}
    &\norm{\mathscr S(h_0',H',\lambda',P,q)-\mathscr S(h_0,H,\lambda,P,q) }_{\mathcal A}\le \mathscr L_{\Delta}(h_0,H,\lambda,h_0',H',\lambda',P,q)^{1/2}\cdot
    \\&\quad \cdot\exp\left\{
\frac{3T}{2}\left[\frac{T}{2(\lambda')^2}\left(
j_1(\lambda') + j_2(H',\lambda')
\right)^2 + \frac{1}{(\lambda')^2}\left(
4\varrho^2 T + \norm{H'}^2_{L^2}
\right)\right]
\right\}\cdot\norm{x'-x}_\X.
\end{split}
\end{equation}
To conclude, we observe that by construction
\begin{equation}\label{eq: increasing}
    \norm{H}_{L^2}< \norm{\hat{H}}_{L^2} + r, \quad \norm{H'}_{L^2}< \norm{\hat{H}}_{L^2} + r,\quad \norm{h_0}_\infty<\norm{\hat{h}_0}_\infty + r,\quad \norm{h'_0}_\infty<\norm{\hat{h}_0}_\infty + r,
\end{equation}
and
\begin{equation}\label{eq: decreasing}
    \hat{\lambda}-r < \lambda,\quad \hat{\lambda}-r < \lambda'.
\end{equation}
After direct inspection of all the functions constituting the whole expression on the right-hand side of \eqref{eq: preGronwall}, we see that this expression is increasing in the variables
\[
\norm{H},\,\,\norm{H'},\,\,\norm{h_0}_\infty,\,\,\norm{h_0'}_\infty
\]
(indeed, we have positive powers thereof), while decreasing in the variables $\lambda$ and $\lambda'$ (indeed, we have negative powers thereof). Thus, upon replacing any appearance of $\norm{H},\norm{H'},\norm{h_0}_\infty$ and $\norm{h_0'}_\infty$ with the corresponding term in \eqref{eq: increasing}, and any appearance of $\lambda$ and $\lambda'$ with the corresponding term in \eqref{eq: decreasing}, we obtain, for a constant $L_\X=L_\X(\hat{h}_0,\hat{H},\hat{\lambda},P,q,r)$ which does not depend on $x$ nor on $x'$ that
\[
\norm{\mathscr S(h_0',H',\lambda',P,q)-\mathscr S(h_0,H,\lambda,P,q) }_{\mathcal A}\le L_\X(\hat{h}_0,\hat{H},\hat{\lambda},P,q,r)
\cdot\norm{x'-x}_\X,
\]
for all $x=(h_0,H,\lambda),x'=(h_0',H',\lambda')\in B^\MM_r(\hat{h}_0,\hat{H},\hat{\lambda})$, $P\in\mathscr P$
and $q\in \R$, showing that $\mathscr S$ is locally Lipschitz on $\MM$ for fixed $(P,q)\in\mathscr P\times \R$.

\end{proof}

\begin{remark}\label{rmk: LX}
    We observe here that, by direct inspection, the expression 
    \[
     L_\X(\hat{h}_0,\hat{H},\hat{\lambda},P,q,r),
    \]
    is increasing with respect to both the variables $\norm{P}_\PP$ and $\abs{q}$, a fact that will turn out to be very useful later.
\end{remark}

\subsection{Strategy of proof of Theorem \ref{thm: stability wrt P} and Theorem \ref{thm: stability wrt q}} The proof of Theorems \ref{thm: stability wrt P} and \ref{thm: stability wrt q} will first rely on Lemma \ref{lemma: apriori estimate for B}, whose aim, as we have already seen, is to provide a more precise estimate for the kernel $B$ in terms of $\lambda$ and $H$, and on Lemma \ref{lemma: estimate for a - a tilde} and \ref{lemma: estimate for a - a tilde 2} respectively. In a second step, we will adapt the Gr\"onwall argument present in the proof of Theorem \ref{thm: main result} and conclude.

\begin{lemma}\label{lemma: estimate for a - a tilde}
Fix $(h_0,H,\lambda)\in\mathscr M$ and $q\in \R$. Then, for any $P,P'\in\mathscr P$ we have
\[
\EE\sup_{t\le T}\abs{a_t-a'_t}^2 \le 
\mathscr L_2(\lambda,H) 
\norm{P-P'}^2_{\mathscr P},
\]
where $\mathscr L_2$ is defined in \eqref{eq: L2}, 
\[
a'_t:= \frac{1}{2\lambda} \left(\EE[P'_t-P'_T |\mathcal F_t] -\tilde{h}_0(t) + \langle \Theta'_t,K_t \rangle_{L^2} + \langle \boldsymbol{\Gamma}_t^{-1}K_t, \mathbbm{1}_{t} (\tilde{h}_0,q)^\top \rangle_{L^2}\right),
\]
and
\[
\Theta'_t(s):= - \left(
    \boldsymbol{\Gamma}_t^{-1}\mathbbm{1}_{\cdot\ge t}\, \EE[P'_\cdot - P'_T|\mathcal F_t]\, \underline{e}_1
    \right)(s).
\]
\end{lemma}

\begin{proof}
We will closely follow the proof of Lemma \ref{lemma: 5}. We have
\begin{equation}\label{eq: lemma estimate for a - a tilde (1)}
    a_t-a'_t=\frac{1}{2\lambda} \left(
\EE[{P}_t-{P}_T |\mathcal F_t]-
\EE[P'_t-P'_T |\mathcal F_t] + \langle \Theta_t-\Theta'_t,K_t \rangle_{L^2} \right),
\end{equation}
and
\[
\EE\sup_{t\le T}\abs{\langle \Theta_t-\Theta'_t,K_t \rangle_{L^2}}^2\le \EE\sup_{t\le T}\int_t^T\abs{\Theta_t(s)-\Theta'_t(s)}^2ds\,\sup_{t\le T}\int_t^T\abs{K(s,t)}^2ds.
\]
Following Lemma \ref{lemma: 5}, we obtain once more
\[
\begin{split}
    &\int_t^T\abs{\Theta_t(s)-\Theta'_t(s)}^2ds
    \\ &\quad \le \int_t^T\abs{\boldsymbol{D}^{-1}_t\left[\1_{\cdot\ge t}
\EE[P_\cdot -P_T |\mathcal F_t]-\1_{\cdot\ge t}
    \EE[\hat{P}_\cdot-\hat{P}_T |\mathcal F_t] \right](s)}^2ds\\
    &\quad\quad + \int_t^T4\phi^2\abs{(\boldsymbol{1}_t\boldsymbol{D}^{-1}_t)\left[\1_{\cdot\ge t}
\EE[P_\cdot -P_T |\mathcal F_t]- \1_{\cdot\ge t}
    \EE[\hat{P}_\cdot-\hat{P}_T |\mathcal F_t] \right](s)}^2ds\\
    &\quad\le \left( \norm{\boldsymbol{D}^{-1}_t}_{op}^2+4\phi^2\norm{\boldsymbol{1}_t\boldsymbol{D}^{-1}_t}_{op}^2 
    \right)\int_t^T 
\EE[P_r -\hat{P}_r -(P_T -\hat{P}_T )|\mathcal F_t]^2dr\\
&\quad\le \frac{1}{2\lambda^2}(1+2\phi^2T^2)\left(
\int_t^T\EE[P_r -\hat{P}_r |\mathcal F_t]^2dr + \EE[P_T-\hat{P}_T|\mathcal F_t]^2(T-t)
\right)\\
&\quad\le \frac{T}{\lambda^2}(1+2\phi^2T^2)\sup_{0\le t\le r\le T}\EE[P_r-P_r'|\mathcal F_t]^2.
\end{split}
\]
Upon taking the supremum in $t$ and then expectation, 
\[
\EE\sup_{t\le T} \int_t^T\abs{\Theta_t(s)-\Theta'_t(s)}^2ds\le 
\frac{T}{\lambda^2}(1+2\phi^2T^2) \norm{P-P' }^2_{\mathscr P}.
\]
We therefore obtain
\[
\EE\sup_{t\le T}\abs{\langle \Theta_t-\Theta'_t,K_t \rangle_{L^2}}^2\le \frac{T}{\lambda^2}(1+2\phi^2T^2)
\left[
2\norm{H}^2_{L^2} +T(1+8\varrho^2)
\right]
\norm{P-P' }^2_{\mathscr P} .
\]
Restarting from \eqref{eq: lemma estimate for a - a tilde (1)}, taking first $\sup_{t\le T}$ and then expectation, we finally arrive at
\[
\begin{split}
    \EE\sup_{t\le T}\abs{a_t-a'_t}^2 &\le 
    \mathscr L_2(\lambda,H)
\norm{P-P' }^2_{\mathscr P}. 
\end{split}
\]
    
\end{proof}

We are ready to prove Theorem \ref{thm: stability wrt P}.  

\begin{proof}[Proof of Theorem \ref{thm: stability wrt P}]
We follow the Gronwall-type argument of Theorem \ref{thm: main result}. For brevity, set $u^\ast_{t}:=\mathscr S(h_{0},H,\lambda,P,q)_t$ and $(u')^\ast_{t}:=\mathscr S(h_{0},H,\lambda,P',q)_t$ with $0\le t \le T$. Fix $t$. Then directly by \eqref{eq: uast}, we have 
\[
    u^\ast_{s}-(u')^\ast_s = a_{s}-a'_s 
    + \int_0^sB(s,r)[u^\ast_{r}-(u')^\ast_r]\,dr,\quad 0\le s\le t.
\]
Set $g(t):=\EE\sup_{0\le s\le t}\abs{u^\ast_s - (u')^\ast_s }^2,\,0\le t \le T$. Then, arguing exactly as previously, we obtain
\[
\frac12 g(t)\le  \EE\sup_{s\le T}\abs{a_s-a'_s}^2 + \sup_{s\le T}\int_0^TB(s,r)^2dr\cdot\int_0^tg(r)\,dr,\quad 0\le t\le T,
\]
and 
\[
g(t)\le 2\EE\sup_{s\le T}\abs{a_s-a'_s}^2\exp\left\{2\sup_{s\le T}\int_0^TB(s,r)^2dr\cdot t  \right\},\quad 0\le t \le T.
\]
Combining this with Lemma \ref{lemma: apriori estimate for B} and \ref{lemma: estimate for a - a tilde} and evaluating at $t=T$, we arrive at 
\[
\norm{\mathscr S(h_{0},H,\lambda,P,q) - \mathscr S(h_{0},H,\lambda,P',q)}_{\mathcal{A}} \le L_\PP(H,\lambda) \norm{P-P' }_{\mathscr P},
\]
with $L_\PP$ defined in \eqref{eq: LP}.
As noted earlier, the functions defining $L_\PP$ are increasing in $\norm{H}_{L^2}$ and decreasing in $\lambda$, and thus \eqref{eq: stability wrt P} easily follows.

\end{proof}

Let us now move to prove another auxiliary lemma, whose proof is similar to that of Lemma \ref{lemma: estimate for a - a tilde}.

\begin{lemma}\label{lemma: estimate for a - a tilde 2}
Fix $(h_0,H,\lambda)\in\MM$ and $P\in\PP$. Then, for any $q,q'\in\R$ we have
\[
\EE\sup_{t\le T}\abs{a_t-a'_t}^2 \le \mathscr L_3(\lambda,H) \abs{q-q'}^2,
\]
where $\mathscr L_3$ is defined in \eqref{eq: L3} and where
\begin{equation}\label{eq: a(q)}
    a'_t:= \frac{1}{2\lambda} \left(\EE[P_t-P_T |\mathcal F_t] -h_0(t) +2\varrho q' + \langle \Theta_t,K_t \rangle_{L^2} + \langle \boldsymbol{\Gamma}_t^{-1}K_t, \mathbbm{1}_{t} (h_0-2\varrho q',q')^\top \rangle_{L^2}\right),
\end{equation}
namely in \eqref{eq: a} we have replaced $q$ with $q'$.
\end{lemma}

\begin{proof}
The proof follows the proof of Lemma \ref{lemma: 6}. We have
\begin{equation}\label{eq: lemma estimate for a - a tilde (2)}
    a_t-a'_t=\frac{1}{2\lambda} \left(2\varrho(q-q') +\langle \boldsymbol{\Gamma}_t^{-1}K_t, \mathbbm{1}_{t} (h_0-2\varrho q,q)^\top - \mathbbm{1}_{t} (h_0-2\varrho q',q')^\top \rangle_{L^2}  \right).
\end{equation}
Notice that this quantity is deterministic. By expanding the scalar product by means of \eqref{eq: expansion of L2 scalar product}, we obtain
\[
\begin{split}
     &\langle \boldsymbol{\Gamma}_t^{-1}K_t, \mathbbm{1}_{t} (h_0-2\varrho q,q)^\top - \mathbbm{1}_{t} (h_0-2\varrho q' ,q' )^\top \rangle_{L^2} = \\
     & \quad\quad =\int_t^T\left[
     (\boldsymbol{D}^{-1}_t2\varrho\1_{t<\cdot})(s) + (\boldsymbol{D}^{-1}_tG(\cdot,t))(s) + (2\phi\boldsymbol{D}^{-1}_t\boldsymbol{1}^\ast_t \1_{t\le\cdot})(s)
     \right][-2\varrho q+2\varrho q' ]\,ds \\
     & \quad\quad\quad +(q-q' )\int_t^T\left[
     (-4\phi\varrho\boldsymbol{1}_t\boldsymbol{D}^{-1}_t\1_{t<\cdot})(s)  + (- 2\phi\boldsymbol{1}_t\boldsymbol{D}^{-1}_tG(\cdot,t))(s)\right]ds \\
     &\quad\quad\quad + 
     (q-q' )\int_t^T\left[  (2\phi\operatorname{id}\1_{t\le\cdot})(s)+ (- 4\phi^2\boldsymbol{1}_t\boldsymbol{D}^{-1}_t\boldsymbol{1}^\ast_t\1_{t\le\cdot})(s)
     \right]ds.
    \end{split}
\]
Arguing in the same spirit of Lemma \ref{lemma: 6}, we arrive at
\[
\begin{split}
    &\abs{\langle \boldsymbol{\Gamma}_t^{-1}K_t, \mathbbm{1}_{t} (h_0-2\varrho q,q)^\top - \mathbbm{1}_{t} (h_0-2\varrho q',q')^\top \rangle_{L^2}}\\
    &\quad\quad\le \left\{
    \frac{2\varrho^2T}{\lambda}+\frac{\varrho\sqrt{T}\norm{H}_{L^2}}{\lambda}+\frac{2\sqrt{2}\phi\varrho T^2}{\lambda} + \frac{\phi T^{3/2}\norm{H}_{L^2}}{\sqrt{2}\lambda} + 2\phi T +\frac{\phi^2 T^3}{\lambda}
    \right\}\cdot\abs{q-q'}.
\end{split}
\]
By restarting from \eqref{eq: lemma estimate for a - a tilde (2)}, we obtain $\EE\sup_{t\le T}\abs{a_t-a'_t}^2 \le \mathscr L_3(\lambda,H) \abs{q-q'}^2$.
\end{proof}

We now prove Theorem \ref{thm: stability wrt q}.  

\begin{proof}[Proof of Theorem \ref{thm: stability wrt q}]
The proof is very close to the proof of Theorem \ref{thm: stability wrt P}. For brevity let $u^\ast_{t}:=\mathscr S(h_{0},H,\lambda,P,q)_t$ and let $(u')^\ast_{t}:=\mathscr S(h_{0},H,\lambda,P,q')_t$ with $0\le t \le T$. Then
\[
    u^\ast_{s}-(u')^\ast_s = a_{s}-a'_s 
    + \int_0^sB(s,r)[u^\ast_{r}-(u')^\ast_r]\,dr,\quad 0\le s\le t.
\]
where $a'_s$ is defined by \eqref{eq: a(q)}. Setting once more $g(t):=\EE\sup_{0\le s\le t}\abs{u^\ast_s - (u')^\ast_s }^2,\,0\le t \le T$, we obtain again
\[
g(T)\le 2\EE\sup_{s\le T}\abs{a_s-a'_s}^2\exp\left\{2\sup_{s\le T}\int_0^TB(s,r)^2dr\cdot T  \right\},
\]
which, combined with Lemma \ref{lemma: apriori estimate for B} and \ref{lemma: estimate for a - a tilde 2}, gives us the sought result
\[
    \norm{\mathscr S(h_{0},H,\lambda,P,q) - \mathscr S (h_{0},H,\lambda,P,q')}_{\mathcal{A}} \le L_\R(H,\lambda) \abs{q-q'},
\]
and \eqref{eq: stability wrt q}, where
\[
L_\R(H,\lambda) := \left(2\mathscr L_3(\lambda,H)\right)^{1/2}
\cdot\exp\left\{
\frac{T^2}{2\lambda^2}\left[
j_1(\lambda) + j_2(H,\lambda)
\right]^2
+\frac{T}{\lambda^2}\left[
4\varrho^2T + \norm{H}^2_{L^2}
\right]
\right\}.
\]
\end{proof}

\subsection{Proof of Theorem \ref{thm: joint continuity}}

In light of this, we are now ready to prove Theorem \ref{thm: joint continuity} as well.

\begin{proof}[Proof of Theorem \ref{thm: joint continuity}]
Fix an arbitrary point $\hat{y}=((\hat{h}_0,\hat{H},\hat{\lambda}),\hat{P},\hat{q})\in \mathscr Y$. Consider the ball $B^\mathscr Y_r(\hat{y})\subset\mathscr Y $ where $r<\frac{\hat{\lambda}}{5+4\sqrt{T}}$. Let $y=((h_0,H,\lambda),P,q)$ and $y'=((h_0',H',\lambda'),P',q')$ be in the open ball $B^\mathscr Y_r(\hat{y})$. 
Then, by repeatedly applying the triangle inequality, we have
\[
\begin{split}
    &\norm{\mathscr S(h_{0},H,\lambda,P,q) -\mathscr S(h_0',H',\lambda',P',q')}_{\mathcal A}  
    \\ &\quad \le \norm{\mathscr S(h'_{0},H',\lambda',P',q') - \mathscr S(h'_{0},H',\lambda',P',q) }_{\mathcal A} 
    \\&\quad\quad+
    \norm{\mathscr S(h'_{0},H',\lambda',P',q) - \mathscr S(h'_{0},H',\lambda',P,q) }_{\mathcal A} 
    \\ &\quad\quad+
    \norm{\mathscr S(h'_{0},H',\lambda',P,q) - \mathscr S(h_{0},H,\lambda_,P,q) }_{\mathcal A}.
\end{split}
\]
By means of Theorem  \ref{thm: stability wrt q} and \ref{thm: stability wrt P},
\[
\begin{split}
    &\norm{\mathscr S(h_{0},H,\lambda,P,q) -\mathscr S(h_0',H',\lambda',P',q')}_{\mathcal A}
    \\ &\quad \le 
    L_\R(H',\lambda')\abs{q-q'} + L_\PP(H',\lambda')
    \norm{P-P'}_{\mathscr P} 
    \\ &\quad\quad
    + \norm{\mathscr S(h'_{0},H',\lambda',P,q) - \mathscr S(h_{0},H,\lambda_,P,q) }_{\mathcal A}
    \\ &\quad \le 
    \{L_\R(H',\lambda') + L_\PP(H',\lambda')\}
    \norm{y-y'}_{\mathscr Y} 
    \\ &\quad\quad
    + \norm{\mathscr S(h'_{0},H',\lambda',P,q) - \mathscr S(h_{0},H,\lambda_,P,q) }_{\mathcal A}.
\end{split}
\]
However, by construction, $(h_0,H,\lambda)$ and $(h'_0,H',\lambda')$ belong to $B^\MM_r(\hat{h}_0,\hat{H},\hat{\lambda})$. Thus, in view of Theorem \ref{thm: main result} we have
\[
\begin{split}
    &\norm{\mathscr S(h_{0},H,\lambda,P,q) -\mathscr S(h_0',H',\lambda',P',q')}_{\mathcal A}
    \\ &\quad \le 
    \{L_\R(H',\lambda') + L_\PP(H',\lambda')\}
    \norm{y-y'}_{\mathscr Y} 
    \\ &\quad\quad
    + L_\X(\hat{h}_0,\hat{H},\hat{\lambda},P,q,r) 
    \norm{ (h'_{0},H',\lambda') - (h_{0},H,\lambda) }_{\X}
    \\ &\quad\le
    \{L_\R(H',\lambda') + L_\PP(H',\lambda') + L_\X(\hat{h}_0,\hat{H},\hat{\lambda},P,q,r)\}\norm{y-y'}_{\mathscr Y}. 
\end{split}
\]
To finish off, we observe that $L_R$ and $L_\PP$ are increasing with respect to $\norm{H'}_{L^2}$ and decreasing with respect to $\lambda'$ (refer to \eqref{eq: LR} and \eqref{eq: LP}). Additionally, in view of Remark \ref{rmk: LX}, we know that $L_\X(\hat{h}_0,\hat{H},\hat{\lambda},P,q,r)$ is increasing with respect to both the variables $\norm{P}_\PP$ and $\abs{q}$. Thus, similarly to the proof Theorem \ref{thm: main result}, upon replacing any appearance of $\lambda'$ with $\hat{\lambda}-r$, $\norm{H'}_{L^2}$ with $\norm{\hat{H}}_{L^2}+r$, $\norm{P}_{\PP}$ with $\norm{\hat{P}}_{\PP}+r$, and $\abs{q}$ with $\abs{\hat{q}}+r$, we obtain 
\[
    \norm{\mathscr S(h_{0},H,\lambda,P,q) -\mathscr S(h_0',H',\lambda',P',q')}_{\mathcal A}
    \le 
     C(\hat{y},r)\norm{y-y'}_{\mathscr Y},
\]
for a constant which depends only on the base point $\hat{y}$ and the radius $r$, and the result is proven.
\end{proof}

\subsection{Proof of Theorem \ref{thm: continuity of J}}

\begin{proof}[Proof of Theorem \ref{thm: continuity of J}]
Fix an arbitrary point $\hat{y}=((\hat{h}_0,\hat{H},\hat{\lambda}),\hat{P},\hat{q})\in\mathscr Y$ and choose $r<\frac{\hat{\lambda}}{5+4\sqrt{T}}$. Let $y=((h_0,H,\lambda),P,q)$ and $y'=((h_0',H',\lambda'),P',q')$ be in the open ball $B_r^\mathscr Y(\hat{y})$. For convenience we write $J^{\ast}(h_0,H,\lambda,P,q)$ as $J_1+\dots +J_4$ where
\begin{align*}
    J_1(h_0,H,\lambda,P,q) &:= \EE\int_0^TS^{u^\ast}_tu^\ast_t\,dt, & J_2(h_0,H,\lambda,P,q) &:= \EE [Q^{u^\ast}_TP_T],\\
    J_3(h_0,H,\lambda,P,q) &:= -\phi\EE\int_0^T(Q^{u^\ast}_t)^2\,dt, & J_4(h_0,H,\lambda,P,q) &:= -\varrho\EE[ (Q^{u^\ast}_T)^2],
\end{align*}
with $u^\ast=\mathscr S(h_0,H,\lambda,P,q)$ and similarly for $(u')^\ast$. Starting with $J_1$ and expanding we have
\[
\begin{split}
    J_1(y) &= \EE\int_0^TP_tu^\ast_t\,dt - \lambda\EE\int_0^T(u^\ast_t)^2dt-\EE\int_0^Th_0(t)u^\ast_t\,dt-\EE\int_0^T(\boldsymbol{H}u^\ast)(t)u^\ast_t\,dt
    \\&=: J_1^{(1)}(y)+J_1^{(2)}(y)+J_1^{(3)}(y)+J_1^{(4)}(y)
\end{split}
\]
where $\boldsymbol{H}$ is the integral operator induced by $\1_{s<t}H(t-s)$. The first and third terms in this expression are essentially the scalar product of $L^2(\Omega\times [0,T])$, which, given the integrability properties of $P$ and $u^\ast$, turn out to be locally Lipschitz continuous in $P$, $u^\ast$, and $h_0$, and therefore in $(h_0,H,\lambda,P,q)$ in force of Theorem \ref{thm: joint continuity} and Proposition \ref{prop: apriori est. for u}. Namely, there exists a constant $C(\hat{y},r)$ (allowed to change from line to line) for which
\[
\abs{ J_1^{(1)}(y)-J_1^{(1)}(y')+ J_1^{(3)}(y)-J_1^{(3)}(y')}\le C(\hat{y},r)\norm{y-y'}_{\X\times\PP\times\R}.
\]
For the term $J_1^{(2)}$, we similarly obtain
\[
\begin{split}
\abs{  J_1^{(2)}(y)-J_1^{(2)}(y') } 
&\le \abs{\lambda-\lambda'}\EE\int_0^T(u^\ast_t)^2dt + \lambda'\abs{\EE\int_0^T[(u^\ast_t)^2 - ((u_t')^\ast)^2]dt }
\\&
\le C(\hat{y},r)\norm{y-y'}_{\X\times\PP\times\R}.
\end{split}
\]
To deal with the term $J_1^{(4)}$, several applications of H\"older's inequality yield
\[
\begin{split}
    &\abs{\EE\int_0^T(\boldsymbol{H}'(u')^\ast)(t)(u')^\ast_{t}\,dt-\EE\int_0^T(\boldsymbol{H}u^\ast)(t)u^\ast_t\,dt}
    \\ &\quad \le \EE\left[\norm{\boldsymbol{H}'(u')^\ast}_{L^2}\norm{(u')^\ast-u^\ast}_{L^2}\right] + \EE\left[\norm{\boldsymbol{H}'(u')^\ast-\boldsymbol{H}u^\ast}_{L^2}\norm{u^\ast}_{L^2} \right]
    \\ &\quad \le \EE\left[\norm{\boldsymbol{H}'}_{op}\norm{(u')^\ast}_{L^2}\norm{(u')^\ast-u^\ast}_{L^2}\right] + \EE\left[\norm{\boldsymbol{H}'((u')^\ast-u^\ast)}_{L^2}\norm{u^\ast}_{L^2} \right] 
    \\ & \quad\quad+ \EE\left[\norm{(\boldsymbol{H}'-\boldsymbol{H})u^\ast}_{L^2}\norm{u^\ast}_{L^2} \right]
    \\ &\quad \le \norm{\boldsymbol{H}'}_{op}\EE\left[\norm{(u')^\ast}_{L^2}\norm{(u')^\ast-u^\ast}_{L^2}\right] + \norm{\boldsymbol{H}'}_{op}\EE\left[\norm{(u')^\ast-u^\ast}_{L^2}\norm{u^\ast}_{L^2} \right] 
    \\ & \quad\quad+ \norm{\boldsymbol{H}' -\boldsymbol{H}}_{op}\EE\left[\norm{u^\ast}_{L^2}^2 \right]
    \\ &\quad \le \norm{\boldsymbol{H}'}_{L^2}\left[\EE\norm{(u')^\ast}_{L^2}^2\right]^{1/2}\left[\EE\norm{(u')^\ast-u^\ast}_{L^2}^2\right]^{1/2} + \norm{\boldsymbol{H}'}_{L^2}\left[\EE\norm{(u')^\ast-u^\ast}_{L^2}^2\right]^{1/2}\left[\EE\norm{u^\ast}_{L^2}^2 \right]^{1/2} 
    \\ & \quad\quad+ \norm{\boldsymbol{H}' -\boldsymbol{H}}_{L^2}\EE\left[\norm{u^\ast}_{L^2}^2 \right] 
    \\ & \quad \le C(\hat{y},r)\norm{y-y'}_{\X\times\PP\times\R},
\end{split}
\]
in virtue once more of Theorem \ref{thm: joint continuity} and Proposition \ref{prop: apriori est. for u}. 

Continuing with $J_2$, by H\"older's inequality once more
\[
\begin{split}
    &\abs{J_2(h'_{0},H',\lambda',P',q')-J_2(h_0,H,\lambda,P,q) } \\
    &\quad \le \abs{\EE\left(q'-q+\int_0^T(u^\ast-(u')^\ast_{s})\,ds\right) P'_{T} } + \abs{\EE\left(q-\int_0^Tu^\ast_{s}\,ds\right) (P'_{T}-P_T) } 
    \\&\quad \le \abs{q'-q}\,\EE\abs{P_T}  + \left[\EE[P^2_T] \right]^{1/2}\left[\EE\abs{\int_0^T(u^\ast_s-(u')^\ast_{s})\,ds}^2 \right]^{1/2} 
    \\&\quad \quad + \abs{q}\EE\abs{P'_{T}-P_T} +
        \left[
        \EE\abs{\int_0^T u^\ast_s\,ds }^2
        \right]^{1/2}
        \left[
        \EE\abs{P'_{T}-P_T}^2
        \right]^{1/2}    
    \\&\quad \lesssim_{q,P} \abs{q'-q} +  \left[\EE\int_0^T(u^\ast_s-(u')^\ast_{s})^2\,ds\right]^{1/2}
    +\EE\abs{P'_{T}-P_T} 
    \\ &\quad\quad + 
    \left[
        \EE\abs{\int_0^T u^\ast_s\,ds }^2
        \right]^{1/2}
        \left[
        \EE\abs{P'_{T}-P_T}^2
        \right]^{1/2}      
    \\&\quad \le C(\hat{y},r)\norm{y-y'}_{\X\times\PP\times\R},
\end{split}
\]
where we have applied repeatedly Theorem \ref{thm: joint continuity} and Proposition \ref{prop: apriori est. for u}. Similarly, for $J_3$ we have
\[
\begin{split}
    &\abs{J_3(h'_{0},H',\lambda',P',q')-J_3(h_0,H,\lambda,P,q)}
    \\&\quad \le \phi \abs{\EE\int_0^T\left[(q')^2-q^2+
    \left(\int_0^t(u')^\ast_{s}\,ds\right)^2 - \left(\int_0^tu^\ast_s\,ds\right)^2 + 2q'\int_0^t(u^\ast_s-(u')^\ast_{s})\,ds
    \right]dt }
    \\&\quad\quad + \phi \abs{\EE\int_0^T
    2(q-q')\int_0^t(u^\ast_s)\,ds \,dt }
    \\&\quad \lesssim_{\phi} 
    \abs{q'^2-q^2}T +    \EE\int_0^T\abs{\int_0^t((u')^\ast_{s}+u^\ast_s)\,ds}\abs{\int_0^t((u')^\ast_{s}-u^\ast_s)\,ds}\,dt 
    \\ &\quad\quad     
    + 2\abs{q'}\EE\int_0^T \abs{\int_0^t((u')^\ast_{s}-u^\ast_s)\,ds}\,dt + 2\abs{q-q'}\EE\int_0^T \abs{\int_0^t u^\ast_s\,ds}\,dt 
    \\&\quad \le \abs{q'^2-q^2}T + \EE\int_0^T\int_0^t\abs{(u')^\ast_{s}+u^\ast_s}\,ds\int_0^t\abs{(u')^\ast_{s}-u^\ast_s}\,ds\,dt + 2\abs{q'}\EE\int_0^T \int_0^t\abs{(u')^\ast_{s}-u^\ast_s}\,ds\,dt 
    \\&\quad\quad + 2\abs{q-q'}\EE\int_0^T\int_0^t\abs{u^\ast_s}ds\,dt
    \\ &\quad \le  C(\hat{y},r)\norm{y-y'}_{\X\times\PP\times\R}.
\end{split}
\]
Finally for $J_4$ we have
\[
\begin{split}
    &\abs{J_4(h'_{0},H',\lambda',P',q')-J_4(h_0,H,\lambda,P,q)}
\\&\quad\lesssim_{\varrho}\abs{\EE\left(\int_0^T(u')^\ast_{s}\,ds\right)^2-\EE\left(\int_0^Tu^\ast_{s}\,ds\right)^2    }+ 2\abs{q'}\abs{\EE\int_0^T(u')^\ast_{s}\,ds-\EE\int_0^Tu^\ast_{s}\,ds    }
\\&\quad\quad + \abs{q'^2-q^2} + 2\abs{q-q'}\abs{\EE\int_0^Tu^\ast_s\,ds}
    \\&\quad\le 
     C(\hat{y},r)\norm{y-y'}_{\X\times\PP\times\R}. 
\end{split}
\]
For the sake of brevity, we have not kept track of the explicit expression of the constant $C(\hat{y},r)$, but this can be derived by means of all the auxiliary functions given in the Appendix, and the theorem is proven.
\end{proof}

\section{Neural-Network Approximation and Numerical Method}\label{sec:neural_network}

\subsection{Neural Approximation of the Solution Map}

In view of the results above, we now see that all the relevant maps for the optimal liquidation problem, namely
\[\mathscr M\times\mathscr P\times\R\ni ((h_0,H,\lambda),P,q)\mapsto \mathscr S(h_0, H,\lambda,P,q)\in\mathcal A,
\]
and
\[
    J^{\ast}(h_0,H,\lambda,P,q) :=J^{h_0,H,\lambda,P,q}\circ\mathscr S(h_0,H,\lambda,P,q),
\]
(with some of the variables possibly frozen) are locally Lipschitz continuous. As we have seen, the domains and codomains of these operators are subsets of suitable normed spaces, and so well-suited to be approximated by appropriate neural architectures, granting a Universal Approximation Theorem of some sort. For instance, the neural architectures introduced in \cite{BDG0} and whose scope has been widely extended in the subsequent work \cite{Non-metric} are suitable for this task because they grant, for instance, the following theorem.
\begin{theorem*}[from \cite{BDG0} (informal version)]\label{Thm:UAT}
    Consider two Banach spaces $(E,\norm{\cdot}_E)$ and $(F,\norm{\cdot}_F)$ and a continuous map $\varphi:E\to F$. Fix a compact set $\mathcal K\subset E$ and an error $\varepsilon>0$. Then there exists a suitable neural architecture $\mathcal{NN}_\theta:E\to F$ specified by a finite dimensional parameter vector $\theta\in\R^P$ for some $P\in\N$ such that
    \[
    \sup_{y\in\mathcal K}\norm{\mathcal {NN}_\theta(y)-\varphi(y)}_F <\varepsilon,
    \]
    namely the neural architecture $\mathcal{NN}_\theta$ is able to approximate the map $\varphi$ with an error at most $\varepsilon$ uniformly on the compact subset $\mathcal K$. 
\end{theorem*}

Therefore, in view of this last result, we may now construct a surrogate model for the relevant maps of the optimal liquidation problem through suitable neural architectures, e.g.
\[
    \sup_{y\in\mathcal K}\norm{\mathcal {NN}_\theta(y) - \mathscr S(y) }_{\mathcal{A}} < \varepsilon,
\]

with $y:=((h_0,H,\lambda),P,q)$ and where $\mathcal{K}$ is a pre-specified compact subset. The merits of this approach (as for any surrogate model) are
\begin{enumerate}
    \item Fast computation of the optimal trading strategies $u^\ast=\mathscr S(h_0,H,\lambda,P,q)$, as all the computational burden has been amortized during the training phase. Once the neural network has been trained, everything boils down to the evaluation of $\mathcal{NN}_\theta$ at the point of interest $y$.
    \item The error incurred by employing the surrogate model is uniform for all inputs $y$ in the compact subset $\mathcal K$ due to the Universal Approximation Theorem stated above. Thus, there is no need to re-run the whole procedure from the beginning if the input does not vary too wildly, and we can potentially overcome the limitations of postulating a specific parametric propagator model in real-world trading applications where market conditions are non-stationary.
\end{enumerate}
In other words, the overall goal of this methodology, based on neural architectures, is to tackle a non-Markovian optimal control problem with unknown state dynamics, where the neural network learns a latent representation in a data-driven and efficient manner.

In subsequent numerical experiments, we set $h_0=0$, and the unaffected price is given by
\[
P_t = P_0 + \int^t_0 \alpha_s ds + \sigma^P W^P_t,
\]
where $\sigma^P \in \R_+$, $W^P$ is a Brownian Motion, and $\alpha$ is an Ornstein-Uhlenbeck process with dynamics
\begin{equation}
d \alpha_t = - \gamma \alpha_t dt + \sigma^{\alpha} d W^{\alpha}_t, \label{OU signal}
\end{equation}
where $\gamma, \sigma^{\alpha} \in \R_+$ and $W^{\alpha}$ are Brownian motions independent of $W^P$. 
With these dynamics, the price process is dependent on the signal and, therefore, can be written as $P^{\alpha}$. Thus, the dependence of the optimal strategy on the signal is already included through its dependence on $P^\alpha$. If the map $\alpha \mapsto P^\alpha$ is continuous, then the map $\mathscr S$ also has continuous dependence on $\alpha$. 

There is also a difference between the theoretical map and the one used for training. $\mathscr S$ returns the whole strategy $u^*$, while the neural network in the numerical experiments returns its value at a given time $t$. We work on a fixed time grid $\{t_k\}_{k=0}^n$ with equal time steps of size $\Delta t = T/n$.
\[
0 = t_0 < t_1 < \cdots < t_n = T,
\]
and the optimal strategy is represented by an $(n+1)$-dimensional vector
\[
(u^*_{t_0}, u^*_{t_1}, \ldots, u^*_{t_n}).
\]
Then the learning problem can be written as 
\[
(t, H, \lambda, P^\alpha, q) \mapsto u^*_t.
\]
The theoretical results provide support and justification for approximating the complete optimal strategy by a neural network, and the numerical experiments use its finite-dimensional implementation on a fixed time grid.

\subsection{Dimensional Reduction}

In this section, we introduce a methodology that helps reduce the complexity of training a neural network to learn the optimal strategy.

\subsubsection{Inventory Rescaling}\label{sec: rescaling}

Recall that the optimal strategy can be written as
\[
u^*_t = a_t + (\mathbf{B}u^*)(t),
\]
or equivalently
\begin{equation}
    u^*_t = \left( (I - \mathbf{B})^{-1}a \right) (t) , \label{u^*}
\end{equation}
where $I$ is the identity. Based on the form of $a$ and $\mathbf{B}$ in \cite{https://doi.org/10.1111/mafi.12465}, we know that $a_t$ depends on the model parameters $(H, \lambda, P^ \alpha, q)$ while $\mathbf{B}$ does not depend on the signal or initial inventory. 

Additionally, $a_t$ has affine dependence on $\alpha_t$ and $q$. Indeed, this is seen by inspection of equations (4.7) and (4.17) in \cite{https://doi.org/10.1111/mafi.12465}, recalling that $\mathbf{\Gamma}^{-1}_t$ and $K_t$ do not depend on $\alpha$ and $q$, and that $\widetilde h_0(t)=h_0(t)-2\varrho q = -2\varrho q$ holds under our assumption of $h_0 \equiv 0$. Thus, we have
\[
a(H, \lambda, P^\alpha, q) = a(H, \lambda, P^\frac{\alpha}{q}, 1)q,
\]
for $q\neq 0$. This relation also holds for the optimal strategy $u^*$ since $\mathbf{B}$ does not depend on $\alpha$ or $q$. We write
\[
\begin{split}
    u^*_t(H, \lambda, P^\alpha, q) &= \left( (I-\mathbf{B})^{-1} a (H, \lambda, P^\alpha, q) \right) (t) \\
    &= \left( (I-\mathbf{B})^{-1} a (H, \lambda, P^\frac{\alpha}{q}, 1)q \right) (t) \\
    &= q\left( (I-\mathbf{B})^{-1} a (H, \lambda, P^\frac{\alpha}{q}, 1) \right) (t) \\
    &= q u^*_t(H, \lambda, P^\frac{\alpha}{q}, 1),
\end{split}
\]
for $q\neq 0$. This motivates defining the normalized strategy
\[
    \hat u_t^*(H,\lambda,P^\alpha)
    :=
    u_t^*(H,\lambda,P^\alpha,1).
\]
Then, for any $q\neq 0$, the strategy can be recovered by
\[
    u_t^*(H,\lambda,P^\alpha,q)
    =
    q\,\hat u_t^*\left(H,\lambda,P^\frac{\alpha}{q}\right).
\]
This relation means that it is only necessary to learn the optimal strategy corresponding to an initial inventory of $q=1$, and implementing the optimal strategy for other values of $q$ requires that the signal be scaled appropriately. This lowers the amount of necessary training data by eliminating dependence on the initial inventory.

\subsubsection{Finite-Dimensional Kernel Representation}

The admissible kernel $H \in \mathcal H$ is an infinite-dimensional input of the optimal strategy. For the purpose of numerical approximation, we need to represent this functional input by finitely many quantities. Let $(e_j)_{j\in\mathbb N}$ be an orthonormal basis of $L^2([0,T])$, and for each $N \in \mathbb N$, denote by $\Pi_N : \mathcal H \to \operatorname{span}\{e_1,\ldots,e_N\}$ the orthogonal projection onto the first $N$ basis functions. The projected kernel is therefore given by
\[
    H^N(t) = (\Pi_N H)(t) = \sum_{j=1}^{N} h_j e_j(t),  \quad h_j = \langle H,e_j\rangle_{L^2([0,T])}.
\]
Since the basis is complete in $L^2([0,T])$, we have
\[
    \|H-H^N\|_{L^2([0,T])} \longrightarrow 0 \quad \text{as } N\to\infty .
\]
However, the projection $H^N$ itself may not be admissible while the operator $\mathscr S$ is defined only on the admissible kernel class, so the projected $H^N$ cannot be inserted directly into \eqref{u^*}. We therefore use the coefficient vector $(h_1,\ldots,h_N)$ as a finite-dimensional encoding of the original admissible kernel $H$. This interpretation is consistent with the neural approximation result stated above. The theorem motivates approximating the continuous solution map $\mathscr S$ by a neural architecture acting on the full functional input. The implemented network is then a finite-dimensional realization of this idea, where the functional input $H$ is represented by its first $N$ basis coefficients. To focus on the finite-dimensional representation of the kernel, let $\psi$ denote the remaining inputs of the solution map. In this sense, we approximate
\[
    \mathscr S(H, \psi) \approx \mathcal{NN}(H, \psi) \approx \mathcal{NN}_\theta(h_1,\ldots,h_N, \psi),
\]
where $\mathcal{NN}$ is the ideal neural architecture acting on the full functional input $H$, while $\mathcal{NN}_\theta$ denotes the implemented finite-dimensional network acting on the coefficient vector. 
Consequently, with the rescaling method described above, the neural network is now trained to learn the map
\[
    \left( t,(h_1,\ldots,h_N),\lambda,P^\alpha \right) \mapsto \hat u_t^*,
\]
where $\hat u^*$ denotes the normalized optimal strategy. 
In the numerical experiments, the original admissible kernel $H$ is still used to compute the reference strategy and the objective; the coefficients are only used as the kernel input to the neural network.

\begin{figure}[H]
\centering
\resizebox{\textwidth}{!}{
\begin{tikzpicture}[
    every node/.style={font=\small},
    arrow/.style={->, thick},
    dashedarrow/.style={->, thick, dashed},
    point/.style={circle, fill, inner sep=1.8pt}
]

\node at (-2.35, 2.40) {$L^2$};

\draw[thick, blue!60, fill=blue!5]
    (-5.95,-2.15) -- (-1.55,-1.55) -- (-0.75,-0.70) -- (-5.15,-1.30) -- cycle;

\node[blue!70] at (-4.55,-2.38)
{$\operatorname{span}\{e_1,\ldots,e_N\}$};

\node[point, blue!70] (O) at (-5.05,-1.50) {};
\node[below left=0.03cm and 0.03cm of O] {$0$};

\draw[thick, green!55!black, fill=green!8]
    (O) -- (-5.05,1.45) -- (-2.35,0.10) -- cycle;

\node[green!45!black] at (-5.25,0.45) {$\mathcal H$};

\node[point, green!55!black] (H) at (-4.15,0.38) {};
\node[above right=0.10cm and -0.06cm of H, inner sep=1pt] {$H$};

\node[point, blue!80] (HN) at (-3.55,-1.45) {};

\draw[arrow, green!60!black]
    (H) -- (HN);
\node[green!45!black, inner sep=1pt] at (-4.08,-0.55) {$\Pi_N$};

\node[black, inner sep=1pt] (HNexp) at (-2.55,-2.05)
{$H^N=\sum_{j=1}^{N} h_j e_j$};

\draw[thin, blue!70]
    ($(HN)+(0.10,-0.08)$) -- ($(HNexp)+(-1.05,0.18)$);

\draw[thick, gray!70] (0.0,-2.45) -- (0.0,2.45);

\node at (2.35, 2.40) {$\mathcal A$};

\node[point, red!70] (SH) at (5.45,1.35) {};
\node[right=0.12cm of SH] {$\mathscr S(H)$};

\node[point, red!70] (NNH) at (5.05,0.05) {};
\node[right=0.12cm of NNH] {$\mathcal{NN}(H)$};

\node[point, red!70] (NNtheta) at (4.05,-1.25) {};
\node[right=0.12cm of NNtheta] {$\mathcal{NN}_{\theta}(h_1,\ldots,h_N)$};

\draw[arrow, red!70]
    (H) to[out=30,in=170]
    node[pos=0.55, above] {$\mathscr S$}
    (SH);

\draw[dashedarrow, red!70]
    (H) to[out=-5,in=180]
    node[pos=0.55, above] {$\mathcal{NN}$}
    (NNH);

\draw[dashedarrow, red!70]
    (HN) to[out=-10,in=195]
    node[pos=0.55, below] {$\mathcal{NN}_{\theta}$}
    (NNtheta);

\draw[<->, thick, gray!70]
    ($(NNtheta)+(0.15,0.15)$) -- node[right] {$\varepsilon$-close} ($(NNH)+(0.15,-0.15)$);

\draw[<->, thick, gray!70]
    ($(NNH)+(0.15,0.15)$) -- node[right] {$\varepsilon$-close} ($(SH)+(0.15,-0.15)$);

\end{tikzpicture}
}

\caption{Finite-dimensional encoding of admissible kernels, with all other inputs of the solution map held fixed and omitted for simplicity. The admissible kernel class $\mathcal H$ is represented as a closed convex cone. For an admissible kernel $H\in\mathcal H$, its projection $H^N=\sum_{j=1}^{N}h_j e_j$ lies in the finite-dimensional subspace $\operatorname{span}\{e_1,\ldots,e_N\}$. $H^N$ may lie outside $\mathcal H$ and therefore cannot be inserted into the solution operator $\mathscr S$ directly. $\mathscr S$ is applied to $H$, while the implemented neural network uses the coefficient vector $(h_1,\ldots,h_N)$ as input.}
\label{fig:kernel_encoding_solution_map}
\end{figure}
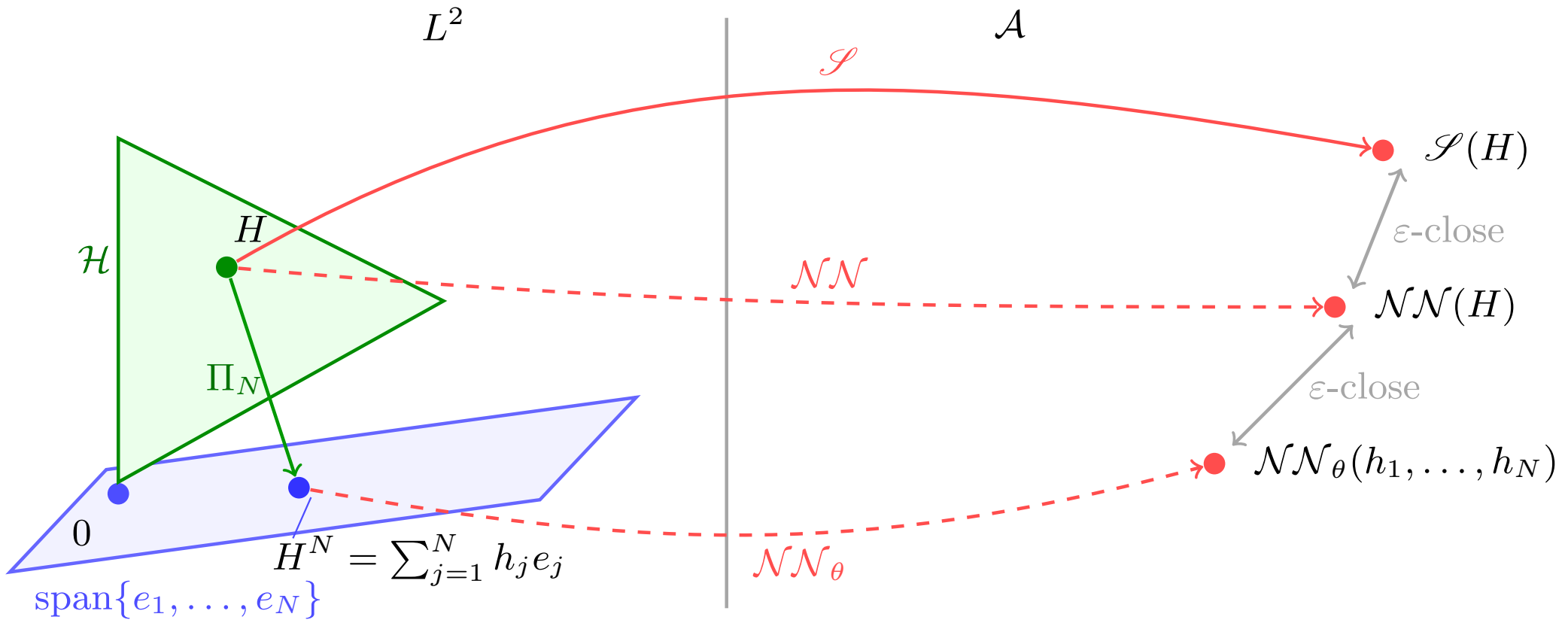

\subsection{Data Generation}

With the representation discussed above, we manage to represent the infinite-dimensional kernels with finitely many coefficients. The main challenge remaining is how to produce admissible kernels as well as what basis and truncation to use in the kernel representation.

\subsubsection{Generation of Admissible Kernels}
\label{sec: KernelSampling}

By Lemma 2.2 in \cite{https://doi.org/10.1111/mafi.12465}, if the kernel $H$ can be represented by
\begin{align}\label{eqn:H_mu}
    H(t) = \int_{\R_+} e^{-xt} \mu(dx), \quad 0 \le t \le T,
\end{align}
where $\mu$ is a nonnegative measure, then $H$ is within the class of admissible kernels. We describe two methods that we use to generate non-negative measures, which are then used to construct $H$ from \eqref{eqn:H_mu}.

\paragraph{\textbf{Basis Method}}

Our first method obtains a kernel $H$ by
\[
    H(t) = \int_{\R_+} e^{-xt} g(x)^2 dx, \quad 0 \le t \le T,
\]
where the function $g$ is given by
\begin{align}
    g(x) = \sum^{Z+1}_{j=1} y_j \beta_j (x),
\end{align}
where $Z$ is a Poisson random variable with parameter $\kappa_{\text{basis}}$, 
 $(\beta_j)_{j\in\mathbb{N}}$ is an orthonormal basis of $L^2([0,\infty))$, and $(y_j)_{j\in\mathbb{N}}$ is a sequence of independent random variables each of which is uniformly distributed on the interval $[-a_j,a_j]$ with $(a_j)_{j\in\mathbb{N}}\in\ell^2(\mathbb{N})$. In the numerical implementation, we choose the bounds in the form
\[
a_j = a_0 d^{j-1}, \quad j \geq 1,
\]
where $d \in (0,1)$ controls how quickly the coefficient range decays. Details about continuity of the map $g \mapsto \mathcal L (g^2) \big|_{[0,T]}$, the compactness of the sampling domain, and the construction of an orthonormal basis are given in Appendix~\ref{app: basis method}.
\begin{algorithm}[H]
\caption{Sampling an admissible kernel via basis method}
\label{alg:basis_method}
\begin{algorithmic}[1]
\Require
Time horizon $T>0$, Poisson intensity $\kappa_{\text{basis}}>0$, coefficient bounds $(a_j)_{j\ge1}\in\ell^2(\mathbb{N})$
\Ensure
A random admissible kernel $H(t)$
\State Choose an orthonormal basis $(e_j)_{j\ge1}$ of $L^2([0,1])$
\State Define $\beta_j(x)=e^{-x/2}e_j(e^{-x})$, yielding an orthonormal basis of $L^2([0,\infty))$
\State Sample $Z\sim\mathrm{Poisson}(\kappa_{\text{basis}})$
\For{$j=1$ to $Z+1$}
    \State Sample $y_j\sim\mathrm{Uniform}([-a_j,a_j])$
\EndFor
\State Define the random function
\[
g(x)=\sum_{j=1}^{Z+1}y_j\beta_j(x), \quad x\ge0
\]
\State Compute the Laplace transform
\[
H(t)=\int_0^\infty e^{-tx}g(x)^2\,dx, \quad t\in[0,T]
\]
\Return $H$
\end{algorithmic}
\end{algorithm}

\paragraph{\textbf{Piecewise Constant Method}}

The second method used to generate kernels defines $H$ by
\[
    H(t) = \int_{\R_+} e^{-xt} f(x) dx, \quad 0 \le t \le T.
\]
where the function $f$ is given by
\[
    f(x) = \sum_{j=1}^{Z+1} c_j \1 _{[b_{j-1},b_j)}(x),
\]
where $Z$ is a Poisson random variable with parameter $\kappa_{\text{piece}}$, $(c_j)_{j\in\mathbb{N}}$ is an iid sequence with $c_j\sim\mathrm{Exp}(\bar h)$, and $b_j = b_{j-1} + \ell_j$ where $(\ell_j)_{j\in\mathbb{N}}$ is an iid sequence with $\ell_j\sim\mathrm{Exp}(\bar \ell)$ and $b_0 = 0$. The parameters $\bar h$ and $\bar \ell$ denote the mean height and length of each interval, respectively. Thus, the function $f$ is piecewise constant over intervals of length $\ell_j$.
\begin{algorithm}[H]
\caption{Sampling an admissible kernel via piecewise constant method}
\label{alg:piecewise_method}
\begin{algorithmic}[1]
\Require
Time horizon $T>0$, Poisson intensity $\kappa_{\text{piece}}>0$, exponential parameters
$\bar \ell>0$ and $\bar h>0$
\Ensure
A random admissible kernel $H(t)$
\State Sample $Z\sim\mathrm{Poisson}(\kappa_{\text{piece}})$
\For{$j=1$ to $Z+1$}
    \State Sample the interval length $\ell_j\sim\mathrm{Exp}(\bar \ell)$
    \State Sample the interval height $c_j\sim\mathrm{Exp}(\bar h)$
\EndFor
\State Define the endpoints of intervals
\[
b_0=0, \quad b_j=\sum_{k=1}^j \ell_k, \quad j=1,\ldots,Z+1.
\]
\State Define the piecewise constant function
\[
g(x)=\sum_{j=1}^{Z+1} c_j \1 _{[b_{j-1},b_j)}(x), \quad x\ge0.
\]
\State Compute the Laplace transform
\[
H(t)=\int_0^\infty e^{-tx}g(x)\,dx, \quad t\in[0,T].
\]
\Return $H$
\end{algorithmic}
\end{algorithm}

\subsubsection{Kernel Encoding}

After generating an admissible kernel $H$, we introduce how to encode it as a finite-dimensional input for the neural network. We use a trigonometric basis on $L^2 ([0,T])$:
\[
\begin{split}
    e_1(t) &= \frac{1}{\sqrt{T}}, \\
    e_{2k}(t) &= \sqrt{\frac{2}{T}} \sin{\frac{2 \pi kt}{T}}, \quad k\in\N, \\
    e_{2k+1}(t) &= \sqrt{\frac{2}{T}} \cos{\frac{2 \pi kt}{T}}, \quad k\in\N,
\end{split}
\]
and the kernel $H$ is approximated by
\[
H^{n_H}(t) = \sum_{j=1}^{n_H} h_j e_j (t),
\]
where $h_j$ denotes the coefficients. In the numerical implementation, $h_j$ is computed on the fixed time grid $0=t_0 < \cdots < t_n = T$ using the trapezoidal rule
\[
\begin{split}
    h_j &= \langle H, e_j \rangle_{L^2([0,T])} \\
    & \approx \sum_{k=0}^{n-1} \frac{\Delta t}{2} \big( H(t_k) e_j(t_k) + H(t_{k+1})e_j(t_{k+1}) \big).
\end{split}
\]
Then we use the vector $(h_1, \ldots, h_{n_H})$ as the finite-dimensional representation of the kernel in the neural network input.

The truncation level $n_H$ controls the trade-off between reconstruction accuracy and input dimensions. To choose a proper truncation, we generated a sample of $500$ admissible kernels, with $250$ kernels obtained from each sampling methods. For each truncation level $n_H$, we computed the mean integrated reconstruction error over the full sample:
\[
    \text{Error}(n_H) = \frac{1}{500} \sum^{500}_{m=1} \int_0^T |H_m(t)-H_m^{n_H}(t)|^2 dt.
\]
As shown in Table~\ref{tab:kernel_reconstruction_error}, the reconstruction error decreases as $n_H$ increases, but the marginal improvement becomes smaller beyond $n_H=30$. In particular, increasing $n_H$ from $20$ to $30$ reduces the mean error by $1.030\times10^{-2}$, whereas the reductions obtained by increasing $n_H$ from $30$ to $40$ and from $40$ to $50$ are $5.240\times10^{-3}$ and $3.150\times10^{-3}$, respectively. We therefore choose $n_H=30$ as a practical compromise between reconstruction accuracy and the dimension of the neural-network input.
\begin{table}[htbp]
\centering
\begin{tabular}{c|c}
\hline
$n_H$ & Error \\
\hline
5  & $1.228\times 10^{-1}$ \\
10 & $6.162\times 10^{-2}$ \\
15 & $4.530\times 10^{-2}$ \\
20 & $3.280\times 10^{-2}$ \\
30 & $2.250\times 10^{-2}$ \\
40 & $1.726\times 10^{-2}$ \\
50 & $1.411\times 10^{-2}$ \\
\hline
\end{tabular}
\caption{Mean integrated reconstruction error for different truncation levels using the trigonometric basis.}
\label{tab:kernel_reconstruction_error}
\end{table}

\subsection{Numerical Reference Strategies}

In the numerical experiments, we compare the neural-network strategy with a reference strategy obtained by discretizing the explicit solution of \cite{https://doi.org/10.1111/mafi.12465}. We now briefly describe how it is implemented in our numerical experiments and what changes we have made.
Recall that the optimal strategy is characterized by the linear Volterra equation
\[
u^*_t = a_t + \int_0^t B(t,s) u^*_s ds,
\]
where $a$ and $B$ are given in equation (4.17) in \cite{https://doi.org/10.1111/mafi.12465}. Then the continuous-time equation can be discretized on the fixed time grid $\{ t_k \}_{k=0}^n$ by
\begin{equation}
u^{(n)} = (I_{n+1} - B^{(n)})^{-1} a^{(n)},
\label{VolterraEquation}
\end{equation}
where $a^{(n)} \in \R^{n+1}$ and $B^{(n)} \in \R^{(n+1) \times (n+1)}$ are given by
\[
\begin{split}
    a^{(n)} &= (a_{t_0}, \ldots, a_{t_n})^\top, \\
    B^{(n)}_{ij} &= \1_{\{ j \le i-1 \}} \int^{t_{j+1}}_{t_j} B(t_i,s) ds, \quad i,j = 0,1,\ldots,n.
\end{split}
\]
In our implementation, the entries of $a^{(n)}$ and $B^{(n)}$ are obtained by a Nystr\"om method discretization of the integral operators appearing in the expressions for $a$ and $B$. Compared with the numerical examples in \cite{https://doi.org/10.1111/mafi.12465}, our implementation also keeps the running inventory penalty $\phi$ in the discretization.
For the experiments with an OU-signal, the conditional expectation term $\EE [P_t - P_T | \mathcal{F}_t]$ entering $a$ is computed explicitly on the same grid using the OU dynamics of the signal. After constructing $a^{(n)}$ and $B^{(n)}$, the numerical reference strategy is obtained by solving \eqref{VolterraEquation}. Then the resulting vector $u^{(n)} = (u^*_{t_0},\ldots,u^*_{t_n})^\top$ is used as the reference strategy. Since the trading speed is implemented over the intervals $[t_k,t_{k+1})$, the neural-network comparison uses only the first $n$ values $(u^*_{t_0},\ldots,u^*_{t_{n-1}})$. The corresponding inventory path is then computed by
\[
Q_{t_k} = q - \Delta t \sum^{k-1}_{j=0} u^*_{t_j}, \quad k=0, \ldots, n.
\]
The neural network is then evaluated using the root mean square errors of strategy and inventory.

\subsection{Neural Network Architecture and Training Procedure}

The approximation result above motivates the use of neural-network surrogates for the solution map. After kernel projection and time discretization, the problem becomes a finite-dimensional, finite-horizon sequence-to-sequence problem. Feedforward computations on unrolled sequences can be represented by recurrent networks \cite{sperduti2015equivalence}, while universal approximation results are available directly for GRUs \cite{hoon2023minimal}: indeed, Corollary 17 therein proves a universal approximation result for GRUs for continuous, past-dependent, finite-length sequence-to-sequence maps on compact domains. 

We therefore employ a GRU whose hidden state carries the path-dependent information induced by the Volterra kernel across the time grid.

\subsubsection{Architecture}

The neural network uses a recurrent architecture to approximate the optimal trading strategy on a fixed time grid. At each time step, the network receives a vector of input features, including the normalized time, the finite-dimensional kernel representation, and, depending on the experiment, the price signal or price process.
The architecture has the form
\[
    \text{Input} \longrightarrow \text{Dense} \longrightarrow \text{GRU} \longrightarrow \text{Dense} \longrightarrow \text{Output}.
\]
The two dense layers use the hyperbolic tangent activation function, while the final output layer is linear. The GRU layer returns a sequence of hidden states so that the network produces one trading speed at each time point of the discretization grid. In other words, for an input sequence defined on
\[
    0=t_0<t_1<\cdots<t_n=T,
\]
the network outputs
\[
    \left(\hat u^*_{t_0},\ldots,\hat u^*_{t_{n-1}}\right),
\]
where $\hat u^*$ denotes the normalized trading strategy.
In the kernel-dependent experiments, the kernel input is given by the coefficient vector
\[
    (h_1,\ldots,h_{n_H}),
\]
which is repeated at each time step. We use $n_H=30$ basis coefficients. The dense layers contain $256$ nodes. The number of GRU nodes is chosen according to the complexity of the experiment: $256$ nodes are used in the no-signal kernel experiment, while $512$ nodes are used in the experiments with an OU signal, where the target strategy is more complex.

\subsubsection{Training and Validation Data}

The training data are generated from the model introduced above. Unless otherwise stated, the model parameters in the numerical experiments are fixed in each experiment for simplicity, as shown in Table~\ref{tab:model-parameters}. We mainly focus on learning the dependence of the optimal strategy on time, the kernel shape, and the realized signal path when the signal is included.
\begin{table}[!htbp]
    \centering
    \small
    \begin{tabular}{cccccccccc}
    \hline
    Parameter & $P_0$ & $\sigma^P$ & $\gamma$ & $\sigma^\alpha$ & $T$ & $n$ & $\lambda$ & $\varrho$ & $\phi$ \\
    \hline
    Value & $100$ & $0.5$ & $0.3$ & $0.5$ & $1$ & $600$ & $0.01$ & $5$ & $0.5$ \\
    \hline
    \end{tabular}
    \caption{Model parameters used in the numerical experiments. The OU parameters $\gamma$ and $\sigma^\alpha$ are used in the experiments with a predictive signal. For each simulated OU signal path, the initial value is sampled independently according to $\alpha_0\sim\operatorname{Unif}[-2,2]$.}
    \label{tab:model-parameters}
\end{table}

In the main experiments, we use the rescaling method described in \ref{sec: rescaling}. The network is trained with a fixed initial inventory $q=1$ and learns the normalized optimal strategy $\hat u^*$. For a general initial inventory $q$, the signal is first scaled down by $q$, and the strategy is then rescaled back by multiplying by $q$. For comparison, in the benchmark experiments, we also train another model where $q$ is directly included as an input variable and is trained with $q \in [1,10]$. We then evaluate the two methods both within the range $q \in [1,10]$ and outside it, at $q \in \{15, 20\}$. The latter experiments test the generalization of the networks.

The input variables depend on the experiment. In the no-signal or no-kernel cases, the corresponding inputs are removed. In the kernel-dependent experiments, each kernel is represented by its first $n_H=30$ basis coefficients, and the same coefficient vector is repeated at every time step. In some benchmark experiments, we also include the current price $P_t$ as an input. In the full OU-signal experiment, we remove $P_t$, because the relevant price information enters the strategy through the signal $\alpha_t$.

Admissible kernels and OU signal paths are generated using the methods described above. The training and validation samples are generated independently from the same sampling procedure. The validation data are used to select the model with the best out-of-sample performance, which is then used in the numerical comparison.

The model is trained with the Adam optimizer by minimizing the negative expected discretized performance functional
\[
\mathcal{L}(\theta) = -\EE \left[ \sum_{k=0}^{n-1} S_{t_k}^{u^\theta} u_{t_k}^{\theta}\Delta t + Q_T^{u^\theta}P_T - \phi\sum_{k=0}^{n-1} \left( Q_{t_k}^{u^\theta} \right)^2 \Delta t - \varrho \left( Q_T^{u^\theta} \right)^2 \right ],
\]
where $u^\theta$ denotes the strategy generated by the neural network. This is the discretized counterpart of the performance functional $J$ in \eqref{eq: J(u)}, with the expectation approximated during training by the mini-batch average. Therefore, this is an unsupervised learning network trained directly through the control objective. The numerical reference strategies are only used later for performance evaluation.

The main training hyperparameters are summarised in Table~\ref{tab:hyperparameters} below.
\begin{table}[htbp]
\centering
\small
\begin{tabular}{c|c|c|c|c|c|c}
\hline
\text{Experiment} & \text{Batch size} & \text{Learning rate} & \text{Dense nodes} & \text{GRU nodes} & \text{Epochs} & \text{Samples/epoch} \\
\hline
\text{Stage 1} & $256$ & $10^{-3}$ & $256$ & $256$ & $600$ & $2,560$ \\
\text{Stage 2} & $256$ & $10^{-3}$ & $256$ & $512$ & $600$ & $5,120$ \\
\text{Stage 3} & $256$ & $10^{-4}$ & $256$ & $256$ & $600$ & $10,240$ \\
\text{Stage 4} & $1,024$ & $10^{-4}$ & $256$ & $512$ & $1,350$ & $65,536$ \\
\hline
\end{tabular}
\caption{Main hyperparameters in training for different stages of experiments.}
\label{tab:hyperparameters}
\end{table}

Stage 1 and Stage 2 are benchmark experiments without transient kernels. Stage 3 includes transient kernels but no signal, and Stage 4 is the full experiment with both transient kernels and an OU signal.

\subsubsection{Evaluation Metrics}
\label{subsubsec:evaluation-metrics}

To evaluate the trained neural network, we compare the RNN strategy with the numerical reference strategy described above. For a test set with $M$ samples, let $u^{\mathrm{RNN},m}_{t_i}$ denote the trading speed generated by the neural network for sample $m$ at time $t_i$, and let $u^{\mathrm{num},m}_{t_i}$ denote the corresponding numerical reference strategy. The strategy error is measured by the normalized root mean square error
\[
    \mathrm{NRMSE}_u = \left( \frac{\sum_{m=1}^{M} \sum_{i=0}^{n-1} \left( u^{\mathrm{RNN},m}_{t_i} - u^{\mathrm{num},m}_{t_i} \right)^2}{\sum_{m=1}^{M} \sum_{i=0}^{n-1} \left( u^{\mathrm{num},m}_{t_i} \right)^2}  \right)^{1/2}.
\]
We also compare the corresponding inventory processes. The $\mathrm{NRMSE}_Q$ is defined by 
\[
    \mathrm{NRMSE}_Q = \left( \frac{\sum_{m=1}^{M} \sum_{i=0}^{n} \left( Q^{\mathrm{RNN},m}_{t_i} - Q^{\mathrm{num},m}_{t_i} \right)^2}{\sum_{m=1}^{M} \sum_{i=0}^{n} \left( Q^{\mathrm{num},m}_{t_i} \right)^2}    \right)^{1/2},
\]
where
\[
    Q^{\mathrm{RNN},m}_{t_i}=q^m-\Delta t \sum_{k=0}^{i-1} u^{\mathrm{RNN},m}_{t_k}, \quad Q^{\mathrm{num},m}_{t_i}=q^m-\Delta t \sum_{k=0}^{i-1} u^{\mathrm{num},m}_{t_k}.
\]
The $\mathrm{NRMSE}_u$ measures the pointwise accuracy of the learned trading speed, while the $\mathrm{NRMSE}_Q$ measures the accumulated error in the resulting liquidation path.

We also compare the value of the performance functional of the two strategies. For each test sample $m$, let $J^{\mathrm{num},m}$ and $J^{\mathrm{RNN},m}$ denote the performance functional of the numerical and neural-network strategies, respectively. Their empirical mean regret is 
\[
    \overline{\mathrm{Regret}} =\frac{1}{M}\sum_{m=1}^{M} \left(J^{\mathrm{num},m}-J^{\mathrm{RNN},m}\right).
\]
When results are compared across different initial inventories, we report the scaled regret
\[
    \mathrm{ScaledRegret}(q) =\frac{\overline{\mathrm{Regret}}}{q^2}.
\]
This normalization is consistent with our rescaling method. When the initial inventory is multiplied by $q$, both the trading speed and the running inventory are also multiplied by $q$. Since the main cost terms are quadratic in the trading speed or inventory, their values are therefore scaled by $q^2$; thus, we divide the mean regret by $q^2$ when comparing different initial inventories. This removes the effect of inventory size and makes the results more comparable.

\section{Numerical Results by Neural Networks}\label{sec:numerical_results}

In this section, we evaluate the performance of the neural network by comparing the RNN strategies against the numerical reference described previously. The experiments increase in difficulty along two dimensions: the presence of a predictive signal and the presence of transient price impact.

We begin with two benchmark models without transient impact. These experiments isolate the network's ability to recover the classical Almgren-Chriss solution and to respond to an Ornstein--Uhlenbeck (OU) signal. We then introduce kernel-dependent models in \ref{sec:kernel-dependent-results} and study kernel misspecification in \ref{sec:kernel-misspecification-results}. The numerical strategy is used as the reference in this section.

\subsection{Validation in Benchmark No Kernel Models}

For the benchmark models, we compare two ways of handling the initial inventory. First, the rescaling network is trained only at $q=1$ and applies the exact rescaling method from \ref{sec: rescaling} during evaluation. Second, the $q$-input network receives $q$ as an additional input and must learn the inventory dependence from the training data. This network is trained with $q\in[1,10]$. We evaluate both networks within the range for $q\in\{1,5,10\}$ and also test out of range for $q \in \{15, 20\}$. For each method and initial inventory, we evaluate the network on the same fixed set of $1{,}000$ test paths and report one value for each metric.

\subsubsection{No Signal}
    
In the absence of both signal and impact kernel, we set $H=0$ and $\alpha=0$ in this section. The unaffected price is then a martingale, and the optimal trading speed is known to be deterministic. 
This provides a basic validation of the recurrent architecture.
\begin{table}[!htbp]
    \centering
    \small
    \begin{tabular}{lrrrr}
        \hline
        Method & $q$ & $\mathrm{NRMSE}_u$ & $\mathrm{NRMSE}_Q$ & Scaled regret \\
        \hline
        Rescaling & 1 & 0.0643 & 0.0222 & $1.41\times 10^{-4}$ \\
        & 5 & 0.0643 & 0.0222 & $1.82\times 10^{-4}$ \\
        & 10 & 0.0643 & 0.0222 & $1.88\times 10^{-4}$ \\
        & $15^{\dagger}$ & 0.0643 & 0.0222 & $1.87\times 10^{-4}$ \\
        & $20^{\dagger}$ & 0.0643 & 0.0222 & $1.89\times 10^{-4}$ \\
        $q$-input & 1 & 0.0939 & 0.0779 & $2.15\times 10^{-4}$ \\
        & 5 & 0.0297 & 0.0125 & $1.01\times 10^{-4}$ \\
        & 10 & 0.1405 & 0.0581 & $8.88\times 10^{-4}$ \\
        & $15^{\dagger}$ & 0.3922 & 0.2640 & $1.62\times 10^{-2}$ \\
        & $20^{\dagger}$ & 0.5866 & 0.5785 & $1.02\times 10^{-1}$ \\
        \hline
    \end{tabular}
    \caption{Stage 1 evaluation of the rescaling and $q$-input networks on the same fixed set of $1{,}000$ previously unseen test paths at different inventory level. Out-of-range evaluations are indicated by $\dagger$. The metrics are defined in Section~\ref{subsubsec:evaluation-metrics}.}
    \label{tab:stage1_paper_results}
\end{table}

Table~\ref{tab:stage1_paper_results} shows that the rescaling network performs consistently across all inventory levels. Its strategy and $\mathrm{NRMSE}_Q$s remain approximately $6.43\%$ and $2.22\%$, respectively. The $q$-input network is more accurate at $q=5$, but its errors increase near and beyond the boundaries of its training range. This shows that rescaling can extend the model to new inventory levels without retraining.

\subsubsection{Inclusion of OU Signal}

We next keep $H=0$ but let the signal follow the OU dynamics specified in \eqref{OU signal}. The optimal strategy is no longer deterministic but depends on the current signal.
For the rescaling network, the signal is divided by $q$ and the resulting normalized strategy is multiplied by $q$ accordingly for evaluation. The $q$-input network receives the unscaled signal and the initial inventory as separate inputs.
\begin{table}[!htbp]
    \centering
    \small
    \begin{tabular}{lrrrr}
        \hline
        Method & $q$ & $\mathrm{NRMSE}_u$ & $\mathrm{NRMSE}_Q$ & Scaled regret \\
        \hline
        Rescaling & 1 & 0.1509 & 0.0533 & $6.77\times 10^{-3}$ \\
        & 5 & 0.1267 & 0.0588 & $1.56\times 10^{-3}$ \\
        & 10 & 0.1252 & 0.0483 & $1.32\times 10^{-3}$ \\
        & $15^{\dagger}$ & 0.1249 & 0.0435 & $1.28\times 10^{-3}$ \\
        & $20^{\dagger}$ & 0.1248 & 0.0413 & $1.27\times 10^{-3}$ \\
        $q$-input & 1 & 0.2099 & 0.0995 & $2.22\times 10^{-2}$ \\
        & 5 & 0.0570 & 0.0404 & $6.03\times 10^{-4}$ \\
        & 10 & 0.1114 & 0.0486 & $6.69\times 10^{-4}$ \\
        & $15^{\dagger}$ & 0.3512 & 0.3476 & $8.95\times 10^{-2}$ \\
        & $20^{\dagger}$ & 0.5666 & 0.7535 & $4.83\times 10^{-1}$ \\
        \hline
    \end{tabular}
    \caption{Stage 2 evaluation of the rescaling and $q$-input networks on the same fixed set of $1{,}000$ previously unseen test paths at different inventory levels. Out-of-range evaluations are indicated by $\dagger$.}
    \label{tab:stage2_paper_results}
\end{table}

Table~\ref{tab:stage2_paper_results} shows that neither method is always better within the training range. The rescaling network performs better at $q=1$, while the $q$-input network performs better at $q=5$ and $q=10$. However, the rescaling network remains stable at the out-of-range values $q=15$ and $q=20$. In contrast, the errors of the $q$-input network increase sharply. This shows that rescaling can extend the model to new inventory levels without retraining, even in the presence of the signal. 

In all subsequent experiments, we fix $q=1$ and it represents one normalised unit of inventory. Other model parameters should be chosen accordingly. Interpreting other initial inventory in physical shares would require the parameters to be calibrated with the chosen units.

\begin{figure}[!htbp]
    \centering
    \includegraphics[width=\textwidth]{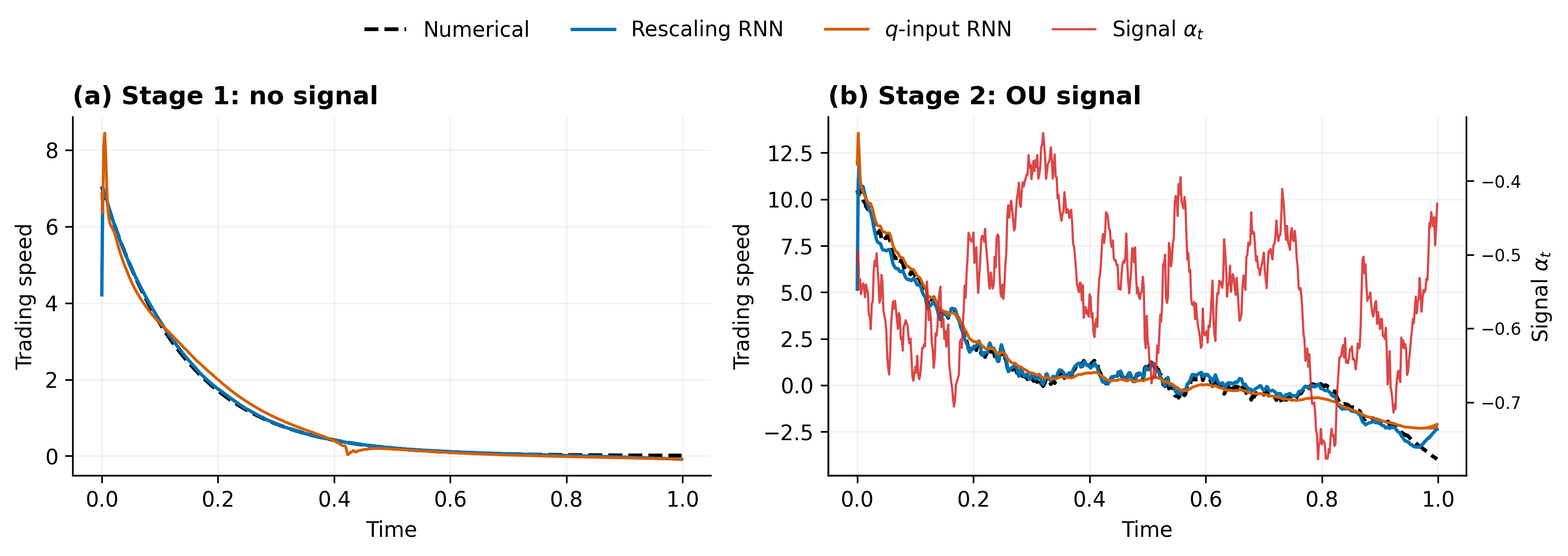}
    \caption{Numerical and neural-network trading speeds in the two benchmark models at $q=1$.  Panel (a) has neither transient impact nor a price signal. Panel (b) adds an OU signal, which is plotted on the right-hand axis.}
    \label{fig:benchmark-strategies}
\end{figure}

Figure~\ref{fig:benchmark-strategies} provides a visual comparison for $q=1$ of the representative strategies for the no-signal and OU-signal experiments.

\subsection{Kernel-Dependent Models}\label{sec:kernel-dependent-results}

We now introduce transient price impact and test the ability of a neural network to learn how the optimal strategy depends on the shape of the kernel. Each kernel $H$ is represented in the network input by its first $n_H=30$ trigonometric basis coefficients $(h_1,\ldots,h_{n_H})$. The original kernel is used to calculate both the numerical reference strategy and the performance criterion.

We consider three kernel families. The basis and piecewise families are generated using the two sampling methods introduced in Section~\ref{sec: KernelSampling}. We also consider exponential kernels of the form
\[
H(t)=e^{-\rho t},
\]
with different values of the decaying speed $\rho$. Exponential kernels are not included during training and therefore provide an additional test of generalization outside the two kernel sampling methods used to construct the training data.

The benchmark results in the previous subsection have already tested the model's dependence on the initial inventory. We therefore only use the rescaling network and fix $q=1$ throughout this subsection, which allows us to focus on the effects of the kernel and the signal. The same set of kernels is used in the following no-signal and OU-signal experiments.

The sampling parameters used for these kernels are shown in Table~\ref{tab:kernel-dependent-sampling-parameters}. Here $\kappa$ denotes the Poisson intensity used in the kernel sampler. 
\begin{table}[!htbp]
    \centering
    \small
    \renewcommand{\arraystretch}{1.1}
    \begin{tabular}{@{}p{0.18\textwidth}p{0.20\textwidth}p{0.52\textwidth}@{}}
    \hline
    Kernel family & Construction & Parameters \\
    \hline
    Basis
    & Basis sampling
    & $\kappa_{\text{basis}}=10$, $a_0=1.0$, $d=0.8$ \\
    Piecewise
    & Piecewise sampling
    & $\kappa_{\text{piece}}=10$, $\bar\ell=1.0$, $\bar h=0.3$ \\
    Exponential
    & $H(t)=e^{-\rho t}$
    & $\rho\in\{0.2,0.5,1.0,2.0\}$ \\
    \hline
    \end{tabular}
    \caption{Kernel parameters used in the kernel-dependent experiments. In the basis method, $d$ is the coefficient decay. In the piecewise
    method, $\bar\ell$ and $\bar h$ denote the mean interval length and height.}
    \label{tab:kernel-dependent-sampling-parameters}
\end{table}

\subsubsection{No Signal}

We first introduce transient price impact kernels while keeping the price signal still equal to zero. The neural network then learns the map
\[
(h_1,\ldots,h_{n_H}) \mapsto \hat u^*_t.
\]
This experiment tests whether the finite-dimensional coefficient representation contains enough information for the neural network to learn the map.
\begin{table}[!htbp]
    \centering
    \small
    \begin{tabular}{lrrr}
    \hline
    Kernel family & $\mathrm{NRMSE}_u$ & $\mathrm{NRMSE}_Q$ & Mean regret \\
    \hline
    Basis & 0.0850 & 0.0395 & $4.42\times 10^{-4}$ \\
    Piecewise & 0.1953 & 0.0197 & $1.24\times 10^{-3}$ \\
    Exponential & 0.1220 & 0.0557 & $1.05\times 10^{-3}$ \\
    \hline
    \end{tabular}
    \caption{Stage 3 evaluation at $q=1$ on $100$ previously unseen kernels from each family ($300$ test cases in total).}
    \label{tab:stage3_paper_results}
\end{table}

Table~\ref{tab:stage3_paper_results} reports the results on $100$ previously unseen kernels from each family.  The basis family
has the lowest $\mathrm{NRMSE}_u$, while the piecewise family is the most challenging at the level of pointwise trading speed, but it has the lowest $\mathrm{NRMSE}_Q$. Across all three families, the $\mathrm{NRMSE}_Q$ is always below $5.6\%$, and mean regret is at most $1.25\times10^{-3}$. Thus, the network closely reproduces the accumulated liquidation path and objective value, even when local differences in trading speed are more visible.
\begin{figure}[!htbp]
    \centering
    \includegraphics[width=\textwidth]{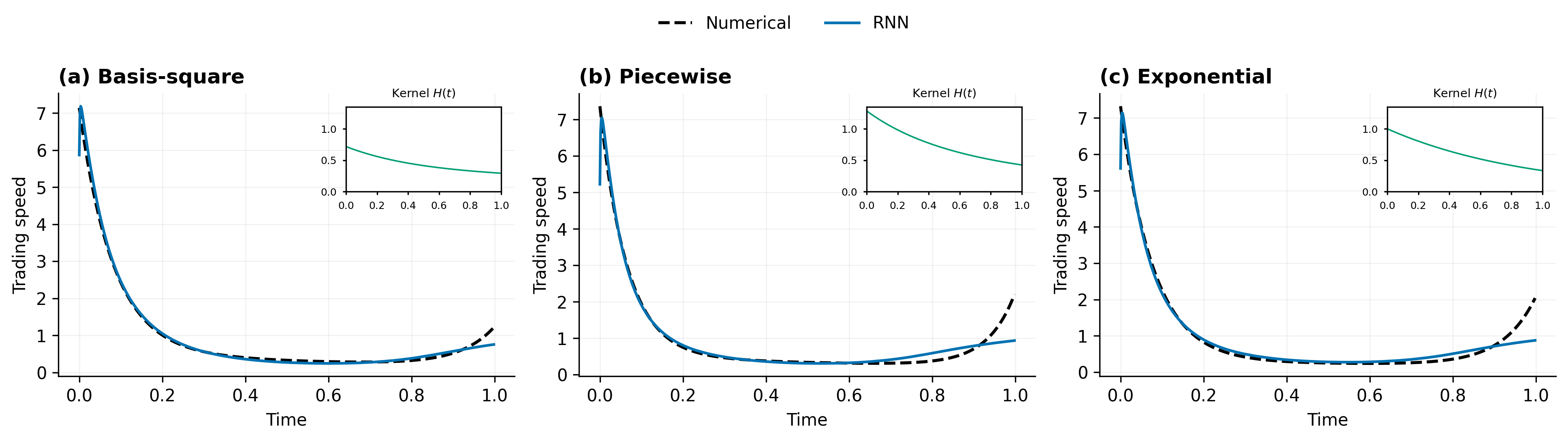}
    \caption{Numerical and neural-network trading speeds in the Stage~3 no-signal experiment at $q=1$. Panels (a)--(c) show representative
    basis, piecewise, and exponential kernels, respectively. The corresponding kernel is displayed in each inset.}
    \label{fig:stage3-strategies}
\end{figure}

Figure~\ref{fig:stage3-strategies} shows one representative case from each kernel family.

\subsubsection{With an OU Signal}

Finally, we combine transient price impact with the OU signal. The strategy now depends jointly on the kernel coefficients and causally on the realized signal path up to each decision time. 
We still use the $100$ previously unseen kernels from each family, as in the no-signal experiment, and generate $20$ OU-signal paths for every kernel. This gives $2,000$ test cases per family and $6,000$ cases in total. Since this is the target experiment, we use a larger training run on a high-performance computer. 
The hyperparameters are reported in Table~\ref{tab:hyperparameters}
\begin{table}[!htbp]
    \centering
    \small
    \begin{tabular}{lrrr}
    \hline
    Kernel family & $\mathrm{NRMSE}_u$ & $\mathrm{NRMSE}_Q$ & Mean regret \\
    \hline
    Basis & 0.0822 & 0.0186 & $9.59\times 10^{-4}$ \\
    Piecewise & 0.1100 & 0.0113 & $2.98\times 10^{-4}$ \\
    Exponential & 0.0820 & 0.0174 & $4.49\times 10^{-4}$ \\
    \hline
    \end{tabular}
    \caption{Stage~4 evaluation at $q=1$ on $100$ previously unseen kernels from each family and $20$ OU-signal paths per kernel, giving $2,000$ test cases per family.}
    \label{tab:stage4_paper_results}
\end{table}

Table~\ref{tab:stage4_paper_results} shows that the network remains accurate after the signal is introduced. The $\mathrm{NRMSE}_u$ lies between $8.20\%$ and $11.00\%$, while the $\mathrm{NRMSE}_Q$ remains below $1.9\%$ for all three kernel families, and mean regret is at most $9.59\times10^{-4}$. The piecewise family has the largest $\mathrm{NRMSE}_u$ but the smallest $\mathrm{NRMSE}_Q$ and mean regret. Thus, as in the no-signal experiment, relatively large errors in trading speed do not necessarily lead to larger errors in the accumulated inventory or the performance functional.
\begin{figure}[!htbp]
    \centering
    \includegraphics[width=\textwidth]{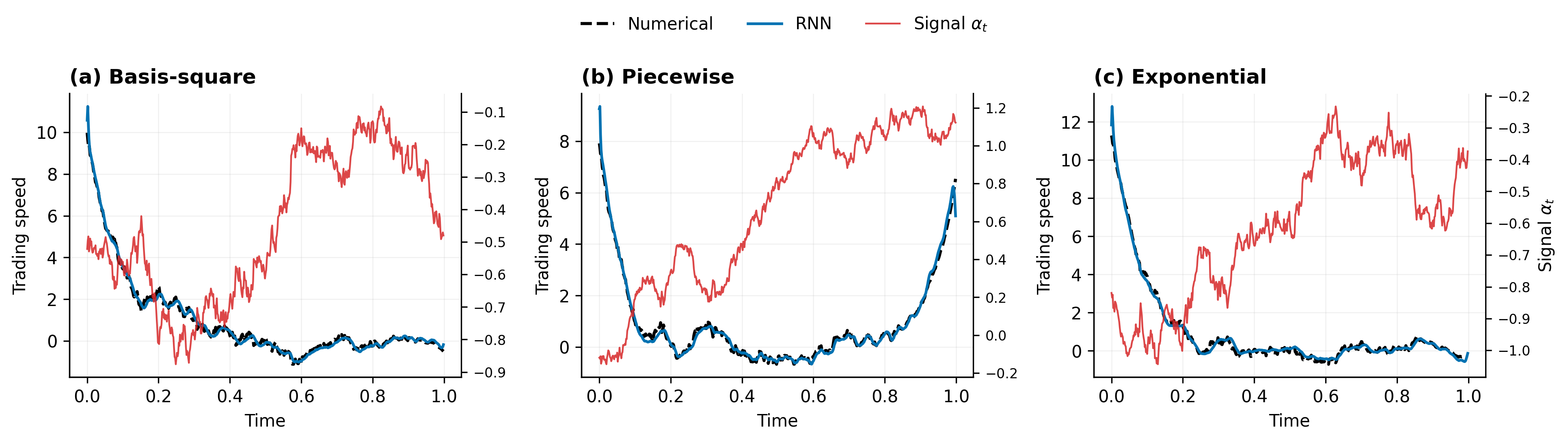}
    \caption{Numerical and neural-network trading speeds in the Stage 4 OU-signal experiment at $q=1$. Panels (a)--(c) use the representative basis, piecewise, and exponential kernels from the Stage 3 figure. The corresponding OU signal is plotted in red on the right-hand axis.}
    \label{fig:stage4-strategies}
\end{figure}

Figure~\ref{fig:stage4-strategies} illustrates one case for each kernel family. To make the no-signal and OU-signal figures comparable, we retain the three representative kernels selected in Figure~\ref{fig:stage3-strategies}. The signal is plotted in red on the right-hand axis. 

\subsection{Kernel Misspecification}\label{sec:kernel-misspecification-results}

The previous experiments mainly test whether the neural architecture can learn the optimal trading strategy given the true kernel. However, the exact kernel may be hard to estimate in practice, and a trader may therefore use a kernel that has the wrong shape. In this section, we consider a trader who does not know the true kernel $H$ and instead uses an exponential kernel of best fit taking the form
\[
    \widehat H(t)=A e^{-\rho t}.
\]
The performance of the optimal strategy corresponding to both $H$ and $\widehat{H}$ is then tested when the dynamics are determined by $H$; thus, this experiment endeavors to determine the financial cost of using the misspecified kernel $\widehat{H}$ instead of the true kernel $H$.

\subsubsection{Kernel families and distance}

We employ four different distributions of kernels that determine the true dynamics of price impact. The four true-kernel families and their sampling parameters are summarized in Table~\ref{tab:kernel-misspecification-families}, where each row defines a distribution of kernels.
\begin{table}[!htbp]
    \centering
    \small
    \renewcommand{\arraystretch}{1.1}
    \begin{tabular}{@{}p{0.20\textwidth}p{0.09\textwidth}p{0.27\textwidth}p{0.32\textwidth}@{}}
    \hline
    Family & Method & Sampling parameters & Main features \\
    \hline
    basis\_smooth & Basis & $\kappa_{\text{basis}}=3$, $a_0=1$, $d=0.45$ & Few basis terms; smaller, smoother kernels. \\
    basis\_rough & Basis & $\kappa_{\text{basis}}=20$, $a_0=1$, $d=0.95$ & Many basis terms; larger scale and more variation. \\
    piecewise\_coarse & Piecewise & $\kappa_{\text{piece}}=3$, $\bar\ell=2.0$, $\bar h=0.3$ & Few wide density pieces; moderate initial impact and fast relative decay. \\
    piecewise\_spiky & Piecewise & $\kappa_{\text{piece}}=20$, $\bar\ell=0.35$, $\bar h=0.7$ & Many short, high density pieces; large initial impact and a steep fall. \\
    \hline
    \end{tabular}
    \caption{The four kernel families used in the misspecification experiment. Here $\kappa_{\text{basis}}$ and $\kappa_{\text{piece}}$ control the expected number of sampled components, $d$ is the basis-coefficient decay, and $\bar\ell$ and $\bar h$ are the mean piece length and height.}
    \label{tab:kernel-misspecification-families}
\end{table}

We specify the distance between two kernels $H_1$ and $H_2$ to be given by their $L^2$ distance
\begin{equation}
    d_{L^2}(H_1,H_2) =\left(\int_0^T\left[H_1(t)-H_2(t)\right]^2\,dt\right)^{1/2},
    \label{kernel distance}
\end{equation}
and performance regret remains the main economic measure.

\subsubsection{Best-fit exponential kernel}

For each true kernel $H$, we fit
\[
    \widehat H(t)=A e^{-\rho t}, \quad A \geq 0, \rho \geq 0,
\]
by considering the quantity
\[
    E(A,\rho)=\int_0^T \left[ H(t)-\widehat H(t) \right]^2\,dt.
\]
The fit is performed using two different methods. The first, which we denote \emph{match-$H(0)$}, fixes $A=H(0)$ and minimizes $E(A,\rho)$ with respect to $\rho$ only, resulting in the preservation of initial price impact between the true kernel and the fitted exponential. The second method, which we denote \emph{free-$A$}, minimizes $E(A,\rho)$ with respect to both parameters $A$ and $\rho$. In both cases, the minimizing parameters are computed numerically.

\subsubsection{Regret evaluation and numerical results}

We use the neural network trained in Stage~4 with $600$ time steps and $q=1$. For each of the four families, we sample $100$ true kernels and use $20$ OU-signal paths per kernel. The network receives the fitted exponential kernel $\widehat H$ and produces $u^{\mathrm{RNN}}(\widehat H)$. However, the resulting strategy is always evaluated under the true kernel $H$ to align with the real situation. Denoting by $J(u;H)$ the objective value of strategy $u$ when market impact follows $H$, we report the regret
\[
    \mathrm{Regret}_{\mathrm{RNN}} =J \left( u^{\mathrm{num}}(H); H \right) - J \left( u^{\mathrm{RNN}} (\widehat H); H \right).
\]
where $u^{\mathrm{num}}(H)$ is the numerical reference strategy obtained using the true kernel. Thus, the assumed exponential $\widehat H$ is only used to produce the RNN strategy, while the true kernel $H$ is used both to calculate its performance and to construct the numerical benchmark $u^{\mathrm{num}}(H)$.

To facilitate economic interpretation, we express the regret in basis points of the initial notional value:
\[
\mathrm{Regret}_{\mathrm{bps}} = 10^4 \frac{\mathrm{Regret}_{\mathrm{RNN}}}{qP_0}.
\]
Since we set $q=1$ and $P_0=100$ in the numerical implementation, the basis-point values are obtained by multiplying $\mathrm{Regret}_{\mathrm{RNN}}$ by $100$. Inventory is expressed in normalized units. For illustration, if $q=1$ represents an initial position of $\$1$ million, then one basis point corresponds to $\$100$. This monetary interpretation is illustrative since the model parameters are not calibrated to a particular asset. 

\begin{table}[!htbp]
    \centering
    \small
    \begin{tabular}{llrrr}
    \hline
    True kernel family & Fit & Mean $L^2$ error
    & Mean regret (bps) & 95th percentile (bps) \\
    \hline
    Basis smooth & Match $H(0)$ & 0.006819 & $-0.0822$ & 0.0590 \\
    Basis smooth & Free $A$ & 0.004493 & 0.0062 & 0.2093 \\
    Basis rough & Match $H(0)$ & 0.045324 & 0.7659 & 3.4877 \\
    Basis rough & Free $A$ & 0.029743 & 0.9243 & 4.1155 \\
    Piecewise coarse & Match $H(0)$ & 0.099540 & 0.2275 & 0.5732 \\
    Piecewise coarse & Free $A$ & 0.076375 & 0.4188 & 1.0201 \\
    Piecewise spiky & Match $H(0)$ & 0.210353 & 0.4277 & 1.0500 \\
    Piecewise spiky & Free $A$ & 0.157529 & 0.6624 & 1.4062 \\
    \hline
    \end{tabular}
    \caption{RNN best-fit exponential results at $q=1$. Mean $L^2$ error is the fitting error over $[0,T]$. Each family uses $100$ true kernels and $20$ economic paths per kernel.}
    \label{tab:kernel-misspecification-best-fit}
\end{table}
Table~\ref{tab:kernel-misspecification-best-fit} summarizes the numerical results of this experiment. Mean RNN regret is below $0.93$ basis points in every case, while the largest 95th-percentile is $4.12$ basis points. Under the illustrative $\$1$ million normalization, these values correspond to approximately $\$93$ and $\$412$, respectively. Averaging equally across the four kernel families, the mean gap is approximately $0.33$ basis points for the match-$H(0)$ method and $0.50$ basis points for the free-$A$ method. Although the free-$A$ method produces a smaller relative $L^2$ fitting error for every family, it does not produce a smaller economic gap. Hence, a closer fit in the $L^2$ norm does not necessarily lead to a better-performing liquidation strategy.

As an additional robustness check, we vary $\lambda$, $\phi$, $\varrho$, $\gamma$, $\sigma_S$, and $\sigma^\alpha$ one at a time and retrain the network for each setting. The mean gap remains below $1.10$ basis points in every setting except $\lambda=0.005$, for which it is approximately $2.83$ basis points under match-$H(0)$ and $2.79$ basis points under free-$A$. The results are therefore broadly stable across the parameter settings considered, with greater sensitivity when temporary impact is low. The complete results are reported in Table~\ref{tab:rnn-regret-robustness} in the Appendix.

\section{Conclusion}\label{sec:conclusion}

In this paper we have performed analysis on the solution map for an optimal liquidation problem when considered as a function of underlying model parameters. Our main theoretical results establish that the optimal trading strategy is locally Lipschitz with respect to the fundamental price process, the temporary price impact parameter, the transient price impact kernel, and the number of shares to liquidate. Additionally, when the performance criteria is evaluated at the optimal trading strategy, the resulting map is also locally Lipschitz with respect to the same model parameters. This type of continuity gives justification for approximating the solution map with an RNN.

We have proposed a methodology for both parameterizing the infinite-dimensional model inputs and producing samples of impact kernels for training purposes with the goal of training an RNN to learn the optimal trading strategy. We then demonstrated the effectiveness of the RNN when applied to models both with and without a trade signal. Finally, we investigated the effects of kernel misspecification by quantifying the magnitude of performance loss when the strategy is computed assuming a kernel which differs from the one implemented in generating the test data.

\section{Appendix}\label{sec: Appendix}

\subsection{Some additional technical results}
In this subsection, we enunciate and prove a few results on Hilbert-Schmidt and integral operators for which we have not been able to find a reference in the literature. Besides, we prove a few technical lemmas related to the convergence of Volterra-type kernels, self-adjoint operators, and more.

Let $HS$ be the space of Hilbert-Schmidt operators from $L^2[0,T]$ into itself, endowed with the inner product
\[
\langle A,B\rangle_{HS}:=\sum_{j=1}^\infty\langle Ae_j,Be_j \rangle_{L^2},\quad A,B\in HS,
\]
where $(e_j)_{j\in\N}$ is any basis for $L^2([0,T])$: the scalar product does not depend on the choice of the basis, and it makes $HS$ a separable Hilbert space. We have:

\begin{lemma}\label{lemma: HS iso to L2}
    Let $\mathcal I:L^2([0,T]^2)\to HS$ be the map that associates to any $G\in L^2([0,T]^2)$ the induced integral operator
    \[
(\boldsymbol{G}f)(t):=\int_0^T G(t,s)f(s)\,ds,\quad  f\in L^2([0,T]),\,0\le t\le T.
\]
Then $\mathcal I$ is an isomorphism of Hilbert spaces from $L^2([0,T]^2)$ onto $HS$.    
\end{lemma}
\begin{proof}
    We observe that $\mathcal I$ is well-defined, linear and $\norm{\mathcal I G}_{HS}=\norm{G}_{L^2([0,T]^2)},G\in L^2([0,T]^2)$ by \cite[page 267]{conway2010}. To prove surjectivity, take $A\in HS$: then by \cite[page 268]{conway2010} there exists a sequence $(A_n)_n$ of finite-rank operators such that $A_n\stackrel{HS}{\to} A$. Being finite rank, it is easy to see that they must be integral operators, namely $A_n=\mathcal Ia_n,n\in\N$ for some $a_n\in L^2([0,T]^2)$. But $\norm{A_n-A_m}_{HS}=\norm{a_n-a_m}_{L^2([0,T]^2)},n,m\in\N$, and so $(a_n)_n$ is Cauchy with some limit $a$. Thus, $\norm{A_n-\mathcal I a}_{HS}=\norm{\mathcal I a_n-\mathcal I a}_{HS}=\norm{a_n-a}_{L^2([0,T]^2)}\to 0$, and therefore $A=\mathcal I a$. Finally, by polarization, $\langle\mathcal I G_1,\mathcal I G_2 \rangle_{HS}=\langle G_1,G_2\rangle_{L^2([0,T]^2)},G_1,G_2\in L^2([0,T]^2)$ and the Lemma is proven.
\end{proof}

The next lemma provides an upper bound for the norm of the inverse of a self-adjoint operator in terms of a lower bound for the norm of the operator itself.
\begin{lemma}\label{lemma: bound for the inverse}
    Let $(E,\langle\cdot,\cdot\rangle)$ be a real Hilbert space. Let $A:E\to E$ be a self-adjoint linear operator for which there exists $c>0$ such that $\langle Ax,x\rangle\ge c\langle x,x\rangle,\,x\in E$. Then $A$ is invertible and $\norm{A^{-1}}_{op}\le c^{-1}$.
\end{lemma}
\begin{proof}
We know that there exists a unique square root $A^{1/2}$ which is self-adjoint and $\langle A^{1/2}x,x\rangle\ge 0$. Therefore, for all $x\in E$
\begin{equation}\label{eq: bound for the inverse 1}
    \norm{A^{1/2}x}^2 = \langle A^{1/2}x,A^{1/2}x\rangle=\langle A^{1/2}A^{1/2}x,x\rangle = \langle Ax,x\rangle \ge c\norm{x}^2,
\end{equation}
showing that $\ker A^{1/2}$ is trivial. By Proposition 2.9 \cite{conway2010}, $\{0\}=(\operatorname{ran}(A^{1/2})^\ast)^\perp=(\operatorname{ran}A^{1/2} )^\perp$, namely $A^{1/2}$ is invertible. From here, we immediately deduce that $A$ is invertible too. Additionally, from Proposition 2.6 \cite{conway2010}, $(A^{1/2})^{-1}$ is also self-adjoint. From \eqref{eq: bound for the inverse 1}, it now holds for any $y\in E$
\[
\begin{split}
    \norm{y}^2&=\langle y,y\rangle \ge c \norm{(A^{1/2})^{-1}y}^2 = c\langle (A^{1/2})^{-1}y, (A^{1/2})^{-1}y\rangle \\&= c\langle (A^{1/2})^{-1}(A^{1/2})^{-1}y, y\rangle = c\langle A^{-1}y,y\rangle,
\end{split}
\]
namely $c^{-1}\ge \langle A^{-1}y,y\rangle$ for any $y$ with $\norm{y}=1$. Since $A^{-1}$ is self-adjoint too, by Proposition 2.13 \cite{conway2010} we obtain $\norm{A^{-1}}_{op}\le c^{-1}$.    
\end{proof}

Regrettably, the Proof of Lemma 7.1 \textbf{iii.} in \cite{https://doi.org/10.1111/mafi.12465} contains a small mistake in the application of the Gr\"onwall's inequality. By implication, it is no longer clear from that argument that $\EE\left[\sup_{t\le T}(u^\ast_t)^2\right]$ is finite. However, this crucial property can be proven, as in the following.

\begin{lemma}\label{lemma: fix to Gronwall}
    Assume the same setting of Lemma 7.1 in \cite{https://doi.org/10.1111/mafi.12465}. Then $\EE\left[\sup_{t\le T}(u^\ast_t)^2\right]<\infty$.
\end{lemma}
\begin{proof}
The solution $u^\ast$ of \eqref{eq: uast} can be equivalently written in terms of the resolvent $R^B$ of $B$, namely
\begin{equation}\label{eq: uast RB}
u^\ast_t = a_t +\int_0^tR^B(t,s)a_s\,ds,\quad 0\le t\le T,
\end{equation}
where $R^B\in L^2([0,T]^2)$ is the unique solution to
\[
R^B = B + B\star R^B,\quad B\star R^B = R^B\star B.
\]
Refer to \cite[Corollary 9.3.16]{Gripenberg_Londen_Staffans_1990}. In terms of integral operators, this translates as
\[
\boldsymbol{R}^B=\boldsymbol{B}+\boldsymbol{B}\boldsymbol{R}^B,\quad \boldsymbol{BR}^B = \boldsymbol{R}^B\boldsymbol{B}.
\]
We observe that a priori $R^B$ satisfies only
\[
\sup_{\norm{\alpha}_{L^2}\le 1,\,\norm{\beta}_{L^2}\le 1}\int\int\alpha(t)R^B(t,s)\beta(s)\,ds<\infty.
\]
However, $\boldsymbol{B}$ is Hilbert-Schmidt, because $B\in L^2([0,T]^2)$ and Lemma \ref{lemma: HS iso to L2}, and so $\boldsymbol{BR}^B=\boldsymbol{R}^B\boldsymbol{B}$ is Hilbert-Schmidt. We conclude that $\boldsymbol{R}^B$ is also Hilbert-Schmidt, and in force of Lemma \ref{lemma: HS iso to L2} again that $R^B\in L^2([0,T]^2)$. Further, from
\[
(B\star R^B)(s,u)=\int_0^TB(s,z)R^B(z,u)\,dz,
\]
we have
\[
\begin{split}
    \int_0^T\abs{(B\star R^B)(s,u)}^2du&\le \int_0^T\left(
\int_0^T\abs{B(s,z)}^2dz\right)\left(\int_0^T\abs{R^B(z,u)}^2dz\right)\,du
\\&\le \int_0^T\abs{B(s,z)}^2dz \,\norm{R^B}^2_{L^2}.
\end{split}
\]
Therefore, 
\[
    \sup_{s\le T} \int_0^T\abs{(B\star R^B)(s,u)}^2du \le \sup_{s\le T} \int_0^T\abs{B(s,z)}^2dz\,\norm{R^B}^2_{L^2}
 < \infty
\]
in virtue of Lemma \ref{lemma: apriori estimate for B}, and
\[
    \frac12 \sup_{t\le T} \int_0^T\abs{R^B(t,s)}^2ds \le \sup_{t\le T} \int_0^T\abs{B(t,s)}^2ds + \sup_{t\le T} \int_0^T\abs{B(t,s)}^2ds \,\norm{R^B}^2_{L^2}
    <\infty.
\]
Clearly, $R^B$ is deterministic. Hence, $\abs{\int_0^tR^B(t,s)a_sds}\le \int_0^t\abs{R^B(t,s)}ds\,\cdot\sup_{s\le T}\abs{a_s}$ and by Jensen's inequality,
\[
\abs{\int_0^tR^B(t,s)a_sds}^2\le t\int_0^t\abs{R^B(t,s)}^2ds\,\cdot\sup_{s\le T}\abs{a_s}^2.
\]
Then,
\[
\begin{split}
    \sup_{t\le T} \abs{\int_0^tR^B(t,s)a_sds}^2&\le T\sup_{t\le T}\int_0^t\abs{R^B(t,s)}^2ds\,\cdot\sup_{s\le T}\abs{a_s}^2
    \\&\le T\sup_{t\le T}\int_0^T\abs{R^B(t,s)}^2ds\,\cdot\sup_{s\le T}\abs{a_s}^2,
\end{split}
\]
and finally
\[
    \EE\sup_{t\le T} \abs{\int_0^tR^B(t,s)a_sds}^2\le T\sup_{t\le T}\int_0^T\abs{R^B(t,s)}^2ds\,\cdot\EE\sup_{s\le T}\abs{a_s}^2<\infty,
\]
where $C$ is coming from a few lines above. From \eqref{eq: uast RB} we obtain the result. 
\end{proof}

\subsection{List of auxiliary functions}\label{sub: list} For convenience of the reader, in this subsection we list in a single place all the auxiliary functions we have used to construct our Lipschitz estimates and that are scattered over the manuscript. In the following, $\delta_1\in\R$, $w_1,w_2\in C([0,T])$, $\sigma_1,\sigma_2\in\mathcal{H}$, $\xi_1,\xi_2\in (0,\infty)$ and $\Pi_1\in\PP$ are dummy variables. 
\begin{align}
    f_2(\xi_1,\Pi_1) &:= \frac{2(1+2\phi^2T^2)T}{\xi_1^2}\norm{\Pi_1}^2_{\mathscr P},\label{eq: f2}\\
    f_1(\xi_1,\sigma_1,\Pi_1) &:= f_2(\xi_1,\Pi_1) \frac{(2+2\sqrt{T})^2}{\xi_1^2} (2\norm{\sigma_1}^2_{L^2} + (1+8\varrho^2)T),\label{eq: f1}\\
    f_3(\sigma_1,\xi_1,\Pi_1) &:= \frac12 f_2(\xi_1,\Pi_1) \biggl(2\norm{\sigma_1}^2_{L^2}+(1+8\rho^2)T \biggr),\label{eq: f3}\\
    g_1(\xi_1,\xi_2,\sigma_1) &:= \frac{2\varrho \sqrt{T} + \norm{\sigma_1}_{L^2}}{2\xi_1} + \frac{\phi T^{3/2}}{\sqrt{2}\xi_2},\label{eq: g1}\\
    g_2(\xi_1,w_1,\delta_1) &:= \frac{(2\varrho\sqrt{T}+1)(1+\sqrt{T})(\norm{w_1}_\infty+2\varrho\abs{\delta_1})}{\xi_1^2},\label{eq: g2}\\
    g_3(\xi_1,w_1,\delta_1) &:= \frac{\sqrt{2}\phi T^{3/2}(1+\sqrt{T})(\norm{w_1}_\infty+2\varrho\abs{\delta_1})}{\xi_1^2},\label{eq: g3}\\
    g_4(\xi_1,\sigma_1,\delta_1) &:= \frac{\sqrt{2}\phi\abs{\delta_1}T^{3/2}(1+\sqrt{T})(2\varrho\sqrt{T} + 2\phi T^{3/2} + \norm{\sigma_1}_{L^2})}{\xi_1^2},\label{eq: g4}\\
    g_5(\xi_1,\delta_1) &:= \frac{\phi\abs{\delta_1}T^{3/2}}{\sqrt{2}\xi_1},\label{eq: g5}\\
    \begin{split}\label{eq: g6}
        g_6(w_1,\sigma_1,\xi_1,\delta_1) &:= \frac{7(4\varrho^2T+2\phi^3T^3+\norm{\sigma_1}^2_{L^2})}{2\xi_1^2}(\norm{w_1}^2_\infty+4\varrho^2\delta_1^2)\\
        &\quad +\frac{7T^2\phi^2\delta_1^2}{2\xi_1^2}(4\varrho^2T+\norm{\sigma_1}^2_{L^2}) + 28T\phi^2\delta_1^2\biggl(1+\frac{\phi^2T^4}{4\xi_1^2}\biggr),
    \end{split}
\end{align}
\begin{align}
    \begin{split}\label{eq: L1}
        &\mathscr{L}_1(w_1,\sigma_1,\xi_1,w_2,\sigma_2,\xi_2,\Pi_1,\delta_1) := \frac{6}{\xi_2^2}\biggl(
                1+ f_1(\xi_1,\sigma_2,\Pi_1) + f_2(\xi_1,\Pi_1)
                \biggr)\\
                &\quad\quad +\frac{6}{\xi_2^2} \biggl(
        g_1(\xi_2,\xi_1,\sigma_2) + g_2(\xi_1,w_1,\delta_1) + g_3(\xi_1,w_2,\delta_1) + g_4(\xi_1,\sigma_2,\delta_1) + g_5(\xi_1,\delta_1)
        \biggr)^2\\
        &\quad\quad + \frac{2}{\xi_1^2\xi_2^2} \biggl(
            4\norm{\Pi_1}^2_\PP+2\norm{w_1}_\infty^2+8\varrho^2\delta_1^2+ f_3(\sigma_1,\xi_1,\Pi_1) + g_6(w_1,\sigma_1,\xi_1,\delta_1)
            \biggr),
    \end{split}\\
    &\mathscr L_2(\xi_1,\sigma_1) :=\frac{1}{\xi_1^2}\biggl(
    2+\frac{T}{2\xi_1^2}(1+2\phi^2T^2)\left[
    2\norm{\sigma_1}^2_{L^2}+T(1+8\varrho^2)
    \right]
    \biggr),\label{eq: L2}\\
    \begin{split}
    &\mathscr L_3(\xi_1,\sigma_1) := 
    \frac{1}{2\xi_1^2}\biggl(
       4\varrho^2 + 
        \biggl[
        \frac{2\varrho^2T + \varrho\sqrt{T}\norm{\sigma_1}_{L^2} + 2\sqrt{2}\phi\varrho T^2 + \phi^2T^3 }{\xi_1}\\ &\hspace{80mm} + \frac{\phi T^{3/2}\norm{\sigma_1}_{L^2}}{\sqrt{2}\xi_1} + 2\phi T 
        \biggr]^2
       \biggr),\label{eq: L3}
   \end{split}
\end{align}

\begin{align}
    \begin{split}\label{eq: j0}
        j_0(\sigma_1,\xi_1,\xi_2) &:= \frac{1+\sqrt{T}}{\xi_1^2} \biggl( 4\varrho^2T+4\phi T^2(\varrho+\phi T)+4\phi\varrho T^{5/2}\\
        & \hspace{25mm} + \left[\norm{\sigma_1}_{L^2}+ 4\sqrt{T}(\varrho^2+\phi T)\right]\norm{\sigma_1}_{L^2}\biggr)\\
        & \hspace{40mm} + \frac{1}{\xi_2}\left(2\sqrt{T}(\varrho^2+\phi T)+\norm{\sigma_1}_{L^2}\right),
    \end{split}\\
    j_1(\xi_1) &:= 2\phi T + \frac{T(2\varrho^2 +\phi\varrho T^{3/2}+2\phi\varrho T +\phi^2T^2 )}{\xi_1},\label{eq: j1}\\
    j_2(\sigma_1,\xi_1) &:=\frac{\biggl(4\sqrt{T}(\varrho+\phi T)+ \norm{\sigma_1}_{L^2}\biggr)\norm{\sigma_1}_{L^2} }{2\xi_1},\label{eq: j2}\\
    j_3(\sigma_1,\xi_1) &:= \exp\left\{ \frac{T^2}{\xi_1^2}\biggl(j_1(\xi_1) + j_2(\sigma_1,\xi_1) \biggr)^2 + \frac{2T}{\xi_1^2}\biggl(4\varrho^2 T + \norm{\sigma_1}^2_{L^2}\biggr)
        \right\},\label{eq: j3}
\end{align}

\begin{align}
    \begin{split}\label{eq: LDeltaB}
        &\mathscr L_{\Delta B}(\sigma_1,\xi_1,\xi_2) := \frac{1}{\xi_2^2}\biggl( Tj_0^2(\sigma_1,\xi_1,\xi_2)+1 \biggr) \\
        & \hspace{40mm} + \frac{1}{\xi_1^2\xi_2^2} \biggl( T (j_1(\xi_1)+j_2(\sigma_1,\xi_1))^2 +8\varrho^2T+2\norm{\sigma_1}^2_{L^2}
        \biggr),
     \end{split}\\
     \begin{split}\label{eq: M}
        &M(w_1,\sigma_1,\xi_1,\Pi_1,\delta_1) := \frac{2j_3(\sigma_1,\xi_1)}{\xi_1^2}\biggl(4\norm{\Pi_1}^2_\PP+2\norm{w_1}_\infty^2\\
        & \hspace{40mm} +8\varrho^2\delta_1^2+ f_3(\sigma_1,\xi_1,\Pi_1) + g_6(w_1,\sigma_1,\xi_1,\delta_1)\biggr),
    \end{split} \\
    \begin{split}\label{eq: LDelta}
        &\mathscr L_{\Delta}(w_1,\sigma_1,\xi_1,w_2,\sigma_2,\xi_2,\Pi_1,\delta_1) := 3\mathscr{L}_1(w_1,\sigma_1,\xi_1,w_2,\sigma_2,\xi_2,\Pi_1,\delta_1)\\
        & \hspace{60mm} +3T\mathscr L_{\Delta B}(\sigma_1,\xi_1,\xi_2)M(w_1,\sigma_1,\xi_1,\Pi_1,\delta_1),
    \end{split}\\
    &L_\PP(\sigma_1,\xi_1) := \left(2\mathscr L_2(\xi_1,\sigma_1)\cdot j_3(\sigma_1,\xi_1)\right)^{1/2}, \label{eq: LP}\\
    &L_\R(\sigma_1,\xi_1) := \left(2\mathscr L_3(\xi_1,\sigma_1)\cdot j_3(\sigma_1,\xi_1)\right)^{1/2}.\label{eq: LR}
\end{align}

\subsection{Details on the Basis Sampling Method}\label{app: basis method}

Consider $g \in L^2([0, \infty ))$ and define the Laplace transform of its square, i.e.
\[
\mathcal{L}(g^2)(t) := \int^{\infty}_0 e^{-tx} g^2(x)dx, \quad 0 \leq t < \infty.
\]

It is clearly well-defined for all $t \geq 0$, since
\[
\int^{\infty}_0 \abs{e^{-tx} g^2(x)} dx \leq \int^{\infty}_0 \abs{g^2(x)} dx < \infty.
\]

Thus we see $0 \leq \mathcal{L}(g^2) \leq \norm{g}^2_{L^2([0,\infty))}$, for $0 \leq t < \infty$. Observe that $\mathcal{L}(g^2) \in C([0,\infty))$. Indeed,
\begin{description}
    \item[(I)] By Prop 1.4 of \cite{Laplace}, we see that $\mathcal{L}(g^2) = C^{\infty}((0,\infty))$,
    \item[(II)] Continuity at zero: $(t_n)_n \subset (0, \infty), \quad t_n \to 0$. 
\end{description}

Then $\lim_{n \to \infty} e^{-t_n x} = 1, \quad \forall x \geq 0$, and $\abs{e^{-t_n x}} \leq 1 \in L^1(g^2 dx)$. By the Dominated Convergence Theorem, we conclude
\[
\lim_{n \to \infty} \mathcal{L}(g^2)(t_n) = \mathcal{L}(g^2)(0).
\]

From now on, we will consider only values of $t \in [0,T]$ so we have
\[
L: L^2([0,\infty)) \to C([0,T]),
\]
\[
g \stackrel{L}{\mapsto} \mathcal{L}(g^2) \big| _{[0,T]}.
\]
\begin{lemma}
    $L$ is locally Lipschitz continuous.
\end{lemma}
\begin{proof}
    Let $g \in L^2([0,\infty))$ be arbitrary and pick $g_1, g_2 \in B^{L^2}_1 (g)$. Then
    \[
    \begin{split}
        \norm{L(g_2)-L(g_1)}_{\infty} &= \sup_{0 \leq t \leq T} \abs{\mathcal{L}(g^2_2)(t) - \mathcal{L}(g^2_1)(t)} \\
        &= \sup_{0 \leq t \leq T} \abs{\int^{\infty}_0 e^{-tx} [g^2_2(x)-g^2_1(x)] dx } \\
        &\leq \sup_{0 \leq t \leq T} \int^{\infty}_0 e^{-tx} \abs{g_2(x)+g_1(x)}\abs{g_2(x)-g_1(x)} dx \\
        &\leq \norm{g_2+g_1}_{L^2} \norm{g_2-g_1}_{L^2} \leq 2\norm{g_2-g_1}_{L^2} .
    \end{split}
    \] 
\end{proof}

By embedding $C([0,\infty)) \hookrightarrow L^2([0,T])$, we see that $L$ can be viewed as a map from $L^2([0,\infty))$ into $L^2([0,T])$, which is still locally Lipschitz continuous, i.e. $\forall g \in L^2([0,\infty))$ and $\forall g_1,g_2 \in B^{L^2}_1 (g)$:
\[
\norm{L(g_2)-L(g_1)}_{L^2([0,T])} \leq \norm{L(g_2)-L(g_1)}_{\infty} \leq 2 \norm{g_2 - g_1}_{L^2([0,\infty))}.
\]

Being continuous implies, in particular, that it maps compact subsets of $L^2([0,\infty))$ into compact subsets of $L^2([0,T])$. Finally, by construction, $L(g) \in \mathcal{H}, \forall g \in L^2([0,\infty))$.

\subsubsection{Compactness of the Sampling Domain} 

Theorem~\ref{Thm:UAT} gives a uniform approximation on compact subsets of the input space. Therefore, we now explain why the basis sampling method produces kernels belonging to a compact subset of $L^2([0,\infty))$.

We construct compact subsets $K$ of $L^2([0,\infty))$. Then $L(K) \subset \mathcal{H} \subset L^2([0,T])$ will be compact. We then put a probability measure $\mu$ on $K$, i.e.
\[
\mu: \mathcal{B}(K) \to [0,1].
\]

Since $L$ is continuous, it is $\mathcal{B}(L^2([0,\infty))) / \mathcal{B}(L^2([0,T]))$ measurable. Then $\mu_L := \mu \circ L^{-1}$ defines a probability measure on $(L(K),\,  \mathcal{B}(L(K))$.

Now let $(\beta_j)^{\infty}_{j=1}$ be an orthonormal basis of $L^2([0,\infty))$. Choose $a=(a_j)^{\infty}_{j=1} \in \ell^2(\mathbb{N})$ with $a_j \geq 0$. Then
\[
K_a := \{g \in L^2([0,\infty)) \big| \abs{\langle g,\beta_J \rangle} \leq a_J, \, J \in \mathbb{N} \}
\]
is compact.

\subsubsection{Construction of Orthonormal Basis for $L^2([0,\infty))$}

To construct an orthonormal basis $(\beta_j)^{\infty}_{j = 1}$ for $L^2([0,\infty))$, we can start from any orthonormal basis $(e_j)^{\infty}_{j = 1}$ of $L^2([0,1])$, such as sine/cosine functions. We then apply the unitary map
\[
U: L^2([0,1]) \to L^2([0,\infty)), \quad (Uf)(x) := e^{-x/2} f(e^{-x}), \ x \ge 0.
\]
Since $U$ is unitary, we have
\[
\langle Uf_1,Uf_2 \rangle_{L^2([0,\infty))} = \langle f_1,f_2 \rangle_{L^2([0,1])}, \quad f_1,f_2 \in L^2([0,1]),
\]
and therefore if $(e_j)^{\infty}_{j=1}$ is an orthonormal basis for $L^2([0,1])$, $(Ue_j)^{\infty}_{j=1}$ is an orthonormal basis for $L^2([0,\infty))$.

In practice, we will proceed by considering for each $j \in \mathbb{N}$ a random variable $Y_j$ uniformly distributed on ${[-a_j,a_j]}$. Let $Z$ follow a Poisson distribution with a given intensity parameter. Assume $\{ Y_j; j \in \mathbb{N} \} \cup \{ Z \}$ is an independent collection of random variables.

Define the $L^2([0,\infty))$-valued random variable
\[
X(\omega) := \sum^{Z+1}_{j=1} Y_j(\omega) \beta_j, \quad \omega \in \Omega,
\]
and set $\mathbb{P}_X := \mathbb{P} \circ X^{-1}: \mathcal{B}(L^2([0,\infty))) \to [0,1]$. 

Clearly, $\mathbb{P}_X [K_a] = 1$. Therefore, if $B \in \mathcal{B}(L^2 ([0,T]))$,
\[
\mathbb{P}_{L \circ X} [B] = \mathbb{P}_X [L^{-1}(B)] = \mathbb{P}_X [L^{-1}(B) \cap K_a].
\]

Pick $B=L(K_a) \in \mathcal{B}(L^2([0,T]))$ to get 
\[
\begin{split}
    \mathbb{P}_{L \circ X} [L(K_a)] &= \mathbb{P}_X [L^{-1}(L(K_a)) \cap K_a] \\
    &= \mathbb{P}_X[K_a] \\
    &= 1,
\end{split}
\]
i.e. $\mathbb{P}_{L \circ X}$ is supported on $L(K_a)$.

Finally, observe that we do not need to specify the whole orthonormal basis $(\beta_j)^{\infty}_{j=1}$. We can use classical choices or even ``construct'' our basis, for example, by considering $N$ functions $g_1, \ldots, g_N \in L^2 ([0,\infty))$ such that $\Supp (g_j)_{j=1}^{N}$ is compact and $\Supp (g_j) \cap \Supp (g_{j'}) = \varnothing$ for $j \neq j'$. 

Clearly $\int^{\infty}_0 g_j(x) g_{j'}(x)dx = \delta_{j,j'} \norm{g_j}^2_{L^2([0,\infty))}$, so upon re-normalising, $\{g_1,\ldots,g_N\}$ are orthonormal. From elementary Hilbert space theory, we can find a basis $(\beta_j)^{\infty}_{j=1}$ which contains our ${g_1,\ldots,g_N}$, but we do not need to specify it because it is enough to work with the original ${g_1,\ldots,g_N}$ as long as the rate of the Poisson r.v. $Z$ is chosen so that $\mathbb{P}[Z>N-1] \ll 1$.

Thus, we have the following algorithm.
\begin{description}
    \item[(I)] Sample ${g_1,\ldots,g_j} \in K_a$ according to $\mathbb{P}_X$;
    \item[(II)] Compute their Laplace transform and get the $H_1, \ldots, H_J$, i.e.
    \[
    \begin{split}
        H_1(t) = \mathcal{L}(g_1^2)(t) &= \int^{\infty}_0 e^{-tx}g_1^2(x)dx, \\
        &\vdots\\
        H_J(t) = \mathcal{L}(g_J^2)(t) &= \int^{\infty}_0 e^{-tx}g_J^2(x)dx, \quad 0 \le t \le T;
    \end{split}
    \]
    \item[(III)] Obtain the kernels :
    \[
    \begin{split}
        G_1(t,s) &= \1_{s<t} H_1(t-s), \\
        &\vdots \\
        G_J(t,s) &= \1_{s<t} H_J(t-s). \\
    \end{split}
    \]
\end{description}

\subsection{Complete Results of Robustness Check of Kernel Misspecification}

In Table \ref{tab:rnn-regret-robustness} are the complete results of the robustness check of kernel misspecification.
\begin{table}[h!]
    \centering
    \small
    \setlength{\tabcolsep}{4pt}
    \begin{tabular}{lrrlrr}
    \hline
    Parameter & Baseline & Tested & Fit & Mean regret (bps) & 95th percentile (bps) \\
    \hline
    Baseline model & -- & -- & Match $H(0)$ & 0.4922 & 2.0514 \\
    Baseline model & -- & -- & Free $A$      & 0.4961 & 1.6308 \\

    $\lambda$ & 0.010 & 0.005 & Match $H(0)$ & 2.8319 & 7.7011 \\
    $\lambda$ & 0.010 & 0.005 & Free $A$      & 2.7941 & 6.7893 \\
    $\lambda$ & 0.010 & 0.020 & Match $H(0)$ & 0.3704 & 0.9352 \\
    $\lambda$ & 0.010 & 0.020 & Free $A$      & 0.3145 & 1.0677 \\

    $\phi$ & 0.500 & 0.250 & Match $H(0)$ & 1.0202 & 3.7731 \\
    $\phi$ & 0.500 & 0.250 & Free $A$      & 1.0914 & 4.3295 \\
    $\phi$ & 0.500 & 1.000 & Match $H(0)$ & 0.2684 & 0.6087 \\
    $\phi$ & 0.500 & 1.000 & Free $A$      & 0.2673 & 0.6509 \\

    $\varrho$ & 5.000 & 2.500 & Match $H(0)$ & 0.4825 & 1.9864 \\
    $\varrho$ & 5.000 & 2.500 & Free $A$      & 0.4702 & 1.9471 \\
    $\varrho$ & 5.000 & 10.00 & Match $H(0)$ & 0.7781 & 1.7621 \\
    $\varrho$ & 5.000 & 10.00 & Free $A$      & 0.8114 & 1.9495 \\

    $\gamma$ & 0.300 & 0.100 & Match $H(0)$ & 0.3564 & 0.9751 \\
    $\gamma$ & 0.300 & 0.100 & Free $A$      & 0.3468 & 1.1955 \\
    $\gamma$ & 0.300 & 1.000 & Match $H(0)$ & 0.1800 & 0.7117 \\
    $\gamma$ & 0.300 & 1.000 & Free $A$      & 0.2209 & 0.8954 \\

    $\sigma^P$ & 0.500 & 0.250 & Match $H(0)$ & 0.3400 & 1.0254 \\
    $\sigma^P$ & 0.500 & 0.250 & Free $A$      & 0.3041 & 0.9377 \\
    $\sigma^P$ & 0.500 & 1.000 & Match $H(0)$ & 0.7620 & 3.1284 \\
    $\sigma^P$ & 0.500 & 1.000 & Free $A$      & 0.7363 & 2.7987 \\

    $\sigma^\alpha$ & 0.500 & 0.250 & Match $H(0)$ & 0.2830 & 0.8374 \\
    $\sigma^\alpha$ & 0.500 & 0.250 & Free $A$      & 0.2997 & 1.1883 \\
    $\sigma^\alpha$ & 0.500 & 1.000 & Match $H(0)$ & 0.4744 & 1.3216 \\
    $\sigma^\alpha$ & 0.500 & 1.000 & Free $A$      & 0.4728 & 1.2277 \\
    \hline
    \end{tabular}
    \caption{RNN regret under one-at-a-time changes in the model parameters. The baseline column gives the parameter value used in the main experiments, while the tested column gives its value in the robustness experiment. All other parameters remain fixed at their baseline values. The 95th percentile is calculated across the kernel-level mean regrets. Each setting uses $80$ true kernels and $50$ economic paths per kernel.}
    \label{tab:rnn-regret-robustness}
\end{table}

\bibliography{literature}

@article{chen2024parametric,
  title={On parametric optimal execution and machine learning surrogates},
  author={Chen, Tao and Ludkovski, Mike and Vo{\ss}, Moritz},
  journal={Quantitative Finance},
  volume={24},
  number={1},
  pages={15--34},
  year={2024},
  publisher={Taylor \& Francis}
}

@article{macri2024reinforcement,
  title={Reinforcement learning for optimal execution when liquidity is time-varying},
  author={Macr{\`\i}, Andrea and Lillo, Fabrizio},
  journal={Applied Mathematical Finance},
  volume={31},
  number={5},
  pages={312--342},
  year={2024},
  publisher={Taylor \& Francis}
}

@article{gueant2019deep,
  title={Deep reinforcement learning for market making in corporate bonds: beating the curse of dimensionality},
  author={Gu{\'e}ant, Olivier and Manziuk, Iuliia},
  journal={Applied Mathematical Finance},
  volume={26},
  number={5},
  pages={387--452},
  year={2019},
  publisher={Taylor \& Francis}
}

@article{ning2021double,
  title={Double deep q-learning for optimal execution},
  author={Ning, Brian and Lin, Franco Ho Ting and Jaimungal, Sebastian},
  journal={Applied Mathematical Finance},
  volume={28},
  number={4},
  pages={361--380},
  year={2021},
  publisher={Taylor \& Francis}
}

@article{mounjid2024improving,
  title={Improving reinforcement learning algorithms: Towards optimal learning rate policies},
  author={Mounjid, Othmane and Lehalle, Charles-Albert},
  journal={Mathematical Finance},
  volume={34},
  number={2},
  pages={588--621},
  year={2024},
  publisher={Wiley Online Library}
}

@article{cartea2021deep,
  title={Deep reinforcement learning for algorithmic trading},
  author={Cartea, {\'A}lvaro and Jaimungal, Sebastian and S{\'a}nchez-Betancourt, Leandro},
  journal={Available at SSRN 3812473},
  year={2021}
}

@article{ekren2019portfolio,
	title={Portfolio choice with small temporary and transient price impact},
	author={Ekren, Ibrahim and Muhle-Karbe, Johannes},
	journal={Mathematical Finance},
	volume={29},
	number={4},
	pages={1066--1115},
	year={2019},
	publisher={Wiley Online Library}
}

@article{chen2019portfolio,
  title={Portfolio liquidation under transient price impact--theoretical solution and implementation with 100 NASDAQ stocks},
  author={Chen, Ying and Horst, Ulrich and Tran, Hoang Hai},
  journal={arXiv preprint arXiv:1912.06426},
  year={2019}
}

@article{bank2019optimal,
	title={Optimal investment with transient price impact},
	author={Bank, Peter and Vo{\ss}, Moritz},
	journal={SIAM Journal on Financial Mathematics},
	volume={10},
	number={3},
	pages={723--768},
	year={2019},
	publisher={SIAM}
}

@article{almgren2012optimal,
	title={Optimal trading with stochastic liquidity and volatility},
	author={Almgren, Robert},
	journal={SIAM Journal on Financial Mathematics},
	volume={3},
	number={1},
	pages={163--181},
	year={2012},
	publisher={SIAM}
}

@article{barger2019optimal,
	title={Optimal liquidation under stochastic price impact},
	author={Barger, Weston and Lorig, Matthew},
	journal={International Journal of Theoretical and Applied Finance},
	volume={22},
	number={02},
	pages={1850059},
	year={2019},
	publisher={World Scientific}
}

@article{chan2015optimal,
	title={Optimal trading with predictable return and stochastic volatility},
	author={Chan, Patrick and Sircar, Ronnie},
	journal={Available at SSRN 2623747},
	year={2015}
}

@article{cheridito2014optimal,
  title={Optimal trade execution under stochastic volatility and liquidity},
  author={Cheridito, Patrick and Sepin, Tardu},
  journal={Applied Mathematical Finance},
  volume={21},
  number={4},
  pages={342--362},
  year={2014},
  publisher={Taylor \& Francis}
}

@article{cartea2016incorporating,
	title={Incorporating order-flow into optimal execution},
	author={Cartea, {\'A}lvaro and Jaimungal, Sebastian},
	journal={Mathematics and Financial Economics},
	volume={10},
	number={3},
	pages={339--364},
	year={2016},
	publisher={Springer}
}

@article{cartea2018enhancing,
	title={Enhancing trading strategies with order book signals},
	author={Cartea, {\'A}lvaro and Donnelly, Ryan and Jaimungal, Sebastian},
	journal={Applied Mathematical Finance},
	volume={25},
	number={1},
	pages={1--35},
	year={2018},
	publisher={Taylor \& Francis}
}

@article{lehalle2019incorporating,
	title={Incorporating signals into optimal trading},
	author={Lehalle, Charles-Albert and Neuman, Eyal},
	journal={Finance and Stochastics},
	volume={23},
	number={2},
	pages={275--311},
	year={2019},
	publisher={Springer}
}

@article{lehalle2017limit,
	title={Limit order strategic placement with adverse selection risk and the role of latency},
	author={Lehalle, Charles-Albert and Mounjid, Othmane},
	journal={Market Microstructure and Liquidity},
	volume={3},
	number={01},
	pages={1750009},
	year={2017},
	publisher={World Scientific}
}

@article{sirignano2019deep,
  title={Deep learning for limit order books},
  author={Sirignano, Justin A},
  journal={Quantitative Finance},
  volume={19},
  number={4},
  pages={549--570},
  year={2019},
  publisher={Taylor \& Francis}
}

@article{cont2014price,
	title={The price impact of order book events},
	author={Cont, Rama and Kukanov, Arseniy and Stoikov, Sasha},
	journal={Journal of Financial Econometrics},
	volume={12},
	number={1},
	pages={47--88},
	year={2014},
	publisher={Oxford University Press}
}

@article{biais1995empirical,
	title={An empirical analysis of the limit order book and the order flow in the Paris Bourse},
	author={Biais, Bruno and Hillion, Pierre and Spatt, Chester},
	journal={The Journal of Finance},
	volume={50},
	number={5},
	pages={1655--1689},
	year={1995},
	publisher={Wiley Online Library}
}

@article{almgren2005direct,
	title={Direct estimation of equity market impact},
	author={Almgren, Robert and Thum, Chee and Hauptmann, Emmanuel and Li, Hong},
	journal={Risk},
	volume={18},
	number={7},
	pages={58--62},
	year={2005},
	publisher={Citeseer}
}

@article{bechler2017order,
	title={Order flows and limit order book resiliency on the meso-scale},
	author={Bechler, Kyle and Ludkovski, Mike},
	journal={Market Microstructure and Liquidity},
	volume={3},
	number={03n04},
	pages={1850006},
	year={2017},
	publisher={World Scientific}
}

@article{almgren2001optimal,
	title={Optimal execution of portfolio transactions},
	author={Almgren, Robert and Chriss, Neil},
	journal={Journal of Risk},
	volume={3},
	pages={5--40},
	year={2001}
}

@article{bertsimas1998optimal,
  title={Optimal control of execution costs},
  author={Bertsimas, Dimitris and Lo, Andrew W},
  journal={Journal of Financial Markets},
  volume={1},
  number={1},
  pages={1--50},
  year={1998},
  publisher={Elsevier}
}

@article{https://doi.org/10.1111/mafi.12465,
author = {Abi Jaber, Eduardo and Neuman, Eyal},
title = {Optimal Liquidation With Signals: The General Propagator Case},
journal = {Mathematical Finance},
volume = {n/a},
number = {n/a},
pages = {},
doi = {https://doi.org/10.1111/mafi.12465},
url = {https://onlinelibrary.wiley.com/doi/abs/10.1111/mafi.12465},
eprint = {https://onlinelibrary.wiley.com/doi/pdf/10.1111/mafi.12465}
}

@book{conway2010,
author = {Conway, John B.},
address = {New York},
edition = {2nd},
isbn = {9781441930927},
language = {eng},
publisher = {Springer Science+Business Media},
series = {Graduate {T}exts in {M}athematics; 96},
title = {A {C}ourse in {F}unctional {A}nalysis},
year = {2010}
}

@article{BDG0,
  title={Neural networks in {F}r\'echet spaces},
  author={Fred Benth and Nils Detering and Luca Galimberti},
  journal={Annals of Mathematics and Artificial Intelligence},
  year={2023},
  volume={91},
  pages={75-103}
}

@book{Gripenberg_Londen_Staffans_1990, place={Cambridge}, series={Encyclopedia of Mathematics and its Applications}, title={Volterra Integral and Functional Equations}, publisher={Cambridge University Press}, author={Gripenberg, G. and Londen, S. O. and Staffans, O.}, year={1990}, collection={Encyclopedia of Mathematics and its Applications}}

@misc{meng2025incontextoperatorlearninglinear,
      title={In-Context Operator Learning for Linear Propagator Models}, 
      author={Tingwei Meng and Moritz Voß and Nils Detering and Giulio Farolfi and Stanley Osher and Georg Menz},
      year={2025},
      eprint={2501.15106},
      archivePrefix={arXiv},
      primaryClass={q-fin.TR},
      url={https://arxiv.org/abs/2501.15106}, 
}

@ARTICLE{Non-metric,
  author = {Galimberti, Luca},
  journal={Analysis and Applications}, 
  title = {Neural networks in non-metric spaces}, 
  year={2025},
  volume={0},
  pages={1-49},
  doi = {10.1142/S0219530525500514}}

@article{hoon2023minimal,
  title={Minimal width for universal property of deep RNN},
  author={hoon Song, Chang and Hwang, Geonho and ho Lee, Jun and Kang, Myungjoo},
  journal={Journal of Machine Learning Research},
  volume={24},
  number={121},
  pages={1--41},
  year={2023}
}

@inproceedings{sperduti2015equivalence,
  title={Equivalence Results between Feedforward and Recurrent Neural Networks for Sequences.},
  author={Sperduti, Alessandro and others},
  booktitle={IJCAI},
  pages={3827--3833},
  year={2015}
}

@misc{Laplace,
  author = "Ante Mimica",
  title  = "Laplace transform",
  url = {https://web.math.pmf.unizg.hr/~amimica/lt.pdf},
  howpublished = "\url{https://web.math.pmf.unizg.hr/~amimica/lt.pdf}",
  year   = 2015
}

@article{king2022king,
  title={King’s computational research, engineering and technology environment (CREATE)},
  author={{King’s College London}},
  journal={Retrieved March},
  volume={2},
  pages={2022},
  year={2022}
}
\bibliographystyle{abbrv}

\end{document}